\documentclass[12pt]{article}
\usepackage{amsmath}
\usepackage{graphicx}
\usepackage{enumerate}
\usepackage{natbib}
\usepackage{authblk}
\usepackage{url} % not crucial - just used below for the 
\usepackage{titlesec}

\titlespacing*{\section}{0pt}{0\baselineskip}{0\baselineskip}
\titlespacing*{\subsection}{0pt}{0\baselineskip}{0\baselineskip}
\usepackage{natbib}
\usepackage{url}
\usepackage{enumerate}
\usepackage{mathtools}
\usepackage{indentfirst}

\usepackage[colorlinks=true,citecolor=blue,linkcolor=blue]{hyperref}
\usepackage{setspace}
\usepackage{caption}
\newcommand{\vtt}[1]{%
  \text{\normalfont\ttfamily\detokenize{#1}}%
}
\usepackage[noend]{algorithmic}
\newlength\myindent
\newcommand\bindent{%
  \begingroup
  \setlength{\itemindent}{\myindent}
  \addtolength{\algorithmicindent}{\myindent}
}
\usepackage{subfig}
\usepackage{tikz}
\usetikzlibrary{decorations.pathreplacing}

\newcommand\eindent{\endgroup}
\usepackage{mathrsfs}
\usepackage{bbm}
\usepackage{tabularx}      
\usepackage{multirow}
\usepackage{amsmath,amsfonts,amssymb}
\usepackage{algorithm}
\usepackage{amsthm}
\usepackage[utf8]{inputenc}
\def\spacingset#1{\renewcommand{\baselinestretch}%
{#1}\small\normalsize} \spacingset{1.9}
\newtheorem{lemma}{Lemma}
\newtheorem{theorem}{Theorem}

\newcommand{\blind}{1}

\expandafter\def\expandafter\normalsize\expandafter{%
    \normalsize%
    \setlength\abovedisplayskip{5pt}%
    \setlength\belowdisplayskip{5pt}%
    \setlength\abovedisplayshortskip{-5pt}%
    \setlength\belowdisplayshortskip{2pt}%
}

\begin{document}

\def\spacingset#1{\renewcommand{\baselinestretch}%
{#1}\small\normalsize} \spacingset{1}

%%%%%%%%%%%%%%%%%%%%%%%%%%%%%%%%%%%%%%%%%%%%%%%%%%%%%%%%%%%%%%%%%%%%%%%%%%%%%%

\if1\blind
{
  \title{\bf Gaussian Copula Models for Nonignorable Missing Data Using Auxiliary Marginal Quantiles}
  \author[1]{Joseph Feldman}
  \author[1]{Jerome P. Reiter}
  \author[2,3]{Daniel R. Kowal\footnote{The findings and conclusions in this paper are those of the
authors and do not necessarily represent the views of the North Carolina
Department of Health and Human Services, Division of Public Health} 
  }
  \affil[1]{Department of Statistical Science, Duke University}
  \affil[2]{Department of Statistics and Data Science, Cornell University}
\affil[3]{Department of Statistics, Rice University}
\date{}
  \maketitle
} \fi

\if0\blind
{
  \bigskip
  \bigskip
  \bigskip
  \begin{center}
    {\LARGE\bf Gaussian Copula Models for Nonignorable Missing Data Using Auxiliary Marginal Quantiles}
\end{center}
  \medskip
} \fi

\bigskip
\begin{abstract}
  We present an approach for modeling and imputation of nonignorable missing data. Our approach uses Bayesian data integration to combine (1) a Gaussian copula model for all study variables and missingness indicators, which allows arbitrary marginal distributions, nonignorable missingess, and other dependencies, and (2) auxiliary information in the form of marginal quantiles for some study variables. We prove that, remarkably, one only needs a small set of accurately-specified quantiles to estimate the copula correlation consistently. The remaining marginal distribution functions are inferred nonparametrically and jointly with the copula parameters using an efficient MCMC algorithm. We also characterize the (additive) nonignorable missingness mechanism implied by the copula model. Simulations confirm the effectiveness of this approach for multivariate imputation with nonignorable missing data. We apply the model to analyze associations between lead exposure and end-of-grade test scores for 170,000 North Carolina students. Lead exposure has nonignorable missingness: children with higher exposure are more likely to be measured. We elicit marginal quantiles for lead exposure using statistics provided by the Centers for Disease Control and Prevention. Multiple imputation inferences under our model support stronger, more adverse associations between lead exposure and educational outcomes relative to complete case and missing-at-random analyses.
\end{abstract}

\noindent%
{\it Keywords:} Bayesian inference; imputation; MNAR; data integration; nonresponse
\vfill

\newpage
\spacingset{1.9} % DON'T change the spacing!

\section{Introduction}\label{sec:intro}

The Gaussian copula is a flexible joint distribution for multivariate data. 
%continuous data. 
%While the model relies on simple primitives, it 
The model can characterize complex dependencies %among variables 
while also  capturing non-Gaussian marginal distributions.
%It is particularly effective for prediction and inference for non-Gaussian data: the model \eqref{cop-sample}-\eqref{xform} demonstrates its ability to capture 
%heterogeneous marginal distributions while the copula correlation is an interpretable parameter determining multivariate relationships. 
Methodological advances have made the Gaussian copula compatible with mixed data types \citep{hoff2007extending, feldman2022bayesian}, increased its scalability to high-dimensional variable sets \citep{murray2013bayesian}, and improved its ability to capture non-linearity and interactions \citep{feldman2022nonparametric}.  Because of these appealing features, it has been deployed in numerous applications,  
%including, 
for example, in economics and finance \citep{fan:patton}, marketing and management \citep{eckert2023addressing}, and political science \citep{chiba:etal}.

The Gaussian copula can be readily implemented when data have missing values. Indeed, researchers \citep[e.g., ][]{kaarik, dilascioetal, gcimputeR, hollenbachetal,sbgcop, mdgc} have suggested Gaussian copulas as multivariate, joint modeling engines for (multiple) imputation. Existing methods, however, tend to assume that missingness among the study variables is  missing completely at random (MCAR) or missing at random (MAR) \citep{rubin76}. We are not aware of methodology for itemwise nonignorable missingness 
%\citep{linero2018bayesian} 
with Gaussian copula models.

In this article, we develop a framework to handle nonignorable missing data in Gaussian copula models. Our general strategy is to specify a Bayesian Gaussian copula model that includes both the study variables and the missingness indicators, coupled with auxiliary information on the marginal distributions of the study variables that we integrate directly into the model.  
%As part of the developments, we prove that 
%the marginal distribution functions of the study variables are the key piece of identifying information which enables estimation of the Gaussian copula under nonignorable missingness. We also connect the posited 
%The auxiliary information is crucial; without it, estimates of the copula correlation can be badly biased in the presence of nonignorable missing data, as we demonstrate.  In fact, we
Using this strategy, we prove that
 %This then provides the rationale for our primary methodological and theoretical result: under AN missingness, 
 analysts need to know only a small, arbitrary set of auxiliary marginal quantiles to enable consistent estimation of the copula correlation; entire distributions are not necessary.
% We provide algorithms for using the quantiles to approximate the Gaussian copula and impute  missing values under nonignorable missingness.  
\textcolor{black}{Further, we show that the copula model encodes a specific identifying assumption about the missingness mechanism, namely that it is   
%When the data copula model These results hold for 
a type of additive nonignorable (AN) mechanism \citep{Hirano2001, sadinle2019sequentially}.}
%\textcolor{black}{which identifies the parameters of the Gaussian copula under nonignorable missingness}. 
%This mechanism allows the reason for missingness in a variable potentially to depend on the value of the variable itself. 
We develop algorithms for estimating the copula correlation that result in significant computational gains relative to alternative copula models, which allows fitting to scale to data with (moderately) large sample sizes in reasonable time using typical computational setups. We also present strategies for estimating other quantiles of the marginal distributions beyond those in the auxiliary information. Our approach provides uncertainty quantification for the unknown marginals, which we utilize for multiple imputation \citep{rubin1987multiple}.
%with nonignorable missing data.
%while requiring no further model assumptions and minimal computation.

% We demonstrate empirically that the approximation can have minimal impact on the accuracy of  estimates of the copula model parameters or the quality of  imputations of missing values.
%In particular, we show that that, under certain conditions and with known marginal distributions, the Gaussian copula implies an additive nonignorable missingness mechanism \citep{Hirano2001, sadinle2019sequentially}. 

We develop the methodology for settings where marginal quantiles for variables subject to missingness are available in  
%auxiliary information about marginal distributions in 
external data sources.   For example, previewing the application in Section \ref{sec:realdat}, suppose an analysis involving health measurements suffers from nonignorable missingness, e.g., people likely to have unconcerning values are not measured. For many health measurements, there exists information on marginal quantiles from national benchmark surveys or administrative databases.  We seek to use the auxiliary quantiles to adjust for nonignorable missingness in the analysis at hand.  When auxiliary quantiles are not available or known precisely, 
%These facts also enable a convenient approach to sensitivity analysis: 
analysts  can posit different plausible marginal quantiles and assess the sensitivity of ultimate analyses under those specifications, e.g., via multiple imputation analyses.  We note that the use of known marginal distributions for imputation of nonignorable missing data also has been proposed for categorical data models  \citep{pham, akande2021leveraging, DengEtAl2013, si:reiter:hillygus15, si:reiter:hillygus16, tang2024using}, but not for continuous and mixed variables like we do here.

We apply the methodology in data comprising health, socioeconomic, demographic and educational 
%cognitive 
%development 
measurements collected on over 170,000 children residing in the state of North Carolina. Among the study variables are end-of-grade math and reading test scores and a measure of lead exposure, the latter of which is subject to abundant missingness that is likely nonignorable.  The state requires children at high risk of lead exposure to be measured, but not children at low risk of lead exposure. To inform the imputation of the missing values, we leverage marginal quantiles on lead exposure published by the U.S.\ Centers for Disease Control and Prevention. Using this information, we find   
%of the association between lead exposure and cognitive development, as measured by EoG test scores, 
lead exposure apparently is more adversely associated with test scores 
%appears to underestimate the 
%adverse association of lead exposure with test scores 
than suggested from complete cases or  MAR analyses.
%that account for the nonignorable missing data using the Gaussian copula model. 

The remainder of this article is organized as follows. In Section \ref{sec:gcmd}, we present the Gaussian copula model for nonignorable missing data which leverages information on marginal distributions. Here we present two theoretical results, namely (i) that  fully specified marginal distributions provide
 posterior consistency of the copula correlation in the presence of nonignorable missingness, and (ii) a derivation that the Gaussian copula implies a version of the AN missingness mechanism. In Section \ref{auxapprox}, we modify the copula model and develop imputation strategies for settings where the auxiliary marginal information constitutes an incomplete set of marginal quantiles. \textcolor{black}{Here we present our main theorem: under the AN missingness mechanism, this limited auxiliary information still allows consistent estimation of the copula correlation.}
%in lieu of full distribution.
In Section \ref{sec:sims}, we present simulations studying the effect of differing amounts of auxiliary quantiles and repeated sampling properties of multiple imputation inferences with the model. %compared to using a Gaussian copula that uses a MAR mechanism. 
In Section \ref{sec:realdat}, we present the analysis of the North Carolina lead exposure data.  In Section \ref{sec:concl}, we summarize and suggest research directions. \textcolor{black}{Supplementary material includes proofs of all theorems and additional results from simulations and the North Carolina data analysis.} Codes for all analyses are available at \href{https://github.com/jfeldman396/EHQL-Impute}{https://github.com/jfeldman396/EHQL-Impute}.  

\section{Gaussian Copula with Nonignorable Missing Data} \label{sec:gcmd}

For $i=1, \dots, n$ individuals, let $\boldsymbol y_i = (y_{i1}, \dots, y_{ip})$ comprise measurements on $p$ study variables.  Let $\boldsymbol y = \{\boldsymbol y_i: i = 1, \dots, n\}$. 
%We assume $\boldsymbol Y$ has missing values.  Let 
%$R_{ij}$ denote the missingness indicator for the corresponding study variable $Y_{ij}$, where 
When $\boldsymbol{y}$ contains missing values, let %the random variable  
$r_{ij} = 1$ when $y_{ij}$ is missing, and $r_{ij}=0$ otherwise.  Let  $\boldsymbol r_i = \{r_{i1}, \dots, r_{ip}\}$ for  $i=1, \dots, n$, and  $\boldsymbol r = \{\boldsymbol r_{i}: i = 1, \dots, n\}$.  We refer to  the study variables using $Y=(Y_1, \dots, Y_p)$ and nonresponse indicators using $R=(R_1, \dots, R_p)$. 
When the missingness is nonignorable, we require a model for the joint distribution of $(\boldsymbol y, \boldsymbol r)$. To aid specification of this distribution, we partition the data into observed  and missing components, $\boldsymbol y = (\boldsymbol y^{obs}, \boldsymbol y^{mis})$, where  $\boldsymbol{y}^{obs} = \{y_{ij}: r_{ij}=0; i=1, \dots, n; j=1, \dots,p\}$ and  $\boldsymbol{y}^{mis} = \{y_{ij}: r_{ij}=1; i=1, \dots, n; j=1, \dots,p\}$.     
%We favor a  Bayesian joint model, which  specifies a likelihood for the \textit{observed data} ($\boldsymbol Y^{obs}, \boldsymbol R$).
%and follow the  framework of \cite{linero2018bayesian}.  
Then, the two principal modeling tasks are to specify the 
%working 
distribution of the observed data,  %$(\boldsymbol{Y}^{obs}, \boldsymbol R)$,
 \begin{equation}\label{work}
     p(\boldsymbol y^{obs}, \boldsymbol r\mid \boldsymbol \gamma) = \int p(\boldsymbol{y}^{obs}, \boldsymbol y^{mis}, \boldsymbol r \mid \boldsymbol \gamma) d\boldsymbol y^{mis},
 \end{equation}
and to impose some identifying restriction on 
%on the conditional distribution of the missing study variables, 
$p(\boldsymbol{y}^{mis} \mid \boldsymbol y^{obs}, \boldsymbol r, \boldsymbol \gamma)$, also known as the extrapolation model \citep{linero2018bayesian}. \textcolor{black}{Here, $\boldsymbol \gamma$ are  parameters of the model for $(\boldsymbol y, \boldsymbol r)$}. 
%The working model is for the observed data $(\boldsymbol{Y}^{obs}, \boldsymbol R)$, 
%We accomplish these tasks by modifying the Gaussian copula with a simple and flexible approach. 
%First, we expand \eqref{cop-sample} and \eqref{xform} to jointly model $\boldsymbol Y^{obs}, \boldsymbol R$. 
We accomplish these tasks via a Gaussian copula specification, which we now describe.

%for $(\boldsymbol Y, \boldsymbol R)$.  
%Before presenting the distributions for the model, we describe how it corresponds to data generation.  For each %Gaussian copula models each 
%$\boldsymbol y_i$ is realized by first simulating latent  $\boldsymbol{z}_i$ from a multivariate normal distribution with some correlation 
%$\boldsymbol{C}_{\boldsymbol Y}$.
%$\boldsymbol{C}_{\boldsymbol \theta}$.
%Each component of the latent vector is transformed to the observed data scale by applying the standard normal cumulative distribution function $\Phi$ and the inverse of marginal distribution function for $Y_{j}$. 
For each $(\boldsymbol y_{i}, \boldsymbol r_i)$, let $\boldsymbol z_{i} = (\boldsymbol z_{\boldsymbol y_i}, \boldsymbol z_{\boldsymbol r_i})$ be a $2p \times 1$ vector of latent variables.  Here, $\boldsymbol z_{\boldsymbol y_i} = (z_{i1}, \dots, z_{ip})$ and $\boldsymbol z_{\boldsymbol r_i} = (z_{i(p+1)}, \dots, z_{i(2p)})$.  %For each $R_{ij}$, we introduce standard normal latent random variables  
 Let $\boldsymbol \alpha = (\mathbf{0}_{p}, \alpha_{r_1}, \dots, \alpha_{r_p})$ be a $2p \times 1$ vector, where  $\mathbf{0}_{p}$ is a vector of $p$ zeros corresponding to  $\boldsymbol z_{\boldsymbol y_i}$ and the next $p$ elements correspond to  $\boldsymbol z_{\boldsymbol r_i}$.  Let $\boldsymbol C_{\boldsymbol \theta}$ be a $2p \times 2p$ copula correlation matrix. With generality, we let $\boldsymbol \theta$ include parameters that generate a copula correlation; see  %The prior distribution on $\boldsymbol \theta$ induces a prior over $\boldsymbol C_{\boldsymbol \theta}$.  We present the prior specification for our analyses in
 Section \ref{sec:sims} for the specification in our analyses. For $j=1, \dots, p$, let $F_{j}$ be the marginal distribution for $Y_j$. To begin, we assume that each $Y_j$ is continuous. Modifications for discrete $Y_j$ are introduced in Section \ref{auxapprox}. 
 %and \textcolor{black}{%for categorical variables  in Section D of the supplement}.  
 The data generating model is then
\begin{align}
    %\boldsymbol z_{\boldsymbol y_i}, \boldsymbol z_{\boldsymbol r_i}
    \boldsymbol z_i \sim &N_{2p}(\boldsymbol \alpha, \boldsymbol C_{\boldsymbol \theta})\label{YRcop}\\
    y_{ij} = F_{j}^{-1}\{\Phi(z_{y_{ij}})\}&;  \quad  r_{ij} = \mathbbm{1}_{z_{i(p+j)} >0}, \label{xformYR}
    % \begin{cases}
    % \begin{cases}
    %     1 & z_{i(p+j)} >0\\
    %     0 & z_{i(p+j)}\leq 0.
    % \end{cases}\label{xformYR}
\end{align}
where $\mathbbm{1}_{e}=1$ when the expression $e$ in its index is true and $\mathbbm{1}_{e}=0$ otherwise.
The $\boldsymbol C_{\boldsymbol \theta}$ captures multivariate dependence among the study variables themselves as well as between study variables and nonresponse indicators. \textcolor{black}{The latter correlations can encode potentially nonignorable missingness, as discussed in Section \ref{sec:AN}.} The values of  $(\alpha_{r_1}, \dots, \alpha_{r_p})$ model the marginal probabilities of missingness.  
For the study variables, each $z_{ij}$ is transformed to the observed data scale by applying the standard normal cumulative distribution function $\Phi$ and the inverse of the marginal distribution function for $Y_{j}$. For the nonresponse indicators, each $z_{i(p+j)}$ satisfies a probit data augmentation \citep{chib1998analysis}.

% , whereby $r_{ij} =1$ implies $z_{i(p+j)} >0$ and $r_{ij} =0$ implies $z_{i(p+j)} \leq 0$.  
%\textcolor{black}{We note that when missingness in some $Y_j$ is MCAR, we need not include terms for $R_j$ in the model.  INCLUDE THAT SETNENCE HERE? DO } 

%The latter corresponds to a specific nonignorable missingness mechanism, as we show in Section \ref{sec:AN}.

%The data  $\boldsymbol y_i$ for individual $i$ is realized by first simulating latent  $\boldsymbol{z}_i$ from a multivariate normal distribution with correlation $\boldsymbol{C}_{\boldsymbol \theta}$. Each component of the latent vector is transformed to the observed data scale by applying the standard normal cumulative distribution function $\Phi$ and the inverse of marginal distribution function for $Y_{j}$. 
%$F_{j}^{-1}$. 
%that is, first note that the relationship between the study variables and their corresponding latent representation is known for any value $y_{ij}$: $z_{ij} = \Phi^{-1}\{F_{j}(y_{ij})\}$. 

%Thus, a Gaussian copula model is parameterized by the copula correlation matrix $\boldsymbol C_{\boldsymbol \theta}$ which characterizes multivariate dependence, and marginals $\{F_{j}\}_{j=1}^{p}$ which capture univariate features in the data.
For prediction or imputation under copula models with no missing or MCAR data, analysts typically estimate each $F_j$ from $\boldsymbol y^{obs}$ using the empirical CDF \citep{hoff2007extending, feldman2022bayesian,  gcimputeR} or with 
%parametric or nonparametric 
some model \citep{pitt2006efficient, feldman2022nonparametric}.  With nonignorable missing data, however, \textcolor{black}{estimates of
%s of 
$\{F_j\}_{j=1}^p$ computed from $\boldsymbol y^{obs}$ can be biased, which in turn affects the quality of imputations of $\boldsymbol y^{mis}$.}
% %are not identifiable 
% from $\boldsymbol y^{obs}$ alone.  Yet, $\{F_j\}_{j=1}^p$ are requiredthe model needs \textcolor{black}{reliable} marginal information to impute \textcolor{black}{sensible values of } $\boldsymbol y^{mis}$. 
Additionally, inference for $\boldsymbol C_{\boldsymbol \theta}$  from  $(\boldsymbol y^{obs}, \boldsymbol r)$ alone may be biased (Section \ref{sim1}). 

%Therefore, 
\textcolor{black}{We address these potential problems by utilizing 
%for any $Y_j$ subject to nonignorable missingness,  
%analysts must 
%the analyst has 
%find
%some source of 
auxiliary information about each   
%The analyst fixes a set of auxiliary information 
$F_j$.}
%which carries potential implications for the dependence structure under the copula model.
We write this auxiliary information set as $\mathcal{A}_{j}$ and, across study variables, as $\mathcal{A} = \{\mathcal{A}_{j}\}_{j=1}^{p}$. 
As a first step, we presume $\mathcal{A}_{j} = F_{j}$ for all $j$, i.e., the full marginal distributions are known.
%which corresponds to and incorporate this information into the copula model. 
 In this case, we show that $\boldsymbol C_{\boldsymbol \theta}$ can be consistently estimated despite the presence of nonignorable missing data (Section \ref{sec:fullmarginals}) and  
 that the Gaussian copula implies a version of an AN missingness mechanism (Section \ref{sec:AN}). 
 %In addition, analysts can specify different collections of auxiliary quantiles to facilitate sensitivity analyses under different extrapolation models.  
We leverage these results in Section \ref{auxapprox} %for settings where 
when $\mathcal{A}$ comprises incomplete information on the marginal distributions.   

In what follows, we include a $R_j$ for each $Y_j$ in the modeling. When $Y_{j}$ is considered  MCAR or has no missing values, analysts can remove $R_{j}$ from \eqref{YRcop} and \eqref{xformYR}, which
 reduces the dimension of $\boldsymbol z_i$.

\subsection{Results with Complete Marginals}\label{sec:fullmarginals}

%The Gaussian copula model for $(\boldsymbol Y^{obs}, \boldsymbol R)$ simplifies when the data generating margins are known. When $\mathcal{A}_{j} = F_{j}$ for each study variable we have $z_{ij} = \Phi^{-1}\{F_{j}(y_{ij})\}$. 

Let $\boldsymbol z_{\boldsymbol y} = \{\boldsymbol z_{\boldsymbol y_i}: i = 1, \dots, n\}$ and $\boldsymbol{z}_{\boldsymbol{r}} = \{\boldsymbol z_{\boldsymbol r_i}: i = 1, \dots, n\}$ be the 
%collections of 
latent variables for the study variables and nonresponse indicators, respectively. 
For convenience, we partition $\boldsymbol z_{\boldsymbol y} = (\boldsymbol z^{obs}, \boldsymbol z^{mis})$ into observed and missing components corresponding to $(\boldsymbol y^{obs}, \boldsymbol y^{mis})$. 
%Here,  $\boldsymbol{Z}^{obs} = \{\Phi^{-1}\{F_{j}(y_{ij})\}: R_{ij}=0; i=1, \dots, n; j=1, \dots,p\}$ and  $\boldsymbol{Z}^{mis} = \{\Phi^{-1}\{F_{j}(y_{ij})\}: R_{ij}=1; i=1, \dots, n; j=1, \dots,p\}$.  
%Let $\boldsymbol{Z}_{\boldsymbol{R}}  = \{Z_{ij}: i=1, \dots, n; j=(p+1), \dots, 2p\}$.
Define the  set restriction $\mathcal{E}(\boldsymbol r)$ as the condition that $\boldsymbol z_{\boldsymbol r}$ satisfies the probit constraints in \eqref{xformYR} for \textcolor{black}{the realized} $\boldsymbol{r}$  across $i = 1,\dots, n$ and $j = p+1,\dots, 2p$.  \textcolor{black}{Because each $F_j$ is  known from $\mathcal{A}$, the latent variable corresponding to each $y_{ij}^{obs}$ is fixed to $z_{ij}^{obs} = \Phi^{-1}\{F_{j}(y_{ij}^{obs})\}$. Thus, by conditioning on $\{F_{j}\}_{j=1}^{p}$,} 
%$Y_{j}$ is continuous,
 %the unknown parameters are the copula correlation $\boldsymbol C_{\boldsymbol \theta}$ and $\boldsymbol \alpha$, and 
 \eqref{work} becomes
\begin{align}
    p(\boldsymbol{y}^{obs}, \boldsymbol r \mid \boldsymbol  C_{\boldsymbol \theta},  \boldsymbol \alpha, \{F_{j}\}_{j=1}^{p}) &=
       p\{\boldsymbol z^{obs}, \boldsymbol z_{\boldsymbol r} \in \mathcal{E}(\boldsymbol r) \mid \boldsymbol C_{\boldsymbol \theta}, \boldsymbol \alpha\}\label{copwork1}\\
       &= \int \int_{\boldsymbol z_{\boldsymbol r}\in \mathcal{E}(\boldsymbol r)} \phi_{2p}(\boldsymbol{z}^{obs},\boldsymbol{z}^{mis}, \boldsymbol{z}_{\boldsymbol r}; \boldsymbol{C}_{\boldsymbol \theta}, \boldsymbol \alpha) d \boldsymbol z_{\boldsymbol r} d \boldsymbol z^{mis}, \label{integrand}
 \end{align}
where $\phi_{2p}$ is the density of a $2p$-dimensional multivariate normal distribution with covariance $\boldsymbol C_{\boldsymbol \theta}$ and mean $\boldsymbol \alpha$. 

With nonignorable missing data, typically one cannot estimate model parameters consistently unless one knows the full distribution of $\boldsymbol{y}^{mis}$.  However, 
%we need far less information
as stated formally in Theorem \ref{postconsist}, 
when the \textcolor{black}{study variables and missingness indicators} follow a copula model,  
%\textcolor{black}{data and the missingness mechanism is as  described in Section \ref{sec:AN}}: 
knowledge of the true 
%the true marginal distributions 
$\{F_{j}\}_{j=1}^{p}$ 
is sufficient information 
to ensure consistent estimation of the copula correlation.  
%in the presence of nonignorable missing data (assuming the copula model accurately describes the joint distribution).} 
%This is stated in Theorem \ref{postconsist}, which is proved in the supplement.  
For a fixed sample $(\boldsymbol y^{obs}, \boldsymbol r)$, the posterior distribution of $(\boldsymbol C_{\boldsymbol \theta}, \boldsymbol{\alpha})$ with known marginals is 
\begin{equation} \label{workpost}
     p(\boldsymbol C_{\boldsymbol \theta}, \boldsymbol{\alpha}\mid \boldsymbol y^{obs}, \boldsymbol r, \{F_j\}_{j=1}^p)\propto   p(\boldsymbol z^{obs}, \boldsymbol z_{\boldsymbol r} \in \mathcal{E}(\boldsymbol r) \mid \boldsymbol C_{\boldsymbol \theta}, \boldsymbol \alpha)
       p(\boldsymbol C_{\boldsymbol \theta}, \boldsymbol \alpha).
\end{equation}
%W\begin{equation} \label{workpost}
 %    p(\boldsymbol C_{\boldsymbol \theta}|\boldsymbol y^{obs}, \boldsymbol r, \{F_j\}_{j=1}^p)\propto    \int \int_{\boldsymbol z_{\boldsymbol r} \in \mathcal{E}(\boldsymbol r)} \phi_{2p}(\boldsymbol{z}^{obs},\boldsymbol{z}^{mis}, \boldsymbol{z}_{r}; \boldsymbol{C}_{\boldsymbol \theta})d \boldsymbol z^{mis} \boldsymbol d \boldsymbol{z}_{\boldsymbol r}  \times p(\boldsymbol C_{\boldsymbol \theta}).
%\end{equation}
% \textcolor{black}{We obtain the posterior distribution of $\boldsymbol C_{\boldsymbol \theta}$ by integrating out $\boldsymbol \alpha$ in \eqref{workpost}. We refer to this marginal posterior as $\Pi_{n}(\boldsymbol C_{\boldsymbol \theta})$.}
\textcolor{black}{Subsequently, we refer to the marginal posterior of $\boldsymbol C_{\boldsymbol \theta}$ as $\Pi_{n}(\boldsymbol C_{\boldsymbol \theta})$}.
% and consider the following theoretical result:

\begin{theorem}\label{postconsist}
Suppose $\{(\boldsymbol{y}_i, \boldsymbol r_i) \}_{i=1}^{n}\overset{iid}{\sim} \Pi_{0}$ where $\Pi_{0}$ is the Gaussian copula with
    % SHOULD WE DELETE $\boldsymbol{y}^{mis}$?  \textcolor{black}{comprise $n$} independent and identically distributed samples from \eqref{YRcop}-\eqref{xformYR} with  true
    correlation $\boldsymbol C_{0}$ and  marginals $\{F_{j}\}_{j=1}^{p}$ as in \eqref{YRcop}--\eqref{xformYR}, and 
    %\textcolor{black}{where $\boldsymbol{y}^{mis}$ follows a nonignorable missingness mechanism under the copula model.}  
    $\{\mathcal{A}_{j}\}_{j=1}^{p} = \{F_{j}\}_{j=1}^{p}$. Let $p(\boldsymbol \theta)$ be a prior with respect a measure that induces a prior $\Pi$ over the space of all $2p \times 2p$ correlation matrices $\mathbbm{C}$
% $\boldsymbol{\mathbbm{C}_{\boldsymbol \theta}}$
with $\Pi(\boldsymbol C_{\boldsymbol \theta})>0$ for all $\boldsymbol C_{\boldsymbol \theta} \in \mathbbm{C}$.  Then, for all $\epsilon > 0$, $\lim_{n\rightarrow \infty}\Pi_{n}\{\mathcal{U}_{\epsilon}(\boldsymbol C_{0})\} \rightarrow 1$ almost surely $[\Pi_{0}]$, where 
    $\mathcal{U}_{\epsilon}(\boldsymbol C_{0}) = \{\boldsymbol C_{\boldsymbol \theta} \in \mathbbm{C}: \lVert \boldsymbol C_{0} - \boldsymbol C_{\boldsymbol \theta}\rVert_{F} < \epsilon\}$ and $\lVert . \rVert_{F}$ is the  Frobenius norm.
\end{theorem}
%All proofs are available in the supplement.  
%extends typical analyses of posited imputation models with nonignorable missing data, which usually concern implications for model identification.  Here, we claim that fixed and known marginals under the Gaussian copula  marginals ensure posterior consistency of the copula correlation. This 
%inherently demonstrates
Theorem \ref{postconsist} 
implies that $\boldsymbol C_{0}$ 
%the copula correlation
can be estimated consistently using the observed data $(\boldsymbol y^{obs}, \boldsymbol r)$ and the true $\{F_j\}_{j=1}^p$. 
%Further, since we can estimate $\boldsymbol C_0$ consistently using the true $\{F_j\}_{j=1}^p$
\textcolor{black}{With these quantities, we are able to 
%use the true $\{F_j\}_{j=1}^p$ and estimates of $\boldsymbol C_{\boldsymbol \theta}$ to
impute $\boldsymbol y^{mis}$ from \eqref{YRcop}--\eqref{xformYR}. In practice, the full $\{F_j\}_{j=1}^p$ may not  be known.  In this case, Theorem \ref{postconsist} 
suggests} that analysts can specify different $\{F_j\}_{j=1}^p$ to enable interpretable sensitivity analyses. \textcolor{black}{However, as we show in Section \ref{auxapprox}, analysts need far less information about the marginals for consistent estimation of $\boldsymbol C_{0}$ and subsequent implementation of multiple imputation.}

\subsection{Implied Additive Nonignorable Missingness Mechanism} \label{sec:AN}

%We now investigate properties of the Gaussian copula model when the marginals are known. 
%Fixing $\boldsymbol Y^{mis} = \boldsymbol y^{mis}$, $\boldsymbol{C} _{\boldsymbol{\theta}}$ and $\{F_j\}_{j=1}^{p}$, it becomes straightforward to write the explicit model for 

%Of course, Theorem \ref{} presumes a particular form of nonignorable missingness mechanism, which we now describe.

%While flexible, 
The Gaussian copula with known margins implies a specific  nonignorable  missingness mechanism, $p(R_{ij} = 1 \mid \boldsymbol y^{obs}, \boldsymbol y^{mis}, \boldsymbol{C} _{\boldsymbol{\theta}}, \boldsymbol \alpha, \{F_{j}\}_{j=1}^{p})$. 
% Here, $R_{ij}$ is a random variable corresponding to possible realizations of $r_{ij}$. 
%which we now derive. 
% Using the probit data augmentation for $R_{j}$, we introduce a $2p$-dimensional intercept vector $\boldsymbol \alpha$, which is non-zero for the components corresponding to $\boldsymbol R$. This additional parameter encodes marginal probabilities of $R_{j} = 1$, as in the probit regression setting.
To show this, we first note that when 
%$(\boldsymbol y, \boldsymbol r)$ 
$(Y,R)$ are distributed according \eqref{YRcop}--\eqref{xformYR}, any subset of these variables also follows a  Gaussian copula 
%with the corresponding marginals, sub-correlation, and mean
\citep{joe2014dependence}. For the joint distribution of  $(Y, R_j)$, 
%$(\boldsymbol y, r_{j})$, 
let the  corresponding $(p+1) \times (p+1)$ copula correlation matrix be $\boldsymbol C_j^*$. This  comprises the $p \times p$ sub-matrix $\boldsymbol C_{\boldsymbol y}$ \textcolor{black}{of $\boldsymbol C_{\boldsymbol \theta}$} corresponding to the study variables concatenated with the $(p+1) \times 1$ column vector $(\boldsymbol C_{\boldsymbol y r_j}, 1)^t$ and $1 \times (p+1)$ row vector $(\boldsymbol C_{r_j \boldsymbol y}, 1)$. \textcolor{black}{Here,  $\boldsymbol C_{\boldsymbol y r_j}$ comprises the entries of $\boldsymbol C_{\boldsymbol \theta}$ in the column for $R_j$, and $\boldsymbol C_{r_j \boldsymbol y}$ is its transpose.}
% denote the sub-correlation matrix $\boldsymbol C^{*}_j = \begin{bmatrix}
%     \boldsymbol{C}_{\boldsymbol{Y}} &\boldsymbol{C}_{\boldsymbol{Y}R_{j}} \\ 
%     C_{R_{j} \boldsymbol Y}
%     &1
% \end{bmatrix}\label{cstar}$
 We have  
 \begin{align}
    p(R_{ij} = 1 &\mid \boldsymbol y^{obs}, \boldsymbol y^{mis}, \boldsymbol{C} _{\boldsymbol{\theta}}, \boldsymbol \alpha, \{F_{j}\}_{j=1}^{p}) = p( R_{ij} = 1 \mid \boldsymbol y_i^{obs}, \boldsymbol y_i^{mis}, \boldsymbol{C}_{j}^{*}, \alpha_{r_{j}},\{F_{j}\}_{j=1}^{p}) \nonumber\\ &=        p(Z_{r_{ij}} >0 \mid \boldsymbol z_i^{obs}, \boldsymbol z_i^{mis}, \boldsymbol{C}_{j}^{*}, \alpha_{r_{j}}) 
    = 1- \Phi_{\alpha_{ij}^{*}, \sigma_{j}^{2*}}(0),    \label{copmis}
 \end{align} 
%this density, as we now describe.
%With fixed margins, it is possible to express joint probabilities of $(\boldsymbol y^{obs}, \boldsymbol y^{mis}, \boldsymbol r)$ in terms of corresponding latent multivariate Gaussian variables $(\boldsymbol z^{obs}, \boldsymbol z^{mis}, \boldsymbol z_{\boldsymbol r})$ with correlation $\boldsymbol{C} _{\boldsymbol{\theta}}$. As such, conditional probabilities on the observed scale are also multivariate Gaussian.
%Now, consider the missingness mechanism $p(\boldsymbol R = 1 \mid \boldsymbol y^{obs}, \boldsymbol y^{mis},\{F_{j}\}_{j=1}^{p}, \boldsymbol{C} _{\boldsymbol{\theta}})$. 
% Under the copula model, we have $p \times 1$ vector $
% \begin{align}\label{copmis}
%        p(Z_{R_{ij}} >0 \mid \boldsymbol z_i^{obs}, \boldsymbol z_i^{mis}, \boldsymbol{C}_{j}^{*}, \boldsymbol \alpha^{*}) = 1- \Phi_{\alpha_{ij}^{*}, \sigma_{j}^{2*}}(0).
% \end{align}
where $\Phi_{\alpha_{ij}^{*}, \sigma^{2*}}$ is the CDF of a Gaussian distribution with mean and variance 
\begin{equation}
      \alpha_{ij}^{*} = \alpha_{r_{j}} + \boldsymbol{C}_{r_{j}\boldsymbol y}\boldsymbol{C}_{\boldsymbol y}^{-1}(\boldsymbol z_i^{obs}, \boldsymbol z_i^{mis}),\quad \sigma_{j}^{2*} = 1 - \boldsymbol{C} _{r_{j}\boldsymbol y}\boldsymbol{C}_{\boldsymbol y}^{-1}\boldsymbol{C} _{\boldsymbol y r_{j}}. \label{moments}
\end{equation}
%derived from the properties of the multivariate Gaussian distribution. 
Therefore, the marginal missingness mechanism 
%for any $Y_{j}$ 
is a probit regression on $(\boldsymbol z_{i}^{obs}, \boldsymbol z_{i}^{mis})$ \textcolor{black}{with parameters expressed as a function of the copula correlation}.
%ransformed from $(\boldsymbol y^{obs}_i, \boldsymbol y^{mis}_{i})$ using the known marginals.

The expression for $\alpha^{*}_{ij}$ reveals a connection to the AN missingness mechanism. For any generic observation $\boldsymbol x_{i} = (x_{i1}, \dots, x_{ip})$ comprising observed and missing components $(\boldsymbol x_i^{obs}, \boldsymbol x_i^{mis})$, the AN missingness mechanism holds for some $X_j$ when 
%\begin{definition}[Additive Nonignorable Missingness Mechanism]
%A missingness mechanism is additive nonignorable if it can be written in the form 
\begin{equation}\label{AN}
    p(R_{ij} =1 \mid \boldsymbol x_i^{obs}, \boldsymbol x_i^{mis}) = g\bigg(\beta_0 + \sum_{k =1}^{p}\beta_{k} x_{ik}\bigg),
\end{equation}
with $g$ satisfying $\lim_{\beta_0\rightarrow -\infty}g(\beta_0) = 0$ and $\lim_{\beta_0 \rightarrow \infty}g(\beta_0) = 1$.  \textcolor{black}{The key feature of AN mechanisms is the additivity in $g$; that is, \eqref{AN} does not include interactions between $X_j$ and other study variables. This additivity is required for model identification  \citep{Hirano2001, sadinle2019sequentially}}.
%\end{definition}
 Special cases of AN mechanisms include  itemwise conditionally independent nonresponse \citep{sadinle2017itemwise}  when $\beta_{j} = 0$ and MCAR  when $\beta_{k} = 0$ for all $k=1, \dots, p$.  Common link functions $g$ include the logistic and probit \citep{Hirano2001}. %The AN model assumes that the probability a study variable is missing depends on a linear combination of the study variables, including that in question. Therefore, the AN may be extended to model the missingness mechanism for MNAR data. 

Lemma \ref{lemma1} formally connects the model for $R_{ij}$
%under the Gaussian copula 
in \eqref{copmis} to the AN missingness mechanism in \eqref{AN}.
%\textcolor{black}{which implies that the copula correlation is identified by the observed data under the Gaussian copula \eqref{YRcop}--\eqref{xformYR}}. 
Unlike the formulation in \cite{Hirano2001}, the AN missingness mechanism for the Gaussian copula model has additivity on the latent scale.

%Thus, the Gaussian copula provides a model for the missingness which connects closely with the assumption of additive non-ignorability. 
%The difference is that the

%One way to visualize the implications of the missingness mechanism on the original scale of $\boldsymbol Y_j$ for a given  
%visualize the missingness probabilities in   
%observed data as well: 
%given $\boldsymbol{C}_{\boldsymbol{\theta}}$ and $\{F_{j}\}_{j=1}^p$ is to 
%for a a grid of $(\boldsymbol{y}^{obs}, \boldsymbol{y}^{mis})$, missingness probabilities may be visualized by 
%plot $1- \Phi_{\boldsymbol \alpha_{j}^{*}, \boldsymbol C^{*}_{jj}}(0)$
%over a grid of values for $y_{ij}$. 

\begin{lemma}\label{lemma1}
   Suppose $\{(\boldsymbol{y}_i, \boldsymbol r_i) \}_{i=1}^{n}\overset{iid}{\sim} \Pi_{0}$ where $\Pi_{0}$ is the Gaussian copula with
    % SHOULD WE DELETE $\boldsymbol{y}^{mis}$?  \textcolor{black}{comprise $n$} independent and identically distributed samples from \eqref{YRcop}-\eqref{xformYR} with  true
    correlation $\boldsymbol C_{0}$ and  marginals $\{F_{j}\}_{j=1}^{p}$ as in \eqref{YRcop}--\eqref{xformYR}, and 
    %\textcolor{black}{where $\boldsymbol{y}^{mis}$ follows a nonignorable missingness mechanism under the copula model.}  
    $\{\mathcal{A}_{j}\}_{j=1}^{p} = \{F_{j}\}_{j=1}^{p}$. For any value of $(\boldsymbol y^{obs}_{i},\boldsymbol y^{mis}_{i})$,  $p(R_{ij} =1 \mid \boldsymbol y_i^{obs}, \boldsymbol y_i^{mis}, \boldsymbol{C}_{j}^{*}, \alpha_{r_{j}}, \{F_{j}\}_{j=1}^{p})$ satisfies \eqref{AN} with  $x_{ik}=z_{ik}$,
   %= \Phi^{-1}(F_{j}^{-1}(y_{ij}))]_{j=1}^{p}$, 
   $g$ the probit link function in \eqref{copmis}--\eqref{moments}, $\beta_{0} = \alpha_{r_{j}}$, and $\beta_{k}$  the $k$th component of the vector $\boldsymbol{C}_{r_{j} \boldsymbol y}\boldsymbol{C}_{\boldsymbol y}^{-1}$.
\end{lemma}
%To illustrate in the context of our motivating example with missing lead exposure measurements, suppose that the propensity for  missingness for lead exposure level depends on the main effects and interaction between the latent variables for the individual's lead exposure level and the indicator for economic disadvantages; the interaction creates a violation of the AN missingness mechanism. We can interpret the interaction as indicating, for example, that the effect of having a low lead exposure level on the propensity for missingness is even greater for individuals without socioeconomic disadvantages than it is for individuals with socioeconomic disadvantages. This interpretation leverages the monotone relationship between $Z_{j}$ and $Y_{j}$ to translate from the latent to natural scale.
%lead exposure levels to be missing when the person both has high lead is economically advantaged a violation of the AN missingness mechanism could occur when, say, 
%i.e., there is an interaction effect. 
%Related, Lemma \ref{lemma1} does not allow for different signs of the pairwise correlations for any $(Y_j, Y_{j'})$ for distinct missingness patterns.

%\vspace{-6pt}

Of course, it is generally impossible to determine whether any specific missingness mechanism holds in practice \citep{molenberghsetal}.  However, if additive nonignorability on the latent scale does not hold, e.g., the true model for $z_{r_j}$ includes interactions between $z_j$ and other latent variables, the Gaussian copula with fixed margins may not offer reliable inferences or sensitivity analyses. Developing methods for  sensitivity analyses with other nonignorable missingness mechanisms is a topic for future research.

\section{Leveraging Incomplete Auxiliary Quantiles}\label{auxapprox}

% %To this point, we have assumed 
% In some settings, analysts may not be able or willing to specify full marginal distributions for use in imputation and sensitivity analysis.  For example, analysts may be able to elicit reasonable marginal quantiles from domain experts but not necessarily an entire distribution. Or, they may have access to sets of quantiles from other data sources but not entire distributions. 
% %knowledge of each ground truth marginal distribution function $\{F_{j}\}_{j=1}^{p}$ for the study variables, which equates to knowledge of the mapping for any realization $y_{j}$ to its cumulative probability $F_{j}(y_{j}) = \tau \in [0,1]$. Often, data holders and agencies will publish official statistics on specific quantiles of study variables, which is incomplete information. 

%In this section, we present modifications to the Gaussian copula when 
In many contexts, the available information on the marginals does not comprise  full distributions.  For example, analysts may have access to sets of quantiles from other data sources but not entire distributions. Or, they may be able to elicit reasonable marginal quantiles of $F_{j}$ from domain experts but not necessarily an entire distribution.
%We conclude that even sparse information on each study variable is sufficient for accurate estimation and imputation with nonigorable missing data. Finally, we leverage this information within the Gaussian copula framework to develop computationally convenient sampling and imputation algorithms.
To incorporate this information within the Gaussian copula framework, we address two salient issues. First, with incomplete knowledge of any $F_j$, the transformation between $y_{ij}$ and $z_{ij}$ is  unknown, which could complicate estimation of $\boldsymbol C_{\boldsymbol \theta}$. Second, a small set of auxiliary quantiles for each study variable could be insufficient for imputation of $\boldsymbol y^{mis}$.  

%A potential workaround when information about $\{F_{j}\}_{j=1}^{p}$ is incomplete or cannot be extrapolated from published figures is to posit a series of parametric families for each variable. While convenient, parametric margins may not capture complex univariate features of the data like multi-modality and can erode the multivariate flexibility of the copula model. Ultimately, this leads to biased imputation of missing values. 

%To address these challanges, we develop a working model, akin to \eqref{copwork1}-\eqref{integrand}, which directly leverages available information on each margin to simultaneously estimate the copula correlation and intermediate quantiles of each univariate marginal. By encoding the auxiliary quantiles into the model, we capture known features of each study variable which also anchor CDF estimates at other points. In addition, we demonstrate that even sparse information preserves model identification in the presence of non-ignorable missing data. An added benefit is the computational efficiency of the proposed sampling algorithms, especially in comparison to competing algorithms for copula estimation. 

For each $Y_{j}$, suppose we have a finite set of $m_{j}\geq 3$ non-decreasing quantiles of $F_{j}$. Thus, 
    $\mathcal{A}_{j} = \{F_{j}^{-1}(\tau^{1}_{j}),\dots, F_{j}^{-1}(\tau^{q}_{j}),\dots,F_{j}^{-1}(\tau^{m_{j}}_{j})\}$, where $\tau_{j}^{q} \in [0,1]$ for $q  = 1,\dots, m_{j}$. We fix each $\tau^{1}_{j} = 0$ and $\tau^{m_{j}}_{j} = 1$, while the remaining quantiles can differ across $Y_j$. 
    %If $\mathcal{A}_{j}$ is not available via public information, it is specified using subject-matter expertise. 
    We require  $\mathcal{A}_{j}$ include $F_{j}^{-1}(0)$ and $F_{j}^{-1}(1)$. 
    %and at least one intermediate quantile. 
    These bounds can be specified based on domain knowledge, e.g., human ages cannot be negative and are generally $\leq 110$. When an intermediate quantile is not available via external sources, it can be specified using subject-matter expertise as part of a sensitivity analysis, as described below and in Section \ref{sec:realdat}.   

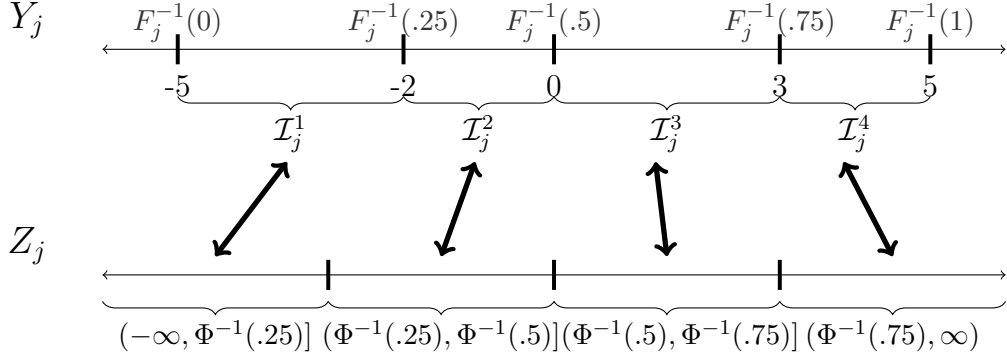
\begin{figure}[h]
\centering
\begin{tikzpicture}[scale=1]
    % Draw the number line
    \draw[<->] (-6,0) -- (6,0)
    node[above, font=\bfseries\large] at (-7,0) {$Y_{j}$};
    % Draw the ticks
    \foreach \x in {-5,-2,0,3,5}
    \draw[line width = 1.5] (\x,0.2) -- (\x,-0.2);
      \def\customlabels{{"$F_{j}^{-1}(0)$","$F_{j}^{-1}(.25)$", "$F_{j}^{-1}(.5)$","$F_{j}^{-1}(.75)$", "$F_{j}^{-1}(1)$"}}
    
    % Draw the ticks and custom labels for the first number line
    \foreach \x/\label [count=\i] in {-5,-2,0,3,5}
    {
        \node[below, font=\bfseries\small, text=black!80] at (\x,0.75) {\pgfmathparse{\customlabels[\i-1]}\pgfmathresult};
    }
    
    % Draw the numbers
    \foreach \x in {-5,-2,0,3,5}
    \node[below] at (\x,-0.2) {\x};
    
    % Add underbraces
    \draw[decorate,decoration={brace,amplitude=5pt,mirror},yshift=-15pt] (-5,-0.1) -- (-2,-0.1) node[midway,below,yshift=-3pt]{$\mathcal{I}^{1}_{j}$};
    \draw[decorate,decoration={brace,amplitude=5pt,mirror},yshift=-15pt] (-2,-0.1) -- (0,-0.1) node[midway,below,yshift=-3pt]{$\mathcal{I}^{2}_{j}$};
    \draw[decorate,decoration={brace,amplitude=5pt,mirror},yshift=-15pt] (0,-0.1) -- (3,-0.1) node[midway,below,yshift=-3pt]{$\mathcal{I}^{3}_{j}$};
    \draw[decorate,decoration={brace,amplitude=5pt,mirror},yshift=-15pt] (3,-0.1) -- (5,-0.1) node[midway,below,yshift=-3pt]{$\mathcal{I}^{4}_{j}$};
    \draw[<->, line width = 2] (-3.55,-1.5) -- (-4.5,-2.75);
    \draw[<->, line width = 2] (-1.05,-1.5) -- (-1.5,-2.75);
    \draw[<->, line width = 2] (1.35,-1.5) -- (1.5,-2.75);
    \draw[<->, line width = 2] (3.85,-1.5) -- (4.5,-2.75);
        % Draw the number line
    \draw[<->] (-6,-3) -- (6,-3)
    node[above, font=\bfseries\large] at (-7,-3)
    {$Z_{j}$};
    % Draw the ticks
    \foreach \x in {-3,0,3}
    \draw[line width = 1.5 ] (\x,-2.8) -- (\x,-3.2);
    
    % % Draw the numbers
    % \def\customlabels{{"","", "$\Phi^{-1}(0.50)$", "$\Phi^{-1}(0.75)$","$\infty$"}}
    
    % % Draw the ticks and custom labels for the first number line
    % \foreach \x/\label [count=\i] in {-6,-3,0,3,6}
    % \node[below] at (\x,-3.1) {\pgfmathparse{\customlabels[\i-1]}\pgfmathresult};
    
    % Add underbraces
    \draw[decorate,font = 
    \small,decoration={brace,amplitude=5pt,mirror},yshift=-10pt] (-6,-3) -- (-3,-3) node[midway,below,yshift=-3pt]{($-\infty,\Phi^{-1}(.25)]$};
    \draw[decorate,font = \small, decoration={brace,amplitude=5pt,mirror},yshift=-10pt] (-3,-3) -- (0,-3.0) node[midway,below,yshift=-3pt]{$(\Phi^{-1}(.25), \Phi^{-1}(.5)]$};
    \draw[decorate,font = \small,decoration={brace,amplitude=5pt,mirror},yshift=-10pt] (0,-3) -- (3,-3)  node[midway,below,yshift=-3pt,xshift = 5pt]{$(\Phi^{-1}(.5), \Phi^{-1}(.75)]$};
    \draw[decorate,font = \small, decoration={brace,amplitude=5pt,mirror},yshift=-10pt] (3,-3) -- (6,-3)  node[midway,below,yshift=-3pt]{$(\Phi^{-1}(.75), \infty)$};
    % Draw the vertical equivalence line
\end{tikzpicture}
\caption{Using auxiliary information $\mathcal{A}_{j}$ to create non-overlapping intervals $\{\mathcal{I}^{q}_{j}\}_{q=1}^{4}$ partitioning the support of $Y_{j}$. For $y_{ij} \in \mathcal{I}^{q}_{j}$, the auxiliary quantiles defining this interval pre-determine the interval containing $z_{ij}$ on the latent Gaussian scale. Here, $\mathcal{A}_{j} = \{F_{j}^{-1}(0) = -5, F_{j}^{-1}(0.25) = -2, F_{j}^{-1}(0.50)= 0, F_{j}^{-1}(0.75) = 3, F_{j}^{-1}(1) = 5\}$.}\label{eql-fig}
\end{figure}

Even though the exact map between each $y_{ij}$ and $z_{ij}$ is unknown, $\mathcal{A}_{j}$ does provide partial information about $z_{ij}$ under \eqref{YRcop}--\eqref{xformYR}. 
%which can be leveraged to estimate $\boldsymbol C_{\boldsymbol \theta}$.
We construct a set of $m_{j} - 1$ non-overlapping intervals that partition the support of $Y_{j}$, 
\begin{equation}\label{int_aux}
\mathcal{I}^{q}_{j} = 
    (F_{j}^{-1}(\tau^{q}_{j}), F_{j}^{-1}(\tau^{q+1}_{j})],\,\, q=1, \dots, m_{j} -1.
\end{equation}
Each $y_{ij}$ belongs to exactly one  $\mathcal{I}^{q}_{j}$. 
%The non-decreasing link between latent and observed data under the Gaussian copula \eqref{xform} demonstrates that 
Further, $y_{ij} \in \mathcal{I}^{q}_{j}$ implies $z_{ij} \in (\Phi^{-1}(\tau^{q}_{j}),\Phi^{-1}(\tau^{q+1}_{j})]$ under \eqref{YRcop}--\eqref{xformYR}; that is, if $y_{ij}$ lies in some quantile interval, then $z_{ij}$ must lie between the same quantiles of a standard Gaussian random variable. 
We visualize this mapping in Figure \ref{eql-fig}. 

%\vspace{-18pt}

\subsection{Estimation of the Copula Correlation}\label{estimEQL-sec}

%Across study variables, t
The partial mapping between $y_{ij}^{obs}$ and $z_{ij}^{obs}$ in Figure \ref{eql-fig} provides the basis for estimating $\boldsymbol C_{\boldsymbol \theta}$.
%the copula correlation.
%when $\mathcal{A}$ comprises only auxiliary marginal quantiles. 
Using the intervals in Figure \ref{eql-fig},
define the binning function
% the effect of binning $\boldsymbol y^{obs}$. We now show that these bins and $\mathcal{A}$ allow us to estimate $\boldsymbol C_{\boldsymbol \theta}$. For any $y_{ij}^{obs}$, define its bin $b_{ij}^{obs}$ as 
%into quantile intervals. To this effect, 
\begin{equation}\label{bins}
    b_{ij}^{obs}(y_{ij}^{obs}) = q \iff y_{ij}^{obs} \in \mathcal{I}_{j}^{q}.
\end{equation}
In what follows, we suppress the dependence of $b_{ij}^{obs}(y_{ij}^{obs})$ on $y_{ij}^{obs}$.  Let $\boldsymbol b_{i}^{obs} = \{b_{ij}^{obs}: r_{ij} = 0, j=1, \dots, p\}$ and $\boldsymbol{b}^{obs} = \{\boldsymbol{b}_{i}^{obs}: i = 1,\dots, n\}$. The binning in \eqref{bins} has the effect of 
coarsening the observed data \citep{heitjan1991ignorability, miller2018robust} based on the  intervals %memberships of $\boldsymbol y^{obs}$ 
defined by $\mathcal{A}$. 
%We leverage the information in the coarsened data $\boldsymbol b^{obs}$ in place of $\boldsymbol y^{obs}$ to learn the copula correlation.
Thus, we require a likelihood for $(\boldsymbol C_{\boldsymbol \theta}, \boldsymbol \alpha)$ using $(\boldsymbol b^{obs}, \boldsymbol{r})$. 

As evident in Figure \ref{eql-fig}, 
%there is a correspondence between each $b_{ij}^{obs}$ and $z_{ij}^{obs}$ conditional on $\mathcal{A}_{j}$: 
whenever $b^{obs}_{ij} = q$, we must have $z^{obs}_{ij} \in (\Phi^{-1}(\tau^{q}_{j}),\Phi^{-1}(\tau^{q+1}_{j})]$.
%under \eqref{YRcop}-\eqref{xformYR}. 
%Across all study variables, define the set
%\begin{align}\label{aux event}
%   \mathcal{D}(\boldsymbol b^{obs}) \coloneqq \{ &\boldsymbol z^{obs} : b_{ij}^{obs} = q  \implies z^{obs}_{ij} \in (\Phi^{-1}(\tau^{q}_{j}),\Phi^{-1}(\tau^{q+1}_{j})\};\\ & i = 1, \dots, n; j =1, \dots, p\}\nonumber.
%\end{align}
\textcolor{black}{We represent this restriction by defining 
%, we define the set
$\mathcal{D}(\boldsymbol b^{obs})$ to be the set of $\boldsymbol z^{obs}$ satisfying the condition that each $z_{ij}^{obs}$
%\in \boldsymbol{z}^{obs}$ to
is in the latent interval defined by corresponding $b^{obs}_{ij}$. }
%marginal quantile interval memberships
%\eqref{aux event} specifies that 
% if $b_{ij}^{obs} = q$, then 
% %(e.g., $y^{obs}_{ij} \in \mathcal{I}_{j}^{q}$), 
% $z^{obs}_{ij}$ is in the corresponding quantile interval 
%defined by $\boldsymbol b^{obs}$ 
%and $\mathcal{A}_{j}$ on the latent scale.} 
By conditioning on $\mathcal{A}$ rather than $\{F_{j}\}_{j=1}^p$, we have
\begin{align}
    &p(\boldsymbol b^{obs}, \boldsymbol r \mid \boldsymbol C_{\boldsymbol{\theta}}, \boldsymbol{\alpha}, \mathcal{A}) =  p(\boldsymbol z^{obs} \in \mathcal{D}(\boldsymbol b^{obs}), \boldsymbol z_{\boldsymbol r} \in \mathcal{E}(\boldsymbol r) \mid \boldsymbol C_{\boldsymbol \theta}, \boldsymbol \alpha) \label{marg-work}\\
    %&p\{\boldsymbol Z^{obs} \in \mathcal{D}(\boldsymbol Y^{obs}), \boldsymbol Z_{\boldsymbol R} \mid \boldsymbol{C} _{\boldsymbol{\theta}}, \mathcal{A}\} \label{marg-work} \\
    &= \int \int_{\mathcal{D}(\boldsymbol b^{obs})} \int_{z_{\boldsymbol r} \in \mathcal{E}(\boldsymbol r)} p(\boldsymbol z^{obs} , \boldsymbol z^{mis}, \boldsymbol z_{\boldsymbol r} \mid \boldsymbol{C}_{\boldsymbol{\theta}},\boldsymbol{\alpha})d \boldsymbol{z}_{\boldsymbol r}d \boldsymbol z^{obs} d \boldsymbol z^{mis} . \label{marg-work-expand}
\end{align}
The equivalence in \eqref{marg-work} is by construction; observing $\boldsymbol b^{obs}$ implies that $\boldsymbol z^{obs}$ must belong to $\mathcal{D}(\boldsymbol b^{obs})$ conditional on $\mathcal{A}$.
We refer to \eqref{marg-work} as the extended quantile likelihood, abbreviated as EQL. %Importantly, this 
The EQL also enables inclusion of 
%mixed continuous and 
%discrete 
\textcolor{black}{discrete} $Y_j$ in the copula model, as $\boldsymbol b^{obs}$ and $\mathcal{A}$ can be constructed from discrete 
%the auxiliary quantiles specified from arbitrary
support. We discuss including unordered categorical  variables in the EQL in the supplement, Section D.

From \eqref{marg-work}, posterior inference for $(\boldsymbol C_{\boldsymbol \theta}, \boldsymbol \alpha)$ under the EQL targets 
\begin{equation} \label{EQLpost}
    %\Pi_{n}^{*}(\boldsymbol C_{\boldsymbol \theta}) 
    p(\boldsymbol C_{\boldsymbol \theta}, \boldsymbol \alpha \mid \boldsymbol z^{obs} \in \mathcal{D}(\boldsymbol b^{obs}), \boldsymbol z_{\boldsymbol r} \in \mathcal{E}(\boldsymbol r), \mathcal{A})  
    \propto  p(\boldsymbol z^{obs} \in \mathcal{D}(\boldsymbol b^{obs}), \boldsymbol z_{\boldsymbol r} \in \mathcal{E}(\boldsymbol r) \mid \boldsymbol C_{\boldsymbol{\theta}}, \boldsymbol{\alpha}) p(\boldsymbol C_{\boldsymbol \theta}, \boldsymbol \alpha).
\end{equation}
\textcolor{black}{We refer to the marginal posterior distribution of $\boldsymbol C_{\boldsymbol \theta}$ under \eqref{EQLpost} as $\Pi_{n}^{*}(\boldsymbol C_{\boldsymbol \theta})$.} 

Remarkably, even when $\mathcal{A}$ comprises only a few marginal quantiles for each study variable, it is still possible to estimate $\boldsymbol C_{\boldsymbol \theta}$ accurately under the EQL in the presence \textcolor{black}{of AN missingness as defined in Lemma \ref{lemma1}, assuming of course} that the full data distribution is a Gaussian copula. 
%under the implied nonignorable missingness mechanism 
This fact is summarized in Theorem \ref{postconsistEQL}.  We empirically examine the concentration of $\Pi_{n}^{*}(\boldsymbol C_{\boldsymbol \theta})$ as a function of the number of auxiliary quantiles 
%specified for each study variable 
in Section \ref{sim1}.
\begin{theorem}\label{postconsistEQL}
   Suppose $\{(\boldsymbol{y}_i, \boldsymbol r_i) \}_{i=1}^{n} \overset{iid}{\sim} \Pi_{0}$ where $\Pi_{0}$ is the Gaussian copula with
    % SHOULD WE DELETE $\boldsymbol{y}^{mis}$?  \textcolor{black}{comprise $n$} independent and identically distributed samples from \eqref{YRcop}-\eqref{xformYR} with  true
    correlation $\boldsymbol C_{0}$ and marginals $\{F_{j}\}_{j=1}^{p}$ as in \eqref{YRcop}--\eqref{xformYR}. For $j = 1, \dots, p$, suppose $\mathcal{A}_{j}$ comprises $m_{j}\geq 3$ auxiliary quantiles of $F_{j}$, including  $F_{j}^{-1}(0)$ and $F_{j}^{-1}(1)$. Let $p(\boldsymbol \theta)$ be a prior with respect a measure that induces a prior $\Pi$ over the space of all $2p \times 2p$ correlation matrices $\boldsymbol{\mathbbm{C}}$ such that $\Pi(\boldsymbol C_{\boldsymbol \theta})>0$ for all $\boldsymbol C_{\boldsymbol \theta} \in \mathbbm{C}$. Then, for any neighborhood $\mathcal{B}$ of $\boldsymbol C_{0}$, $\lim_{n\rightarrow \infty} \Pi^{*}_{n}(\boldsymbol C_{\boldsymbol \theta} \in \mathcal{B}) \rightarrow 1 \ \textrm{almost surely } [\Pi_{0}]$.
    % MAYBE WE SHOULD ADD THE AN MISSINGNESS MECHANISM AS A CONDITION?
\end{theorem}
%Theorem \ref{postconsistEQL} dictates that well-specified auxiliary quantiles is sufficient information to accurately estimate the copula correlation under the EQL in the presence of nonignorable missing data. 
\textcolor{black}{Theorem \ref{postconsistEQL} provides a practically useful result when study variables have nonignorable missing values: analysts need only specify lower/upper bounds and a single intermediate quantile for each study variable to estimate $\boldsymbol C_{\boldsymbol \theta}$ accurately, provided the joint distribution for $(\boldsymbol y, \boldsymbol r)$ is a Gaussian copula and thus the missingness follows the AN mechanism of Section \ref{sec:AN}. Though the result is asymptotic, we demonstrate empirically in Section \ref{sim1} that under the conditions of Theorem \ref{postconsistEQL}, $\Pi_{n}^{*}(\boldsymbol C_{\boldsymbol \theta})$ can concentrate rapidly with sample sizes of a few hundred.
%and greatly facilitates sensitivity analyses.
Furthermore, estimation of $\boldsymbol C_{\boldsymbol \theta}$ is possible without specifying full marginal distribution models that require parameter updates for each study variable.}
%$j = 1,\dots,p$.}

% This result holds for any Gaussian copula model, including those for mixed continuous and discrete study variables.}

%$\{F_{j}\}_{j=1}^{p}$, 
% In particular, the events $\{\boldsymbol Z^{obs} \in \mathcal{D}(\boldsymbol Y^{obs}), Z_{\boldsymbol R} \in \mathcal{E}(\boldsymbol R)\}$ are independent of the true (in this case unknown) marginals $\{F_j\}_{j=1}^p$ given $\mathcal{A}$ and the probit augmentation.  
%The benefit of both the EQL and RL/RPL is that estimation of $\boldsymbol{C} _{\boldsymbol{\theta}}$ is possible without models for any of the marginal distribution functions $\{F_{j}\}_{j=1}^{p}$. 
%Similar benefits exist for t
\textcolor{black}{The EQL resembles the extended rank (RL) and rank-probit (RPL) likelihoods  \citep{hoff2007extending, feldman2022bayesian}, which target posterior inference for the Gaussian copula correlation by conditioning on the set of latent variables consistent with the multivariate ranks of the observed data values.} 
% Empirical and theoretical analyses \citep{murray2013bayesian, feldman2022nonparametric} have shown that this type of marginal likelihood maintains posterior consistency for the copula correlation in the presence of ignorable (but not  nonignorable) missingness mechanisms. 
However, the EQL and the RL/RPL make different uses of their conditioning events.  Under the EQL event, 
%we obtain identifying information, in that  
$\mathcal{A}$ partially locates $\boldsymbol z^{obs}$.  By comparison, the RL/RPL event does not restrict where $\boldsymbol z^{obs}$ lies in  latent space, as long as the orderings of the individual $z_{ij}^{obs}$ are consistent with the ranks of $y_{ij}^{obs}$. As a result, inferences for $\boldsymbol{C}_{\boldsymbol{\theta}}$ under \textcolor{black}{AN missingness may be biased for the RL/RPL, as we demonstrate in Section \ref{sim1}.}.

To estimate the model, we utilize a Gibbs sampler with data augmentation \citep{chib1998analysis}, alternating sampling $(\boldsymbol z^{obs}, \boldsymbol z^{mis}, \boldsymbol z_{\boldsymbol r}) \mid \boldsymbol C_{\boldsymbol \theta}, \boldsymbol \alpha$ and $\boldsymbol C_{\boldsymbol \theta}, \boldsymbol \alpha \mid \boldsymbol (\boldsymbol z^{obs}, \boldsymbol z^{mis}, \boldsymbol z_{\boldsymbol r})$. 
%The sampler can be efficient sunder a variety of priors for $\boldsymbol \theta$, since the integrand in \eqref{marg-work-expand} is a multivariate  Gaussian likelihood.   
Algorithm \ref{algEQL} summarizes the two steps for an arbitrary specification of $p(\boldsymbol C_{\boldsymbol \theta}, \boldsymbol \alpha)$. We present details for the $p(\boldsymbol C_{\boldsymbol \theta}, \boldsymbol \alpha)$ used in the analyses \textcolor{black}{in Section \ref{sec:sims} and Section E of the supplement.}
%subject to $\boldsymbol Z^{obs} \in \mathcal{D}(\boldsymbol Y^{obs}), \boldsymbol Z_{\boldsymbol R} \in \mathcal{E}(\boldsymbol R)$. This facilitates 
%see Section \ref{sec:sims} for the specification used in our analyses.
%The sampler is summarized in Algorithm \ref{algEQL}. 
In the algorithm, the subscript $(-j)$ in a vector indicates that vector without the $j$th element and in the column (row) index of a matrix indicates exclusion of the elements corresponding to the $j$th column (row) of that matrix.  The subscript $-(jj)$ in a matrix indicates exclusion of all row and column elements for the $j$th variable in that matrix.   
 
%  The imputation step in Algorithm \ref{algEQL} requires linking sampled $\tilde{z}_{ij}^{mis}$ to imputations $\tilde{y}_{ij}^{mis}$ via $\tilde{y}_{ij}^{mis} = \hat{F}_{j}^{-1}\{\Phi(\tilde z_{ij}^{mis})\}$ using an estimator of the marginal distribution, $\hat F_{j}$. In the EQL, for each $Y_j$ we fit a monotone interpolating spline through the auxiliary quantiles.
% %$\{F_{j}^{-1}(\tau^{q}_{j}), \tau^{q}_{j}\}$. 
% This expands the support of each $F_j$ beyond the quantiles in $\mathcal{A}_j$. The interpolations are most reasonable when each $\mathcal{A}_{j}$ provides enough information to outline salient features of each $F_{j}$. We modify the imputation step when $\mathcal{A}_{j}$ has few quantiles as described in Section \ref{MA-EHQL}. 

\begin{algorithm}[t]
\caption{Gibbs sampler for the EQL Gaussian copula}
\begin{algorithmic}\label{algEQL}
     \STATE \textbf{Require:}  prior $p(\boldsymbol C_{\boldsymbol \theta}, \boldsymbol \alpha)$, auxiliary quantiles $\mathcal{A}$.  Let $\boldsymbol C = \boldsymbol C_{\boldsymbol \theta}.$
     \begin{itemize}
         \item  \textbf{Step 1}: Sample $(\boldsymbol z^{obs}, \boldsymbol z^{mis}, \boldsymbol z_{\boldsymbol r}) \mid \boldsymbol C, \boldsymbol \alpha$
     \end{itemize}
       \bindent
       \FOR{$ z_{ij} \in \boldsymbol z^{obs}$}
            \STATE Sample $ z_{ij} \sim \mbox{Normal}(\mu_{ij}, \sigma_{j}^{2}) \mathbbm{1}(\ell_{ij}, u_{ij}]$
                \STATE $\ell_{ij} = \Phi^{-1}(\tau_{ij}^{\ell})$, $\tau_{ij}^{\ell} = \max\{ F_{j}^{-1}(\tau^{q}_{j}) \in \mathcal{A}_{j}: y^{obs}_{ij} > F_{j}^{-1}(\tau_{j}^{q})\}$, 
                \STATE $u_{ij} = \Phi^{-1}(\tau_{ij}^{u})$, $ \tau_{ij}^{u} = \min\{ F_{j}^{-1}(\tau^{q}_{j}) \in \mathcal{A}_{j}: y^{obs}_{ij} \leq F_{j}^{-1}(\tau_{j}^{q})\}$
                % u_{ij} = \Phi^{-1}(\tau^{u})$  
                
                % \max\{\Phi^{-1}(\tau^{\ell}), \max(z_{vj}^{obs}: y_{vj} \in \mathcal{I}^{(q-1)*}_{j}, v=1, \dots, n)\}$ 
                % \STATE $u_{ij} =\min\{\Phi^{-1}(\tau^{u}), \min(z_{vj}^{obs}: y_{vj} \in \mathcal{I}^{(q+1)*}_{j}, v=1, \dots, n)\}$
                \ENDFOR
                
            \FOR{$ z_{ij} \in \boldsymbol z_{\boldsymbol r}$}
            \STATE Sample  $ z_{ij} \sim \mbox{Normal}(\mu_{ij}, \sigma_{j}^{2}) \mathbbm{1}(\ell_{ij},u_{ij})$
                \STATE $\ell_{ij} = 0 - \infty \mathbbm{1}_{r_{ij} =1}, \quad u_{ij} = 0 + \infty \mathbbm{1}_{r_{ij} = 0}$ %\begin{cases}
                % 0 & r_{ij} = 1  \\
                % -\infty & r_{ij} = 0
                % \end{cases}$ \quad
                % $u_{ij} = \begin{cases}
                % \infty & R_{ij} = 1 \\
                % 0 & R_{ij} = 0
                % \end{cases}$
                
                \ENDFOR
            \FOR{$ z_{ij} \in \boldsymbol z^{mis}$}
                \STATE Sample $z_{ij} \sim \mbox{Normal}(\mu_{ij}, \sigma_{j}^{2})$ 
                \ENDFOR
                    
            %         \STATE Compute $\sigma_{j}^{2} = \boldsymbol \theta^{*}_{jj}- \boldsymbol \theta^{*}_{j-j}\boldsymbol \theta^{*^{-1}}_{-j-j} \boldsymbol \theta^{*}_{-jj}$
            %         \STATE Compute $\mu_{ij} = \boldsymbol Z_{i- j}(\boldsymbol \theta^{*}_{j-j}\boldsymbol \theta^{*^{-1}}_{-j-j}), i \in \{1, \dots, n\}$
               
            %     \FOR{$x \in \mbox{unique}\{y^{obs}_{1j}, \dots, y^{obs}_{nj}\}$ and $\{i : Y^{obs}_{ij} = x$\}}

            %         \STATE Sample $Z^{obs}_{ij} \sim N(\mu_{ij}, \sigma_{j}^{2})\mathbbm{1}(z_{l}, z_{u})$
                
            %     \ENDFOR
            %     \FOR{$i : R_{ij} = 1$}
            %     \STATE Sample $Z^{mis}_{ij} \sim N(\mu_{ij}, \sigma_{j}^{2})$
            %     \ENDFOR 
     \STATE where  $\mu_{ij} = \alpha_{j} + \boldsymbol C_{j(-j)} \boldsymbol C^{-1}_{-(jj)}(\boldsymbol z_{i(-j)} - \boldsymbol \alpha_{-j})$, $\sigma_{j}^{2} = \boldsymbol C_{jj}- \boldsymbol C_{j(-j)}\boldsymbol  C^{-1}_{-(jj)}\boldsymbol C_{(-j)j}$
            
     \eindent

            %         \STATE $$
          \begin{itemize}
         \item  \textbf{Step 2}: Sample $\boldsymbol C, \boldsymbol \alpha \sim p(\boldsymbol C, \boldsymbol \alpha \mid \boldsymbol{z}^{obs}, \boldsymbol z^{mis}, \boldsymbol z_{\boldsymbol r})$
              \end{itemize}
              \bindent
         \STATE where $p(\boldsymbol C, \boldsymbol \alpha \mid \boldsymbol{z}^{obs}, \boldsymbol z^{mis}, \boldsymbol z_{\boldsymbol r}) \propto N_{2p}((\boldsymbol{z}^{obs}, \boldsymbol z^{mis}, \boldsymbol z_{\boldsymbol r}); \boldsymbol C, \boldsymbol{\alpha}) p(\boldsymbol C, \boldsymbol \alpha) $
         \eindent
         % \begin{itemize}
         % \item  \textbf{Step 3}: Interpolate Marginal CDFs 
         % \end{itemize}
         % \bindent
         % \FOR{$j \in \{1,\dots,p\}$}
         %    \STATE --If EQL: $\tilde{F}_{j} = \vtt{splinefun}(\mathcal{A}_{j},  \{\tau^{q}_{j}\}_{q=1}^{m_{j}}, \vtt{method = "monoH.FC")})$ 
         %    \STATE --If EHQL:
         %    \bindent
         %        \STATE a) Compute margin adjustment $\{\tilde{F}_{j}(y_{j}^{q})\}_{q=1}^{s_{j}} = \{\Phi(Z_{j}(y_{j}^{k}))\}_{q=1}^{s_{j}}$
         %        \STATE b) $\tilde{F}_{j} = \vtt{splinefun}(\mathcal{A}_{j} \cup \{y_{j}^{q}\}_{q=1}^{s_{j}}, \{\tau^{q}_{j}\}_{q=1}^{m_{j}} \cup \{\tilde{F}_{j}(y_{j}^{q})\}_{q=1}^{s_{j}}$,
         %        \STATE $\,\,\,\,\,\,$ \vtt{method = "monoH.FC")}) \eindent  
         %  \ENDFOR 
         %  \eindent
         %    \begin{itemize}
         % \item  \textbf{Step 4}: Impute $\tilde{y}_{ij}^{mis} = \tilde{F}_{j}^{-1}(\Phi(\tilde{z}_{ij}^{mis}))$
         % \end{itemize}
\end{algorithmic}
\end{algorithm}

%Under the factor model \eqref{factor}, 
%Posterior inference for $\boldsymbol{C}_{\boldsymbol{\theta}}$ is available via 
%\begin{equation}\label{EQLpost}
%     p\{\boldsymbol C_{\boldsymbol \theta} \mid Z^{obs} \in \mathcal{D}(\boldsymbol Y^{obs}), \boldsymbol Z_{\boldsymbol R},\mathcal{A}\} \propto p\{\boldsymbol Z^{obs} \in \mathcal{D}(\boldsymbol Y^{obs}), \boldsymbol Z_{\boldsymbol R} \mid \boldsymbol C_{\boldsymbol \theta}, \mathcal{A}\} \times p(\boldsymbol \theta)
%\end{equation}
%which may be the case if there are few auxiliary quantiles available for each study variable. 

%Posterior inference based on the EQL can also be viewed as a type of coarsening \citep{coarsening}. Instead of updating the posterior through conditioning on an explicit transformation between $y_{ij}$ and $z_{ij}$, the EQL provides that each observation lies in a $p$-dimensional hypercube. 
%This type of learning using indirect information from the observed data has been shown to protect against model mis-specification and provide inferences that are more robust to erroneously collected and outlying data points.

%simulation results \textcolor{black}{probably should make its way into the simulation study -- the first one, in particular}. 

Algorithm \ref{algEQL} offers significant computational advantages over similar samplers for RL/RPL Gaussian copula models. Typically, $m_{j} << n$, so  
% Because the cardinality of each $\mathcal{A}_j$ is typically small relative to $n$,
% %the sample size, 
%i.e., $\lvert \mathcal A_{j} \rvert << n$, 
the upper and lower truncation regions in Step 1 of Algorithm \ref{algEQL} are shared by many observations. Consequently, the data augmentation is computationally efficient: for all $b_{ij}^{obs} = q$, corresponding $z_{ij}^{obs}$ may be sampled simultaneously using truncated normal distributions.
%e.g., via \verb|truncnorm| in \verb|R|.
This enables reasonable computation times for (moderately) large $n$; for example, we fit the copula model to the North Carolina data comprising nearly 170,000 children. By contrast, the computational complexity of RL Gibbs samplers depends on the number of unique marginal ranks for each $Y_j$, which may approach $n$.
%By contrast, the computational complexity of RL Gibbs samplers depends on the marginal ranks for each study variable, which may be unique and thus approach $n$.
\textcolor{black}{Because of these computational benefits,}
%For this reason, 
it can be advantageous to specify auxiliary quantiles for $Y_j$ with no or MCAR missingness, for example, by letting  $\mathcal{A}_j$ for those variables comprise a set of empirical quantiles. We note that empirical quantiles may be biased if the missingness mechanism is not MCAR.  For such $Y_j$, analysts should 
%in large samples. For variables suspected to be MAR, we recommend 
specify a small set of auxiliary quantiles believed to closely approximate their corresponding true quantiles and conduct sensitivity analysis to alternative specifications of $\mathcal{A}_j$. 

\subsection{Imputation with Limited Auxiliary Information}\label{MA-EHQL}

Given posterior samples of $\boldsymbol C_{\boldsymbol \theta}$ and posterior predictive samples of $z^{mis}_{ij}$ from Algorithm 
\ref{algEQL}, missing study variables may be imputed through $y^{mis}_{ij}=\tilde{F}_{j}^{-1}(\Phi^{-1}(z^{mis}_{ij}))$, where $\tilde{F}_{j}$ is an estimator of $F_{j}$. When $\mathcal{A}_{j}$ contains sufficient information to outline salient features of $F_{j}$, one option is to construct $\tilde{F}_{j}$ via a monotone interpolating spline through the  quantiles.
%$\{F_{j}^{-1}(\tau^{q}_{j}), \tau^{q}_{j}\}$. 
This expands the support of each $F_j$ beyond the quantiles in $\mathcal{A}_j$. 
%now consider the case when each $\mathcal{A}_{j}$ is sparse, i.e. 
However, when $\mathcal{A}$ comprises few  quantiles, the interpolation may not accurately approximate the marginals needed for imputation. Furthermore, this strategy does not account for the uncertainty in the resulting estimate of ${F}_j$ at values between the specified auxiliary quantiles.  
With this in mind, we describe a model-based approach for estimating
%and uncertainty quantification for 
intermediate quantiles of $Y_{j}$ not included in $\mathcal{A}_{j}$.  The method applies to any study variable, requires no modifications of the model for $\boldsymbol z$, and maintains the computational benefits of the EQL.

%The goal is %Together, the intermediate quantiles and $\mathcal{A}_{j}$ 
%to produce a more refined discrete approximation of each $F_{j}$.
%which in turn allows for more accurate interpolated estimates are more accurate. 
%Equipped with estimates for more quantiles of the marginal of $Y_{j}$, 
%it is then possible to interpolate $F_{j}$ using the strategy outlined in Section \ref{estimEQL-sec}. 

%We first incorporate unknown quantiles for each study variable into the EQL. 

The basic idea is to augment each $\mathcal{A}_j$ with a finite, increasing set of $s_j$ values, 
$\{y^{q}_{j}\}_{q=1}^{s_j}$, which we use as intermediate quantiles. Each $y_j^q$ 
is distinct from the quantiles in $\mathcal{A}_{j}$. We specify these points to be consistent with the support of $Y_j$; for example, if $Y_{j}$ is discrete, each $y^{q}_{j}$ is  discrete. For $Y_j$ taking on few values, $\{y_j^q\}$ can cover its full support. For $Y_j$ taking on many unique values, using $s_{j} \approx 15$ intermediate quantiles across the range of $\boldsymbol y_{j}^{obs}$  suffices to provide a discrete approximation of $F_{j}$, which we then smooth for imputation. 

\begin{figure}[t]
\centering 
\begin{tikzpicture}[scale=1]
    % Draw the number line
    \draw[<->] (-6,0) -- (6,0)
    node[above, font=\bfseries\large] at (-7,0) {$Y_{j}$};
    % Draw the ticks
    \foreach \x in {-5,-2,0,3,5}
    \draw[line width = 1.5] (\x,0.1) -- (\x,-0.2);
      \def\customlabels{{"$F_{j}^{-1}(0)$","$F_{j}^{-1}(.25)$", "$F_{j}^{-1}(.5)$","$F_{j}^{-1}(.75)$", "$F_{j}^{-1}(1)$"}}
    
    % Draw the ticks and custom labels for the first number line
    \foreach \x/\label [count=\i] in {-5,-2,0,3,5}
    {
        \node[below, font=\bfseries\small, text=black!80] at (\x,0.75) {\pgfmathparse{\customlabels[\i-1]}\pgfmathresult};
    }
    \foreach \x in {-4,-3,-1,1,2,4}
   \color{blue} \draw[line width = 1.5] ( \x,0.2) -- (\x,-0.2);
    % Draw the numbers
    \foreach \x in {-5,-2,0,3,5}
    \node[below] at (\x,-0.3) {\x};
     \foreach \x in {-4,-3,-1,1,2,4}
    \node[below] at (\x,-0.3) {\color{blue}{\x}};
    % Add underbraces
% Add underbraces
    \draw[decorate,decoration={brace,amplitude=5pt,mirror},yshift=-15pt] (-5,-0.25) -- (-4,-0.25) node[midway,below,yshift=-3pt]{$\mathcal{I}^{1}_{j}$};
    \draw[decorate,decoration={brace,amplitude=5pt,mirror},yshift=-15pt] (-4,-0.3) -- (-3,-0.3) node[midway,below,yshift=-3pt]{$\mathcal{I}^{2}_{j}$};
    \draw[decorate,decoration={brace,amplitude=5pt,mirror},yshift=-15pt] (-3,-0.3) -- (-2,-0.3) node[midway,below,yshift=-3pt]{$\mathcal{I}^{3}_{j}$};
    \draw[decorate,decoration={brace,amplitude=5pt,mirror},yshift=-15pt] (-2,-0.3) -- (-1,-0.3) node[midway,below,yshift=-3pt]{$\mathcal{I}^{4}_{j}$};
    \draw[decorate,decoration={brace,amplitude=5pt,mirror},yshift=-15pt] (-1,-0.3) -- (0,-0.3) node[midway,below,yshift=-3pt]{$\mathcal{I}^{5}_{j}$};
    \draw[decorate,decoration={brace,amplitude=5pt,mirror},yshift=-15pt] (0,-0.3) -- (1,-0.3) node[midway,below,yshift=-3pt]{$\mathcal{I}^{6}_{j}$};
    \draw[decorate,decoration={brace,amplitude=5pt,mirror},yshift=-15pt] (1,-0.3) -- (2,-0.3) node[midway,below,yshift=-3pt]{$\mathcal{I}^{7}_{j}$};
    \draw[decorate,decoration={brace,amplitude=5pt,mirror},yshift=-15pt] (2,-0.3) -- (3,-0.3) node[midway,below,yshift=-3pt]{$\mathcal{I}^{8}_{j}$};
    \draw[decorate,decoration={brace,amplitude=5pt,mirror},yshift=-15pt] (3,-0.3) -- (4,-0.3) node[midway,below,yshift=-3pt]{$\mathcal{I}^{9}_{j}$};
    \draw[decorate,decoration={brace,amplitude=5pt,mirror},yshift=-15pt] (4,-0.3) -- (5,-0.3) node[midway,below,yshift=-3pt]{$\mathcal{I}^{10}_{j}$};

    \draw[<->, line width = 2] (-4.5,-1.65) -- (-5.5,-2.75);
    \draw[<->, line width = 2] (-3.5,-1.65) -- (-4.5,-2.75);
    \draw[<->, line width = 2] (-2.5,-1.65) -- (-3.5,-2.75);
    \draw[<->, line width = 2] (-1.5,-1.65) -- (-2.25,-2.75);
    \draw[<->, line width = 2] (-.5,-1.65) -- (-.75,-2.75);
    \draw[<->, line width = 2] (.5,-1.65) -- (.5,-2.75);
    \draw[<->, line width = 2] (1.45,-1.65) -- (1.5,-2.75);
    \draw[<->, line width = 2] (2.5,-1.65) -- (2.5,-2.75);
    \draw[<->, line width = 2] (3.5,-1.65) -- (3.75,-2.75);
    \draw[<->, line width = 2] (4.5,-1.65) -- (5.25,-2.75);
    
        % Draw the number line
    \draw[<->] (-6,-3) -- (6,-3)
    node[above, font=\bfseries\large] at (-7,-3)
    {$Z_{j}$};
    % Draw the ticks
    \foreach \x in {-3,0,3}
    \draw[line width = 1.5] (\x,-2.8) -- (\x,-3.2);
    \foreach \x in {-5,-4,-1.5,1,2,4.5}
    \color{blue} \draw[line width = 1.5] (\x,-2.8) -- (\x,-3.2);
    % % Draw the numbers
    % \def\customlabels{{"","", "$\Phi^{-1}(0.50)$", "$\Phi^{-1}(0.75)$","$\infty$"}}
    
    % % Draw the ticks and custom labels for the first number line
    % \foreach \x/\label [count=\i] in {-6,-3,0,3,6}
    % \node[below] at (\x,-3.1) {\pgfmathparse{\customlabels[\i-1]}\pgfmathresult};
    
    % Add underbraces
    \draw[decorate,font = 
    \small,decoration={brace,amplitude=5pt,mirror},yshift=-10pt] (-6,-3) -- (-5,-3) node[midway,font = \bfseries,below, rotate = 45, anchor = east,shift ={(0,-.5)}]{($\boldsymbol{-\infty,\color{blue}?\color{black}]}$};
    \draw[decorate,font = \small, decoration={brace,amplitude=5pt,mirror},yshift=-10pt] (-5,-3) -- (-4,-3) node[rotate = 45,midway,below, anchor = east, shift ={(0,-.5)}]{$\boldsymbol{(\color{blue}?,?\color{black}]}$};
    \draw[decorate,font = \small,decoration={brace,amplitude=5pt,mirror},yshift=-10pt] (-4,-3) -- (-3,-3)  node[rotate = 45, midway,below, anchor = east,shift ={(0,-.5)}]{$\boldsymbol{(\textcolor{black} ?, \Phi^{-1}(.25)]}$};
    \draw[decorate,font = \small, decoration={brace,amplitude=5pt,mirror},yshift=-10pt] (-3,-3) -- (-1.5,-3)  node[rotate = 45,midway,below,anchor = east,shift ={(0,-.5)}]{$\boldsymbol{(\Phi^{-1}(.25), \textcolor{black}{?}]}$};
    \draw[decorate,font = \small, decoration={brace,amplitude=5pt,mirror},yshift=-10pt] (-1.5,-3) -- (0,-3)  node[rotate = 45,midway,below,anchor = east,shift ={(0,-.5)}]{$ \boldsymbol{(\textcolor{black}{?}, \Phi^{-1}(0)]}$};
    \draw[decorate,font = \small, decoration={brace,amplitude=5pt,mirror},yshift=-10pt] (0,-3) -- (1,-3)  node[rotate = 45,midway,below,anchor = east,shift ={(0,-.5)}]{$\boldsymbol {(\Phi^{-1}(0),\textcolor{black}{?}]}$};
    \draw[decorate,font = \small, decoration={brace,amplitude=5pt,mirror},yshift=-10pt] (1,-3) -- (2,-3)  node[rotate = 45,midway,below,anchor = east,shift ={(0,-.5)}]{$\boldsymbol{(\textcolor{black}{?},\textcolor{black}{?}]}$};
     \draw[decorate,font = \small, decoration={brace,amplitude=5pt,mirror},yshift=-10pt] (2,-3) -- (3,-3)  node[rotate = 45,midway,below,anchor = east,shift ={(0,-.5)}]{$\boldsymbol{(\textcolor{black}{?},\Phi^{-1}(0.75)]}$};
        \draw[decorate,font = \small, decoration={brace,amplitude=5pt,mirror},yshift=-10pt] (3,-3) -- (4.5,-3)  node[rotate = 45,midway,below,anchor = east,shift ={(0,-.5)}]{$\boldsymbol{(\Phi^{-1}(0.75),\textcolor{black}{?}]}$};
                \draw[decorate,font = \small, decoration={brace,amplitude=5pt,mirror},yshift=-10pt] (4.5,-3) -- (6,-3)  node[rotate = 45,midway,below,shift ={(0,-.5)},anchor = east]{$\boldsymbol{(\textcolor{black}{?}, \infty]}$};

    % Draw the vertical equivalence line
\end{tikzpicture}

\caption{Augmenting $\mathcal{A}_{j}$ with intermediate points $\{y^{q}_{j}\} = \{-4,-3,-1,1,2,4\}$.  Thus, $|\mathcal{A}^{*}_{j}| = 11$. 
%The finer partition induces a partial ordering on the latent scale for $Z_{j}$ similar to the event \eqref{aux event}. However, 
For $y_{ij} \in \mathcal{I}^{q}_{j}$ with at least one intermediate endpoint, the unknown bounds on the latent interval containing $z_{ij}$ are 
%unknown, besides their locations relative to known auxiliary quantiles. This property is 
represented by the question marks.
%undRLying the latent intervals created by the mapping between $Y_{j}$ and $Z_{j}$.
}\label{eql-fig2}
\end{figure}
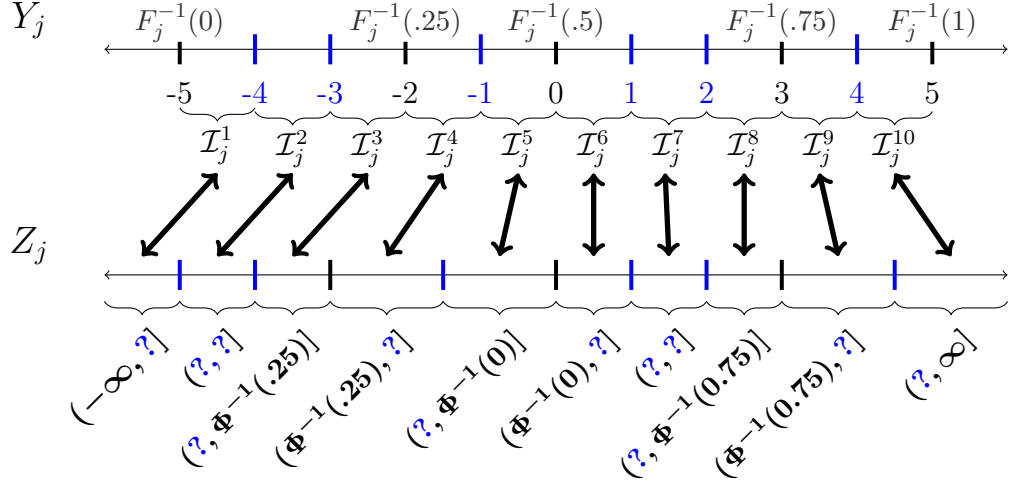

The key step is to coarsen $\boldsymbol y^{obs}$ into bins using both the quantiles in $\mathcal{A}$ and intermediate quantiles in $\{y^{q}_{j}\}_{q=1}^{s_j}$, through which we can relate the latent variables to the binned data. In doing so, we maintain the ordering between auxiliary and intermediate quantiles on the latent scale.
%which enables estimation of $\{F_{j}(y^{q}_{j})\}_{q=1}^{s_j}$ .
%constructing a similar correspondence to the latent variables under the model as in \eqref{marg-work}-\eqref{marg-work-expand}.
Let $\mathcal{A}^{*}_{j} = \mathcal{A}_{j} \cup \{y^{q}_{j}\}_{q=1}^{s_j}$, $a_{j} = \lvert \mathcal{A}^{*}_{j}\rvert$, and $\mathcal{A}^{*} = \{\mathcal{A}^{*}_{j}\}_{j=1}^{p}$. Using $\mathcal{A}^{*}_{j}$, we construct $a_{j} - 1$ disjoint intervals $\{\mathcal{I}^{1}_{j}, \dots, \mathcal{I}^{a_{j}}_j\}$ like those in \eqref{int_aux}, partitioning the support of $Y_{j}$ at the $a_j$ points in $\mathcal{A}_j^*$. We then  define $b_{ij}^{obs}$ similarly to \eqref{bins} but based on the $a_j-1$ intervals. 
%and the partial ordering between $b^{obs}_{ij}$ and $z^{obs}_{ij}$. 
%When $\mathcal{I}^{q}_{j}$ is constructed from one or more points in $\{y^{q}_{j}\}_{q=1}^{s_j}$, the precise latent interval for $z^{obs}_{ij}$ is unknown, except for its location relative to the quantiles in $\mathcal{A}_{j}$.  
The interval for $z_{ij}^{obs}$ is  determined relative to the closest auxiliary quantiles in $\mathcal{A}_{j}$ and adjacent intermediate points in $\{y_{j}^q\}$.
This is illustrated in Figure \ref{eql-fig2}. \textcolor{black}{Using the mapping from the intervals to the latent variables allows us to estimate $\boldsymbol C_{\boldsymbol \theta}$ and ${F}_{j}(y_{j}^{q})$.} 

We first formally describe the method, followed by motivation for why it works.  Whenever
%observing
$b^{obs}_{ij}= q$ (and analogously, $y_{ij}^{obs} \in \mathcal{I}_{j}^{q})$ we have $\ell_{ij}< z_{ij}^{obs} <u_{ij},$ where 
\begin{align}\label{EHQL bounds}
    \ell_{ij} &= \max\{\Phi^{-1}(\tau_{ij}^{\ell}), \max(z_{vj}^{obs}: b_{vj} = q -1 ); v = 1, \dots, n\}\\
    u_{ij} &= \min\{\Phi^{-1}(\tau_{ij}^{u}), \min(z_{vj}^{obs}: b_{vj} = q+1 ); v = 1, \dots, n\}\nonumber
\end{align}
and $\tau_{ij}^{\ell}, \tau_{ij}^{u}$ are defined as in Step 1 of Algorithm \ref{algEQL}. The interval $(\ell_{ij},u_{ij})$ ensures that whenever $b_{ij}^{obs} < b_{vj}^{obs}$ we have $z_{ij}^{obs} < z_{vj}^{obs}, i \neq v$. With intermediate points determined analogously for $(Y_1,\dots, Y_p)$, \textcolor{black}{we define the quantile ordering 
set restriction 
%becomes 
%\begin{equation}
$\mathcal{D}^{*}(\boldsymbol b^{obs})$, which  incorporates the intermediate quantiles to encode the   condition that $z_{ij}^{obs}$ is in the interval \eqref{EHQL bounds}
%$ \in (\ell_{ij}, u_{ij}]$ 
%defined by its 
corresponding to its $b_{ij}^{obs}$.} 
%$b_{ij}^{obs}= q$
%\coloneqq \{ \boldsymbol z^{obs} : b_{ij}^{obs}= q \implies z_{ij} \in (\ell_{ij}, u_{ij}], i = 1,\dots, n; j = 1,\dots, p\}.
%\end{equation}
We replace $\mathcal{A}$ with $\mathcal{A}^{*}$ and  $\mathcal{D}(\boldsymbol b^{obs})$  with $ \mathcal{D}^{*}(\boldsymbol b^{obs})$ in \eqref{marg-work}  and \eqref{marg-work-expand} to estimate  $\boldsymbol C_{\boldsymbol \theta}$.
%and posterior \eqref{EQLpost}.  
We refer to this variation 
%of the working model and resultant posterior is referred to 
as the extended hybrid quantile likelihood (EHQL). Estimating the copula correlation under the EHQL requires minor modifications to Algorithm \ref{algEQL}, namely replacing the truncation bounds for $z_{ij}^{obs}$ in Step 1 with \eqref{EHQL bounds}.
%\textcolor{black}{Using these intervals, we estimate the quantile for each $y_j^q$ on the observed scale, as we now describe. First, since we know the quantiles in $\mathcal{A}_{j}$, for any $y_{j} < F_{j}^{-1}(\tau_{j}^{q})$ we have $F_{j}(y_{j}) < \tau_{j}^{q}$. }
%Furthermore, the Gaussian copula model enables us to express cumulative probabilities on the observed scale in terms of latent variables. Together, we leverage the marginal identifying information in $\mathcal{A}$ with the data coarsening in Section \ref{estimEQL-sec} to incorporate and estimate intermediate quantiles of each marginal within the EQL Gaussian copula framework.}

The modified Algorithm \ref{algEQL} produces draws of each $z_{ij}^{obs}$. 
%which are restricted to an interval on the latent scale corresponding to $b_{ij}^{obs}$ and thus $\mathcal{I}_{j}^{q}$. 
For any interval $\mathcal{I}_{j}^{q}$ constructed using an intermediate point as an upper bound, define 
%constructed with $y^{q}_{j}$ as an upper bound, we define 
%\begin{equation}\label{cutpoint}
    $Z_{j}^{q} = \max\{z_{ij}^{obs}: b_{ij}^{obs} = q\}$.
%\end{equation}
%This is the latent variable corresponding to the largest $y_ij$ in the interval  $\mathcal{I}^{q}_{j}$. 
%Because each $\mathcal{I}^{q}_{j}$ is located relative to the known quantiles in $\mathcal{A}_{j}$, with 
%Under realistic AN missingness mechanisms, w
With enough observations having $b_{ij}^{obs} = q$, Algorithm \ref{algEQL} should generate values of $z_{ij}^{obs}$ that cover much of the interval on the latent scale corresponding to $\mathcal{I}_{j}^{q}$.  When this is the case, we should sample a $Z_j^q$  that is close to the corresponding $\Phi^{-1}(F_{j}(y_{j}^{q}))$ under the copula model. 
%Under the copula model, \eqref{cutpoint} is a random variable bounded above by $\Phi^{-1}(F_{j}(y_{j}^{q}))$. 
%Since \eqref{cutpoint} is monotone in $n$, $Z_{j}^q$ must converge to its upper bound. Furthermore, $\mathcal{A}_{j}$ locates $\mathcal{I}_{j}^q$ such that it is appropriately situated between auxiliary quantiles. e.g., we hope we have enough data points in every interval so that we can roughly approximate the maximum Z in that interval and the corresponding F for that Z. 
 %Algorithm \ref{algEQL} produces posterior draws of $Z_{j}^{q}$, and so 
 Using each sampled $Z_j^q$ from
 %from each iteration of   
 Algorithm \ref{algEQL},
 for all $j$ and $q$, we compute  posterior draws  of
 \begin{equation}\label{marginadjust}
     \tilde{F}_{j}(y_{j}^{q}) = \Phi(Z_{j}^{q}).
 \end{equation}
The draws of \eqref{marginadjust} provide estimates of uncertainty in the intermediate quantiles. 
When few individuals have $b_{ij}^{obs} = q$, which may occur for intervals constructed at quantiles in the tails, we expect higher uncertainty in the draws of $\tilde{F}_{j}(y_{j}^{q})$.  
This is borne out in the simulations of Section \ref{sec:sims}.  Algorithm \ref{algMA} summarizes the process for
estimating the $\{\tilde{F}_j(y_j^q)\}$.

 \begin{algorithm}[t]
\caption{Estimating $F_j$ under the EHQL Gaussian copula}
\begin{algorithmic}\label{algMA}
    \STATE \textbf{Require:}  $\mathcal{A}^{*}$ and one draw of $\boldsymbol z^{obs}$ from Algorithm~\ref{algEQL}
    \STATE \textbf{Return:} One draw  $\{\tilde{F}_{j}(y_{j}^q)\}_{q=1}^{s_{j}}$, $j \in \{1,\dots, p\}$
     \hspace{3em}\FOR{$j \in \{1,\dots, p\}$}
           \STATE Compute $Z_{j}^{q} = \max\{z_{ij}^{obs}: b_{ij}^{obs} = q\}$ for each element in $\{y_{j}^{q}\}_{q=1}^{s_{j}}$
           \STATE Compute $\tilde{F}_{j}(y_{j}^{q})  = \Phi(Z_{j}^{q})$  
               \ENDFOR
\end{algorithmic}
\end{algorithm}
 For imputation, we interpolate between posterior samples of $\{y_{j}^{q}, \tilde{F}_{j}(y_{j}^{q})\}_{q=1}^{s_{j}}$ and the points in  $\mathcal{A}_{j}$ via a monotone spline, fit using the package \vtt{splinefun} in \vtt{R}. Because the upper and lower bounds of $\mathcal{A}^{*}_j$ are fixed, the smoothing step is guaranteed to produce samples of a valid distribution function that pass through the points in $\mathcal{A}^{*}_{j}$. We use these 
%smoothed 
versions of $\tilde{F}_{j}$ to impute each $y^{mis}_{ij}$ at any iteration of Algorithm \ref{algEQL} by setting $y^{mis}_{ij} = \tilde{F}_{j}^{-1}(\Phi(z^{mis}_{ij}))$. 

Using \eqref{marginadjust} in addition to $\mathcal{A}_{j}$ to approximate 
%interpolation of 
$F_{j}$ has advantages over approximating $F_j$ based on 
%interpolating across 
the quantiles in $\mathcal{A}_j$ alone, as done in the EQL. First, $\{y_{j}^{q}\}$ lends more information to the discrete approximation, helping the estimator better capture features of each marginal. Second, the draws of $\Phi(Z_j^q)$ in \eqref{marginadjust} propagate uncertainty about ${F}_{j}(y_{j}^{q})$ to the imputations, 
%which is beneficial for imputation, 
whereas the EQL interpolation of ${F}_{j}(y_{j}^{q})$ is deterministic. For these reasons, we recommend employing the EHQL for copula estimation and Algorithm \ref{algMA} for marginal CDF estimation whenever imputation is needed. We note that \citet{feldman2022nonparametric} use a strategy akin to what we do in \eqref{marginadjust} for the RL Gaussian mixture copula under MAR mechanisms. \textcolor{black}{However, their method cannot leverage auxiliary marginal information and hence does not address the nonignorable missingness setting we consider here.} 

\section{Simulation Studies}\label{sec:sims}
In this section, we present results of simulation studies evaluating (i) the impact of the level of detail in the marginal distributions on the quality of inferences and (ii) the repeated sampling performance of the model as a multiple imputation engine compared to alternative methods that do not use the auxiliary information.
%for imputations.

In the simulations as well as the analysis of the North Carolina data in Section \ref{sec:realdat}, the prior distribution on the parameters $\boldsymbol \theta$ that index $\boldsymbol C_{\boldsymbol \theta}$ is 
%To put this into action,
the factor model,
\begin{equation} \label{factor}
    %(\boldsymbol{z}_i^{obs}, \boldsymbol{z}_i^{mis}, \boldsymbol z_{\boldsymbol{r}_i}) 
    \boldsymbol{z}_i
    = \boldsymbol \alpha + \boldsymbol \Lambda \boldsymbol \eta_i + \boldsymbol \epsilon_i, \quad \boldsymbol \epsilon_i \sim N_{2p}(\boldsymbol 0, \boldsymbol \Sigma).
\end{equation}
Here, $\boldsymbol \Lambda$ is a $(2p)\times k$ matrix of factor loadings possibly with $k << 2p$; $ \boldsymbol \eta_i$ is a $k\times 1$ vector of factors; and, $\boldsymbol \Sigma = diag(\sigma_{1}^{2}, \dots, \sigma_{2p}^{2})$. By specifying 
 $ \boldsymbol{\eta}_i \sim N_{k}(\boldsymbol{0}, \boldsymbol I_{k})$,  where $\boldsymbol I_{k}$ is the $k \times k$ identity matrix, marginally we have  
 %the latent variables satisfy
 %$(\boldsymbol{z}_i^{obs}, \boldsymbol{z}_i^{mis}, \boldsymbol z_{\boldsymbol{r}_i}) 
 $\boldsymbol{z}_i \sim N_{2p}(\boldsymbol \alpha, \boldsymbol \Omega)$, where $\boldsymbol{\Omega} = \boldsymbol{\Lambda} \boldsymbol{ \Lambda}^{\intercal} + \boldsymbol \Sigma$ is the reduced rank covariance. Thus, $\boldsymbol{\theta} = (\boldsymbol \Lambda, \boldsymbol \Sigma)$. Given posterior samples of $\boldsymbol{\Omega}$, samples of $\boldsymbol C_{\boldsymbol \theta}$ are obtained by scaling $\boldsymbol \Omega$ into  correlations.
 %i.e., $\rho_{uv} = \omega_{uv}/ \sqrt{\omega_{uu}\omega_{vv}}$.
 %Modeling the latent variables on covariance scale is convenient
% done for computational simplicity and
 %because correlations are invariant to scaling.  
%The factor model \eqref{factor} is beneficial for moderate-to-high dimensional data. 
The prior on the factor loadings provides shrinkage, automating rank selection \citep{bhattacharya2011sparse}. In addition, the components of  $\boldsymbol{z}_i$ 
%$(\boldsymbol{z}_i^{obs}, \boldsymbol{z}_i^{mis}, \boldsymbol z_{\boldsymbol{r}_i})$ 
are independent conditional on $\boldsymbol \eta_i$, which benefits computation, especially in Step 1 of Algorithm \ref{algEQL}. The full hierarchical specification of \eqref{factor} is available in Section E of the supplement.
%Though the posterior consistency in Theorem \ref{postconsist} holds for a prior over the space of all valid, full-rank correlation matrices, simulations suggest that the factor model still can offer posterior consistency.

% in favor of parsimony.

%that do not use auxiliary margins usiugon a repeated sampling simulation with genuine data.

%\vspace{-12pt}

\subsection{Accuracy with Sparse Auxiliary Information}\label{sim1}

%We now design a simulation study to evaluate how the amount of auxiliary information known for each study variable affects estimation and imputation under the Gaussian copula with nonignorable missing data.
%As highlighted in Section \ref{auxapprox}, w

%the number of auxiliary quantiles known for each study variable.
%This represents a plausible situation in real data analysis.  There may be incomplete knowledge of the marginal distribution of each study variable, e.g., only a few population quantiles are known from external sources.  Or, in a sensitivity analysis, it may be easier for domain experts to specify several quantiles of the marginal distribution than the entire distribution. 

%Recognizing the downstream goal of imputation, the Gaussian copula requires some estimate of each marginal. Consequently, we evaluate the ability of the margin adjustment to estimate intermediate quantiles of diverse marginal distribution functions. In conjunction, we demonstrate that even with little external information and severe missingness, the EHQL enables accurate estimation  of both the copula correlation and marginal distribution functions. 

%To assess these questions, we
%We generate data from the Gaussian copula model and apply our imputation algorithm with different marginals.  Specifically, we 

%Theorem \ref{postconsistEQL} provides posterior consistency for the copula correlation when each $\mathcal{A}_j$ comprises at least $m_{j} \geq 3$ ground truth quantiles. 
We first investigate how sensitive the contraction of the posterior is to the cardinality of $\mathcal{A}_{j}$.  We simulate 
 $2p$-dimensional observations $\{(\boldsymbol y_{i}, \boldsymbol r_i)\}_{i=1}^{n}$ for $n \in \{200,1000,5000\}$ and $p \in \{5,10,20\}$  from Gaussian copulas. For each $(n,p)$, we randomly generate $\boldsymbol C_{0}$ from a scaled inverse-Wishart distribution. Since each entry of $\boldsymbol C_{0}$ is non-zero, 
 %including those for $\boldsymbol R$, 
 this generates nonignorable missingness per Lemma \ref{lemma1}.  We vary the amount of missing data by setting $\Phi(\alpha_{r_{j}}) \in \{0.25,0.50\}$. Thus, marginally, missingness in each study variable is approximately 25\% or  50\%. For $j \leq p$, we vary $F_j$ 
 so that  
 $Y_{j} \sim
\mbox{Gamma}(1,1)$ when $j \in \{1, 4, 7, \dots\}$;
$Y_{j} \sim 
\mbox{t}(\nu = 5, ncp = 2)$ when  $j \in \{2, 5, 8, \dots\}$; and, 
$Y_{j} \sim \mbox{Beta}(1,2)$ when $j \in \{3, 6, 9, \dots\}$. Here, $\nu$ and $ncp$ are a degrees of freedom and non-centrality parameter. 
Each $z_{ij}$ for $j \leq p$ is transformed to $y_{ij}$ via \textcolor{black}{$F_{j}^{-1}(\Phi(\tilde{z}_{ij}))$.}  When $z_{i(p+j)}>0$, we set $r_{ij} =1$ and make $y_{ij}$ missing. 

% For example, with $p = 5$ and 50\% marginal missingness, we generate   
% \begin{align}
%     (\boldsymbol z_{\boldsymbol Y}, \boldsymbol z_{\boldsymbol R}) &\sim N_{10}( \boldsymbol \alpha, \boldsymbol C_{0})\label{lat}\\
% Y_1 \sim \mbox{Gamma}(1,1), Y_2 \sim\mbox{t}(5,2), Y_3 &\sim \mbox{Beta}(1,2),Y_4\sim \mbox{Gamma}(1,1), Y_5 \sim \mbox{t}(5,2)\\
% \boldsymbol \alpha = (\boldsymbol \alpha_{\boldsymbol Y}, \boldsymbol \alpha_{\boldsymbol R}) &= \{\boldsymbol 0, \Phi^{-1}(0.50)\}.
% \end{align}
%Figure \ref{examplesim} displays the observed and missing values for \eqref{lat}. \texcolor{red}{INCLUDE FIGURE?  MIGHT BE NICE TO ENABLE VISUALIZATION.  Also, did we repeat this more than one time?  It is always good practice to repeat simulations more than once to show that we did not get "lucky" with the one draw.  We might redo the simulation one or two more times and put the plots in the supplement as evidence for a statement like, "We repeated the simulation for each combination and observe qualitatively similar results."}   

%Let $\mathcal{A} = (\mathcal{A}_1, \dots, \mathcal{A}_p)$ represent this auxiliary information. 
We estimate the copula using Algorithm \ref{algEQL}  
for different 
%incorporating three 
granularities of $\mathcal{A}$.  %progressively increase the number of known marginal quantiles from 
The first uses the fully specified, true $\{F_j\}_{j=1}^p$, referred to as ``Full.''  The second is significantly more sparse, using just the lower/upper bounds and median,  i.e., each $\mathcal{A}_{j} = \{F_{j}^{-1}(0), F_{j}^{-1}(0.5), F_{j}^{-1}(1)\}$. This is referred to as ``EQL-M.'' The supplement includes results 
%for variants of EQL
where $\mathcal{A}$ comprises deciles and every fourth quantile.
%\textcolor{black}{I don't see the deciles and .04 quantiles in Figure 2.  We should describe only what goes into the figure.  We can say instead that we examined deciles and .04 and saw similar results.  ALternatively, perhaps we should rethink the display to allow more quantiles. Do we need to display all the posterior intervals?  Maybe show a stratified sample in the paper, with full results in the supplement? That might allow us to get more columns. Lastly, can we order the intervals in each plot by $\rho$ instead of bias? That's important for arguing that the results are stable across auxiliary information if the bias for any particular $\rho$ is the same across all sources, which we can't conclude from the current plot ordering.  Also, that might show the highest absolute correlations also have the largest bias.  If so, such differences might be a product of the model fit rather than our missing data method (e.g. maybe the prior shrinks correlations to zero) } 
% The third includes  deciles, i.e., each $\mathcal{A}_{j} = \{F_{j}^{-1}(0), F_{j}^{-1}(0.1),  F_{j}^{-1}(0.2),\dots, F_{j}^{-1}(1)\}$. \textcolor{black}{Delete above sentence if plots exclude deciles.}
%to every fourth quantile (i.e., each $\mathcal{A}_{j} = \{F_{j}^{-1}(0), \\F_{j}^{-1}(0.04),  F_{j}^{-1}(0.08),\dots, F_{j}^{-1}(1)\}$). \textcolor{black}{select only quantiles we keep in plot in main text}.  
We also employ the EHQL, with each  $\mathcal{A}_{j} = \{F_{j}^{-1}(0), F_{j}^{-1}(0.5), F_{j}^{-1}(1)\}_{j=1}^p$ and $s_{j} \approx 15$ intermediate quantiles. We refer to this as ``EHQL-M.''  We construct intermediate quantiles in each simulation run by first specifying 20 evenly spaced bins over the range of $\boldsymbol y^{obs}_j$ and creating $\{\mathcal{I}_{j}^{q}\}$ based on the bins occupied by the observed data. This results in approximately 15 intermediate points in each simulation run.
%This created  and posterior samples of the margin adjustment \eqref{marginadjust} at these points are accumulated using Algorithm \ref{algMA}. 

We generate several datasets for each $(n, p, \boldsymbol \alpha)$ setting. For each dataset and  $\mathcal{A}$, we simulate 1000 posterior samples of  $\boldsymbol C_{\boldsymbol \theta}$. 
%It is important to note that the EQL and EHQL copula models differ from the data generating process. 
We emphasize that fitting EQL-M and EHQL-M in these datasets does not simply parrot the data generating model, as they use only a sparse set of marginal quantiles in $\mathcal{A}$. 
%to estimate $\boldsymbol C_{\boldsymbol \theta}$ and impute $\boldsymbol y^{mis}$. 
%while the latter uses this to additionally approximate each $F_{j}$ via \eqref{marginadjust}. 
We also include comparisons to the copula fit under the RL of \citet{hoff2007extending}, which we estimate using the \vtt{sbgcop} package in \vtt{R}. Although the RL copula is a joint model for $(\boldsymbol y, \boldsymbol r)$, it does not leverage any information beyond the observed data.
%We also extract smoothed posterior samples of the margin adjustment using the interpolation strategy outlined in Section \ref{MA-EHQL}. 

\begin{figure}[t]
    \centering
\includegraphics[width = \textwidth, keepaspectratio]{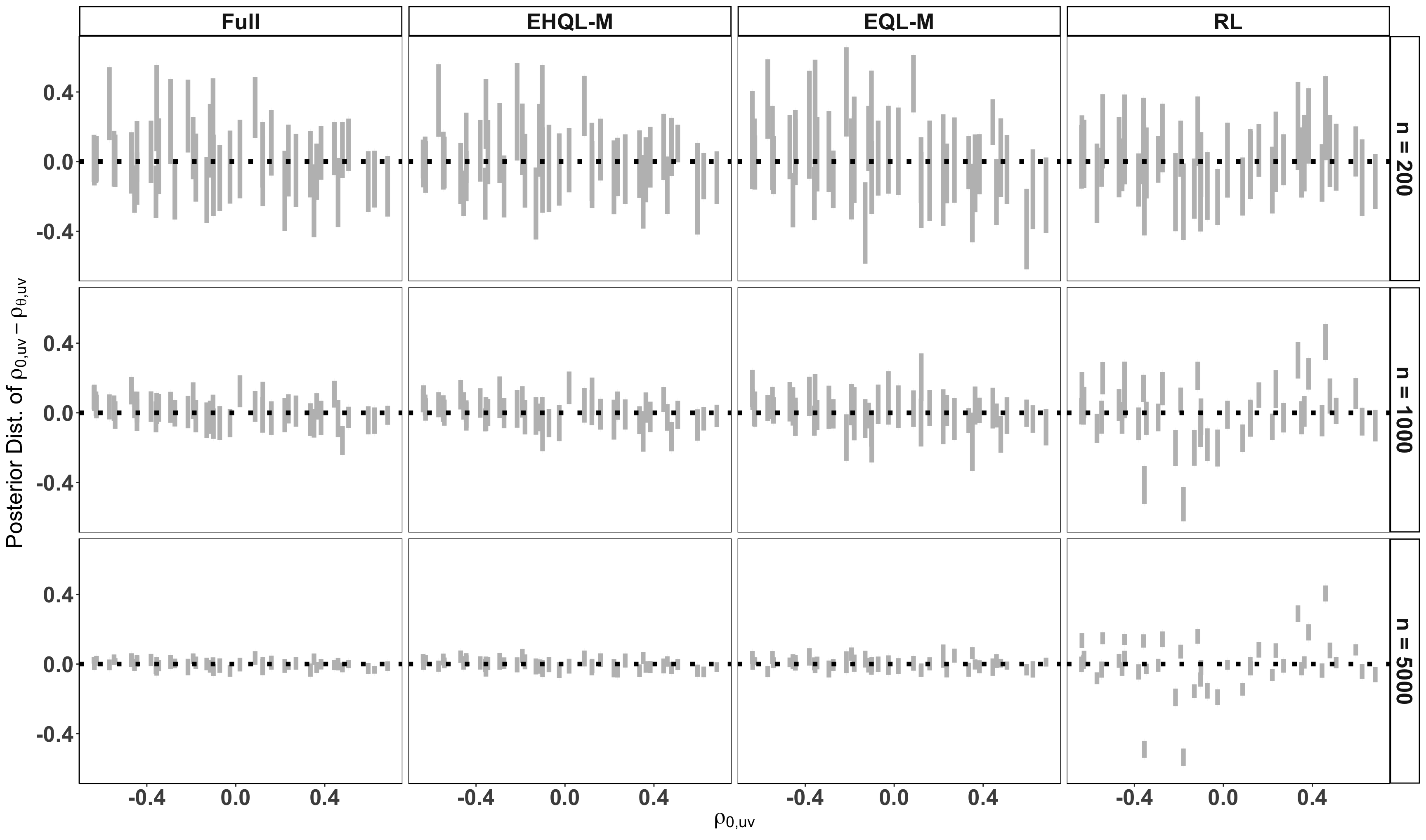}
    \caption{Plots of $95\%$ credible intervals for  $\rho_{0, uv} -\rho_{\boldsymbol \theta, uv}$ for all  unique correlations in $\boldsymbol C_0$ using various $\mathcal{A}$ in scenario with $~50\%$ missing values for each of $p=5$  study variables.
    %    for three sample sizes. 
    %based on 1000 posterior samples of $\rho_{s}^{uv} - \rho_{0}^{uv}$.  
    %In each panel we plot $95\%$ credible intervals based on 1000 posterior samples of $\rho_{s}^{uv} - \rho_{0}^{uv}$. %extracted under the proposed approach and Algorithm \ref{algEQL}. 
    ``Full'' uses full marginals; ``EQL-M'' uses  the lower/upper bounds and median; ``EHQL-M'' augments each $\mathcal{A}_j$ for EQL-M with $s_j\approx 15$ intermediate quantiles. %Results are presented for scenario 
    %with $~50\%$ missing values for each of $p=5$  study variables. 
    The right-most column provides results 
    %for the copula correlation 
    estimated under the RL. For the larger sample sizes, posterior inferences for $\boldsymbol C_{\boldsymbol \theta}$ under the EQL/EHQL are accurate with minor efficiency losses when incorporating fewer auxiliary quantiles. In contrast, the inferences under RL can be biased.} 
    %Fixing the column and comparing across rows, the biases are virtually indistinguishable. In addition, the bias diminishes as the sample size increases. This suggests that the posterior consistency result in Theorem \ref{postconsist} may be more broadly applicable to settings where there is incomplete information about each marginal. Furthermore, the copula estimated under the EHQL (MA + (0,0.5,1)) maintains this consistency. }
    \label{figconsist}
\end{figure}

Let $\rho_{\boldsymbol \theta,uv}$  represent 
%the $s$th draw of 
the element (correlation) 
in row $u$ and column $v$ of $\boldsymbol C_{\boldsymbol \theta}$. Similarly, let $\rho_{0,uv}$ be the corresponding element  in the $\boldsymbol C_0$ used in data generation. 
Figure \ref{figconsist} displays 95\% credible  intervals based on 1000 draws of $\rho_{0,uv} -\rho_{\boldsymbol \theta,uv}$ for the $2p(2p-1)/2$  unique correlation coefficients 
%using different amounts of auxiliary information and proportions of missing data.
%for $\mathcal{A}_{j} = F_{j}^{-1}(0), F_{j}^{-1}(0.5), F_{j}^{-1}(1)\}$ (Median), the same level of auxiliary information plus additional grid points estimated using the margin adjustment (Median + MA), and complete knowledge of each marginal (All Quantiles). 
%The results are shown 
for one randomly selected dataset in the $p = 5$ and 50\% missingness scenario.  Results for other datasets and settings are qualitatively similar; see Section F.1 of the supplement.  

%The posterior consistency result in 
Theorem \ref{postconsistEQL} implies that, under the EQL,  $\rho_{0, uv} -\rho_{\boldsymbol \theta, uv}$ should converge to 0 as sample size increases \textcolor{black} regardless of the level of granularity in $\mathcal{A}$.  This is confirmed in Figure \ref{figconsist}. The credible intervals 
%for the $\rho_{0,uv}$ 
under EQL-M are slightly wider than those under Full, which suggests a loss in efficiency with lower levels of auxiliary information. \textcolor{black}{ However, augmenting sparse auxiliary information with intermediate quantiles (e.g., Figure~\ref{eql-fig2}) corrects this discrepancy}; the intervals for EHQL-M are virtually indistinguishable from those for Full. 
%We emphasize that EHQL-M leverages the same level of auxiliary information as EQL-M, but incorporates unknown intermediate quantiles into the estimation.
Even though EHQL-M and EQL-M leverage the same $\mathcal{A}$, EHQL-M makes greater use of the information in $\boldsymbol y^{obs}$ by better locating the corresponding $\boldsymbol z^{obs}$, which improves precision.   By contrast, for the two larger values of $n$, estimates of $\boldsymbol C_{\boldsymbol \theta}$ under the RL copula are significantly biased.  
%This results from the lack of identifying information in the RL model, i.e., it does not use an $\mathcal{A}$. 
Finally, for $n=200$, the sampling variability is sufficiently large that inferences for all four methods are not obviously different qualitatively.
%For each level of auxiliary information, we are 
%We compute $\rho_{s}^{ij} - \rho_{0}^{ij}, i \neq j$.
%which measures the bias of individual coefficients in $\boldsymbol C_{s}$ relative to the data generating copula correlation $\boldsymbol C_{0}$. In 

\begin{figure}[t]
    \centering
    \includegraphics[width = .85\textwidth, keepaspectratio]{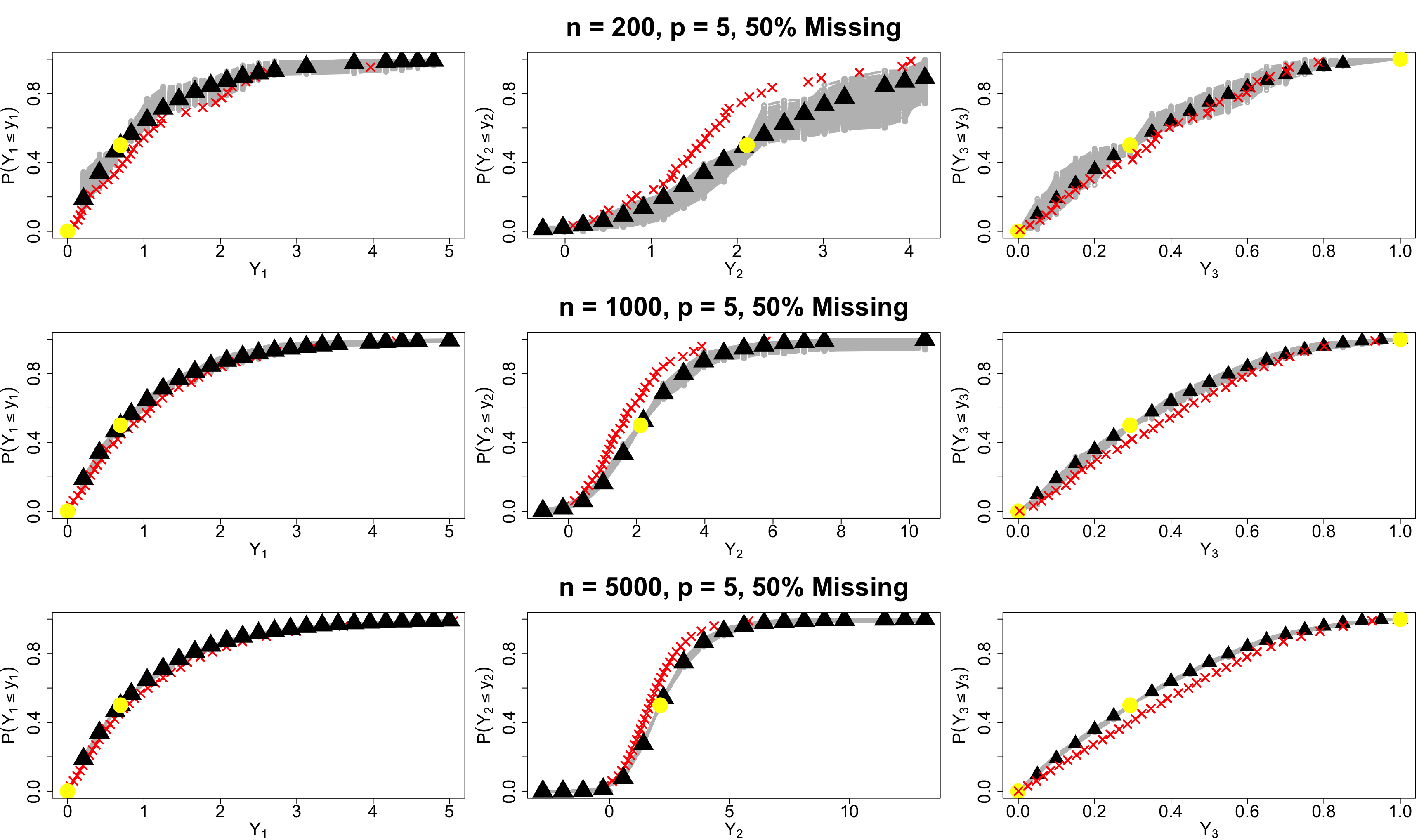}
    \caption{EHQL posterior samples of $\tilde{F}_j$ (lines) compared to true $F_j(y^q_j)$ (triangles) and ECDFs (crosses) computed from $\boldsymbol y^{obs}$. Columns index $j = 1,2,3$ from left to right. Rows index sample sizes. Each $\mathcal{A}^*$ uses the lower/upper bounds, the median (dots), and 15 intermediate quantiles. The EHQL accurately estimates each true CDF and corrects the biases caused by nonignorable missingness.
    %For imputation, we interpolate each posterior sample of the margin adjustment, which is then used for the transformation $\tilde y^{mis}_{ij} = \tilde{F}_{j}^{-1}\{\Phi(\tilde{z}^{mis}_{ij})\}$.
    }
    \label{figconsistMA}
\end{figure}

We next evaluate how closely  \eqref{marginadjust} under the EHQL-M approximates each 
%true 
$F_j$.  Figure \ref{figconsistMA} displays posterior samples of $\tilde{F}_{j}(y^{q}_{j})$ for $j = 1,2,3$ evaluated at the specified intermediate quantile points;
%and overlays the true cumulative probabilities and the auxiliary quantiles (when finite).  
it also displays the empirical CDFs (ECDF) computed from $\boldsymbol y^{obs}$.  These ECDFs exhibit varying degrees of bias caused by the nonignorable missing data.  
%for each setting. 
%for each point at which  margin adjustment is computed. 
%Even with scarce observed data, t
In contrast, each $\tilde{F}_{j}(y^{q}_{j})$  
%corrects the bias in the ECDF and 
accurately approximates the general shape of $F_j$, even though $\mathcal{A}$ only includes bounds and medians. 
%Interpolations through these points would outline each marginal CDF, which would enable imputation outside of the auxiliary and estimated intermediate quantiles. 
%As mentioned in Section \ref{MA-EHQL}, p
As expected, posterior uncertainty with \eqref{marginadjust} 
is highest for regions with relatively small sample sizes.
%since these intervals are sparsely populated. %Interpolation through these points enables imputation which accounts for the uncertainty underlying each $F_{j}$. 
%Together with the accurate estimates of $\boldsymbol C_{\boldsymbol \theta}$, t
% This accuracy suggest that the EHQL Gaussian copula can be effective for imputation of missing values.
%even with scarce auxiliary information on each study variable available.

We run each EQL/EHQL and RL sampler for 10,000 iterations on a 2023 Macbook Pro. When $n = 5000$, the EQL/EHQL samplers average around five minutes to complete, whereas the RL sampler takes nearly four hours. Mixing is also facilitated by introducing $\mathcal{A}$, with apparent convergence of the posterior of $\boldsymbol C_{\boldsymbol \theta}$ after a few hundred samples.

%Finally, for comparisons, Figure \ref{other} displays the posterior distributions of the correlations in $\boldsymbol C$ and the marginal distributions for the simulation with $(n=1000, p=5)$ when we have no missing data and when we have missing data but no margins.  We see that.... \textcolor{black}{Can we add a new figure for this case where we don't use the margins at all and just impute based on MAR? This to show that using the margin helps us do better than not using any margin at all. The model with no margins and no missing data, is another baseline to compare against.}\textcolor{black}{Let's discuss what you mean here - if we don't have missing data, the copula correlation is lower dimensional since we aren't estimating correlations for $\boldsymbol R$. Not sure how to conduct comparisons here.}
%\vspace{-1.5em}
\subsection{Simulation of Repeated Sampling Performance}\label{sim2}
%In this simulation study, w
In this section, we evaluate repeated sampling properties of the EHQL copula with multiple imputation inferences. 
%to commonly employed software for imputation inference. 
We use $p=5$ variables from the North Carolina lead exposure data described in Section \ref{sec:realdat}, namely  Economically Disadvantaged (\vtt{EconDisadv}), Mother's Age (\vtt{MAge}), Mother's Race (\vtt{MRace}), an index of Neighborhood Deprivation (\vtt{NDI}), and end-of-fourth grade (EoG) standardized math test scores (\vtt{Math_Score}). These variables are mixed binary, unordered categorical, count, and continuous variables that have complex univariate and multivariate features. 
We collect all individuals with complete data on these five variables, excluding observations with $\lvert \vtt{NDI} \rvert >5$ to provide stability, 
 which we treat as a finite population comprising approximately 165,000 individuals.  Using this population, we estimate 10th, 50th, and 90th  quantile regressions \citep{koenker} of \vtt{Math_Score} on main effects of the other four variables, which we treat as population quantities. The target models aim to uncover potentially heterogeneous effects of the covariates depending on the level of academic achievement, previewing the analysis in Section \ref{sec:realdat}.
%We first subset the data to complete cases of these five study variables, which yield approximately 140,000 observations. We fit the target quantile regression models to these data and 

%\textcolor{black}{Do we do this repeatedly or just once?}
We take 500 simple random samples of $n=5000$  individuals from this constructed population.  In each sample, we generate nonignorable missingness in $\vtt{NDI}$. Letting $j=1$ index the variable corresponding to $\vtt{NDI}$, we do so by the missingness mechanism,
\begin{equation}\label{sim-mech}
    p(R_{i1} = 1 \mid \boldsymbol{y}) \sim \mbox{Bernoulli}(\Phi(-0.5 -1.3 y^{scale}_{i1})),
\end{equation}
where superscript ${scale}$ indicates that the variable is centered and scaled to unit variance. After deleting any $y_{i1}$ where $r_{i1}=1$, we have approximately 40\% missing values of \vtt{NDI}, with lower values more likely to be missing in the sampled data. We also randomly remove 5\% of the other variables, including \vtt{Math_Score}. Because of the nonignorable missingness, complete case analysis could result in biased estimates of the quantile regression coefficients.
%The nonignorable missingness mechanism results in a biased empirical marginal distribution of \vtt{NDI_Birth}. 
%as children with lower levels are more likely to have missing values. 
%This also impacts quantile regressions between the covariates and \vtt{MathScal1}, as the complete cases are substantially different from the population ground truth.  

%This design is meant to mirror our real data analysis (Section \ref{sec:realdat}), which aims to estimate the association between a different exposure (lead exposure), plausibly subject to non-ignorable missingness, and EoG test scores.

 As auxiliary information, we assume access to selected quantiles of \vtt{NDI}, which we take from its marginal distribution in the constructed population. Here, we let $\mathcal{A}$ include the lower/upper bounds and median.  Section F of the supplement includes results using the lower/upper bounds and 75th quantile, which are qualitatively identical to those presented here, \textcolor{black}{as well as for several scenarios where the median of \vtt{NDI} is inaccurately specified.} We implement the EHQL using 15 evenly spaced quantiles across the range of observed values of \vtt{NDI}. For the other study variables, we use the empirical deciles in the sampled data as the quantiles in $\mathcal{A}$.  We add a missingness indicator for \vtt{NDI} to the Gaussian copula model. We exclude the remaining indicators, which effectively models that data for the corresponding variables are MCAR.
 
 We run the Gibbs sampler in Algorithm \ref{algEQL} for 5,000 iterations. After a conservative burn-in of 2,500 draws, we extract interpolated CDFs for imputation using Algorithm \ref{algMA} and take the completed data in every 125th iteration to create $m = 20$ multiple imputations.  We estimate the targeted quantile regressions  and use the combining rules of \citet{rubin1987multiple} for point estimates and 95\% confidence intervals for the quantile regression coefficients.  We also implement a default application of multiple imputation by chained equations (MICE) based on the \vtt{mice} package in \vtt{R} \citep{micebook}.  This does not use $\mathcal{A}$, although we add $R_1$ as a predictor for MICE. 
 %as an additional covariate to extend this approach to the non-ignorable setting.
 We repeat the entire procedure of sampling 5000 individuals, making missing values, and obtaining multiple imputation inferences 500 times.

 \begin{figure}[t]
     \centering
     \includegraphics[width = .49\textwidth, keepaspectratio]{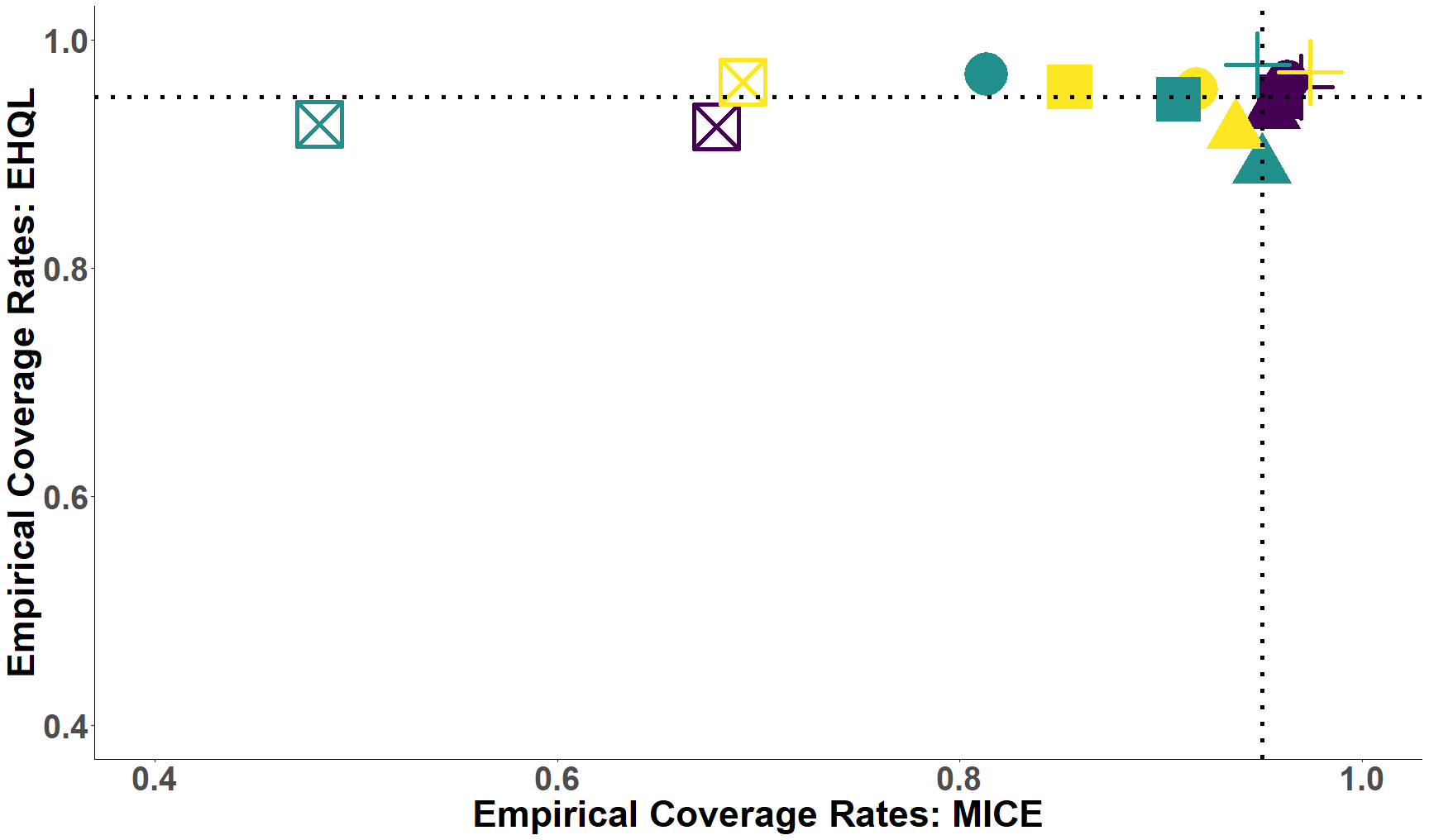}
            \includegraphics[width = .49\textwidth,keepaspectratio]{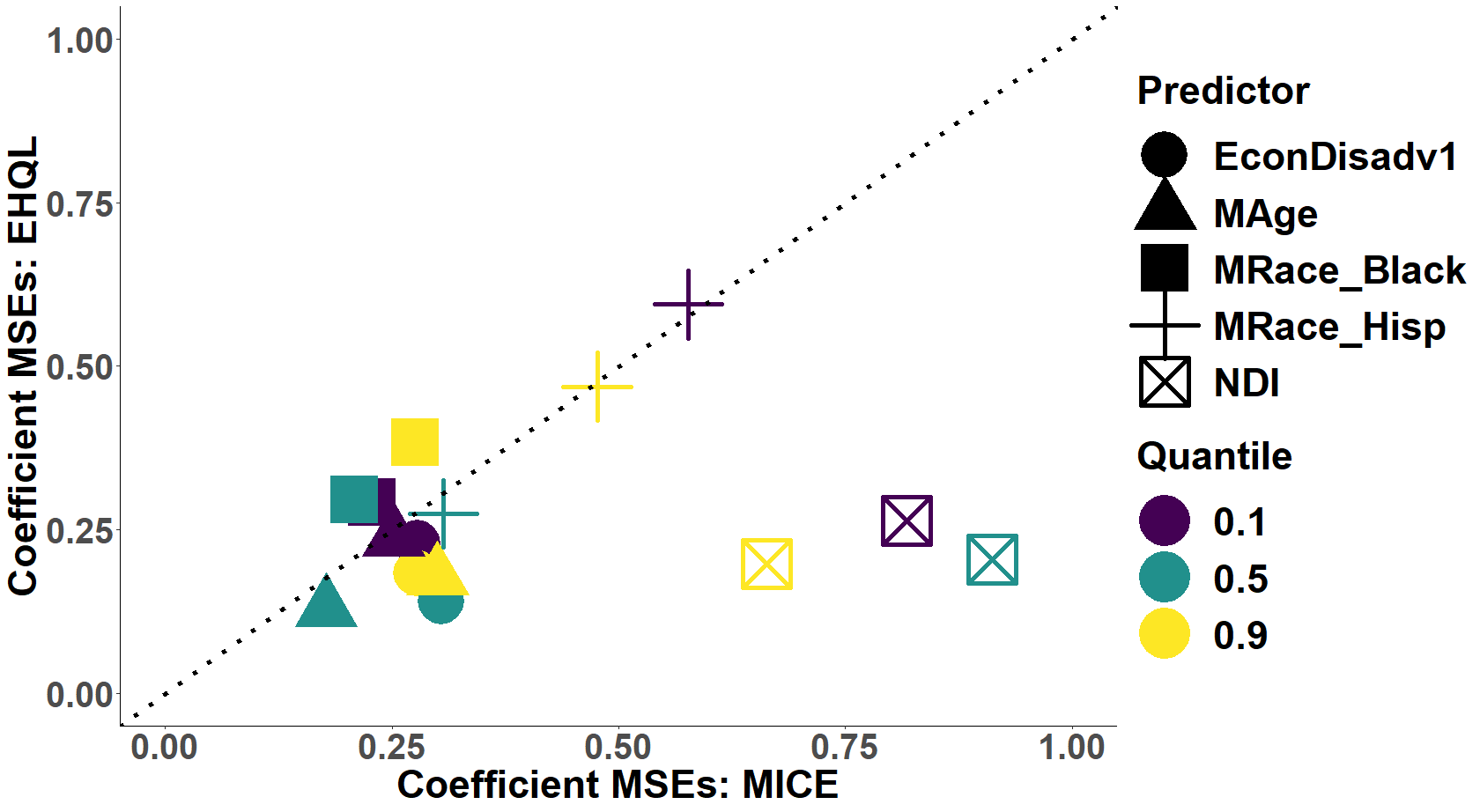}
        \caption{Empirical coverage rates (left) and mean squared error of multiple imputation point estimates (right) for EHQL and MICE imputations in the repeated simulation study. 
        %Both approaches perform similarly 
        For the coefficients of \vtt{NDI},
        %which is subject to nonignorable missingness. For this coefficient, 
        the EHQL provides significantly lower bias and higher coverage rates. While not shown, the average interval widths are similar.}
     % \caption{Imputation under the proposed approach (EHQL Copula) compared to MICE. Across quantiles, the inference is similar between the two methods besides the \vtt{NDI_Birth} variable, which is subject to nonignorable missingness. Clearly, the copula model is capable of capturing the dependence in the data, while incorporation of auxiliary quantiles helps to identify the extrapolation distribution and improve imputation.  }
     \label{impinferencesim}
 \end{figure}

\begin{figure}[t]
    \centering
    \includegraphics[width = .65\textwidth, keepaspectratio]{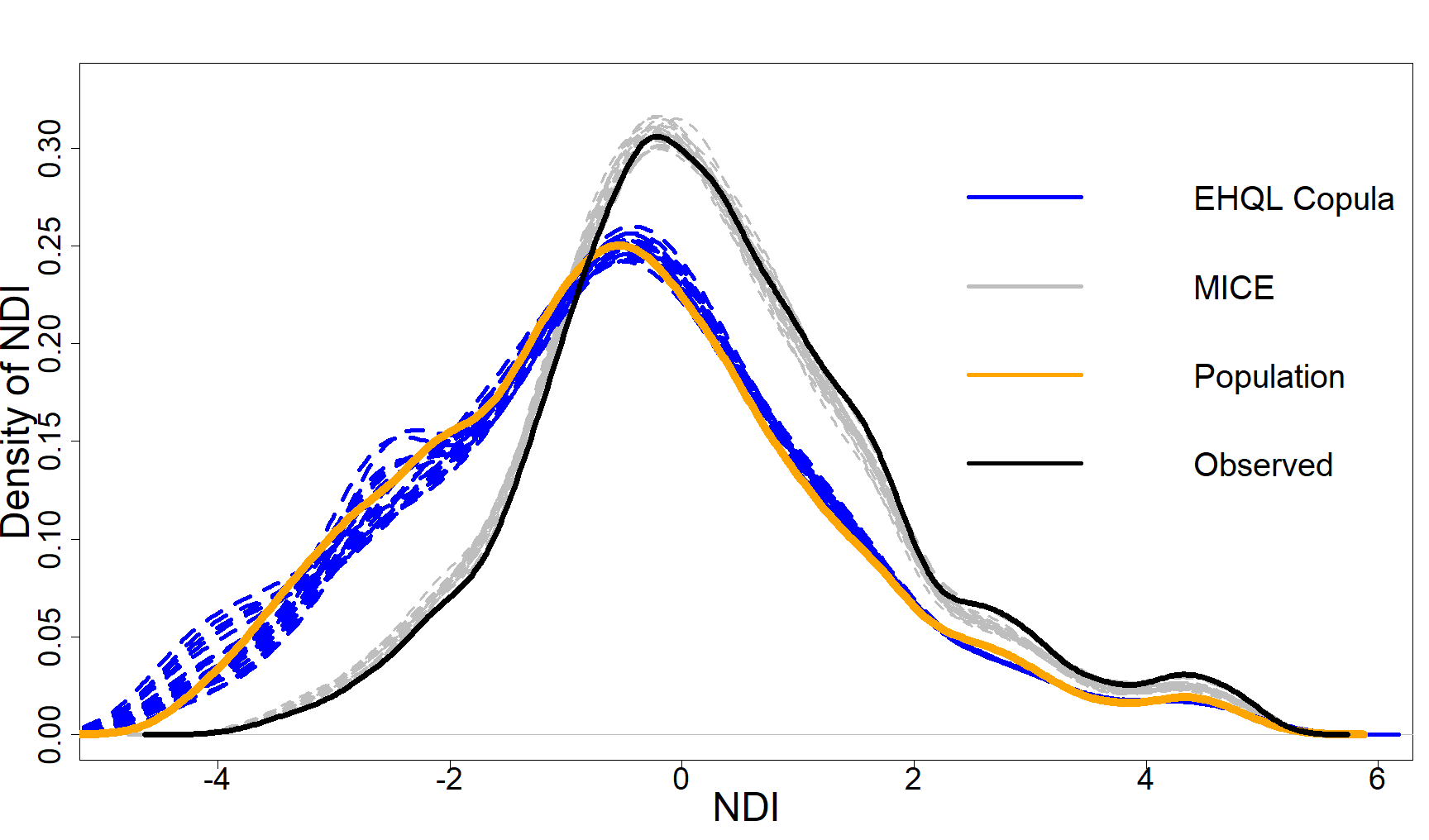}
    \caption{Marginal distribution of \vtt{NDI} in 20 completed datasets produced by EHQL and MICE.  EHQL imputations capture the bi-modality and skewness of the population marginal, whereas  MICE imputations  mimic the distribution of the observed values.}
    \label{NDImarg}
\end{figure}

Figure \ref{impinferencesim} summarizes the multiple imputation inferences over the 500 runs. 
% inferences about the quantile regression coefficients under the EHQL Copula and 
 %to the ground truth (True) and 
 %also include estimates and uncertainty quantification obtained from
The  inferences for the coefficients subject to MCAR missingness are reasonably similar for EHQL and MICE, and generally of high quality. However, we see substantial differences among EHQL and MICE coefficient estimates
%Across quantiles, the inferences are mostly consistent between the proposed approach and MICE when compared to the truth. 
for \vtt{NDI}.  These coefficients are accurately estimated with close to nominal coverage rates under  EHQL, but not MICE. Evidently, 
%The improvement of the EHQL copula model is clear, which is due to both 
the incorporation of $\mathcal{A}$ and the flexible dependence structure under the copula model allow the imputations to reflect the multivariate relationships more accurately than MICE does.  
%By contrast, MICE does not impute using a valid joint model and thus does not preserve the relationship between \vtt{NDI_Birth} and EoG test scores. 
The advantage of EHQL relative to MICE is also evident in the imputations of \vtt{NDI} displayed in Figure \ref{NDImarg}, which match the population marginal for EHQL but do not for MICE. % This is visualized in Figure \ref{NDImarg}.

\section{Analysis of North Carolina Lead Exposure Data}\label{sec:realdat}
Public health research in recent years has concluded overwhelmingly  that lead exposure has adverse impacts on childhood cognitive development \citep[e.g., ][]{bellinger1992low,miranda:kim:etal, kowal2021bayesian,bravo2022racial}. Many studies investigating this topic rely on administrative health and education datasets, linked at the individual child level, to estimate associations between lead exposure and outcomes of interest. %A prevalent issue in such data is that 
Typically, blood-lead measurements are available only for children who are tested for lead exposure, and children are more likely to be tested when there is concern about their exposure 
%Research has shown that tested children are more likely to be lower income and live in older housing 
\citep{kamai2022patterns}. 
Consequently, abundant missingness among lead measurements is commonplace, and individuals who are measured tend to have higher levels of exposure than much of the population.   Accurate imputation of lead measurements is important in this setting, as  selection biases could influence population-level inferences on the associations between educational outcomes and lead exposure (e.g., as simulated in Section \ref{sim2}).

% \textcolor{black}{This is a lot of detail for the datasets.  Do we need this level to satisfy the data provider? Can we trim it down as I commented out below?} \textcolor{black}{We can ask Dan about this, but I think what you have should be fine}

\begin{table}[t]
\centering 
\begin{tabular}{lll}
%\multicolumn{2}{|l|}{\textbf{Birth information (\% Missing)}}\\ 

Variable Name  & %$\,\,\,\,\,\,\,\,\,\,\,\,\,\,\,$
$\bar{Y}^{obs}/\bar{Y}^{mis}$ & Description  \\ \hline
\texttt{Blood\_lead} &%$\,\,\,\,\,\,\,\,\,\,\,\,\,\,\,\,\,\,\,\,$ 
 2.79/NA & Blood lead level (micrograms per deciliter) \\ 
\texttt{Math\_Score} &  %$\,\,\,\,\,\,\,\,\,\,\,\,\,\,\,\,\,\,\,\,$ 
448.9/452.5&  Standardized score on first 4th grade EoG  math test  \\ 
\texttt{Reading\_Score} & %$\,\,\,\,\,\,\,\,\,\,\,$
445.2/448.5& Standardized score on first 4th grade EoG reading test\\ 
\texttt{MEduc }  &  \begin{tabular}[t]{@{}l@{}} $(.25,.56,.21)$/ \\ $(.10,.48,.42)\,$  \end{tabular}   & \begin{tabular}[t]{@{}l@{}}Mother's education  at the time of birth \\(No degree,  High school  degree, College degree) \end{tabular} \\ 
\texttt{MRace } &  \begin{tabular}[t]{@{}l@{}} (.57,.31,.12)/ \\ (.76,.19,.05)\,  \end{tabular}    & \begin{tabular}[t]{@{}l@{}}Mother's race/ethnicity\\  (Non-Hispanic (NH) White,  NH Black, Hispanic)\end{tabular}                                                                                         \\ 
\texttt{BWTpct } &  %$\,\,\,\,\,\,\,\,\,\,\,\,\,\,\,\,\,\,\,\,\,\,\,\,\,\,\,\,\,\,\,\,\,\,\,\,$ 
46.9/51.7     &  Birthweight percentile  \\ 
\texttt{MAge}  & %$\,\,\,\,\,\,\,\,\,\,\,\,\,\,\,\,\,\,\,\,\,\,\,\,\,\,\,\,\,\,\,\,\,\,\,\,\,\,\,\,\,\,\,\,\,$ 
25.9/28.7        & Mother's age at the time of birth                                                                                                                                                                                                    \\ 
\texttt{Gestation} & 
%$\,\,\,\,\,\,\,\,\,\,\,$ 
38.6/38.7& Gestational period (in weeks)
\\ 

\texttt{Male } & %$\,\,\,\,\,\,\,\,\,\,\,\,\,\,\,\,\,\,\,\,\,\,\,\,\,\,\,\,\,\,\,\,\,\,\,\,\,\,\,\,\,\,\,\,\,\,\,\,\,$ 
.50/.50        & Male infant (1/0 = yes/no)                                                                                                                                                                                                               \\ 
\texttt{Smoker} & %$\,\,\,\,\,\,\,\,\,\,\,\,\,\,\,\,\,\,\,\,\,\,\,\,\,\,\,\,\,\,\,\,\,\,\,\,\,\,\,\,\,\,\,\,\,\,$ 
.15/.09 
& Mother smoked (1/0 = yes/no)                                                                                                                                                                                                         \\ 
\texttt{NotMarried} &  %$\,\,\,\,\,\,\,\,\,\,\,\,\,\,\,\,\,\,\,\,\,\,\,\,\,\,\,\,\,\,\,$ 
.46/.22 
& Not married at time of birth (1/0 = yes/no) 

\\ 

%\multicolumn{2}{|l|}{\cellcolor[HTML]{C0C0C0}\textbf{Education/End-of-grade (EoG) test information}}                                                                                                                                                                       \\

%\multicolumn{2}{|l|}{\cellcolor[HTML]{C0C0C0}\textbf{Blood lead surveillance}}                                                                                                                                                                                     \\ \hline

%\multicolumn{2}{|l|}{\cellcolor[HTML]{C0C0C0}\textbf{Social/Economic status}}                                                                                                                                                                                     \\ \hline

\texttt{EconDisadv} &
%$\,\,\,\,\,\,\,\,\,\,\,$
.61/.32       & \begin{tabular}[t]{@{}l@{}} Economically disadvantaged, per participation in \\Child Nutrition Lunch Program (1/0 = yes/no)
%(1 = yes, 0 = no)
\end{tabular} \\ 
\texttt{RI}& %$\,\,\,\,\,\,\,\,\,\,\,\,\,\,\,\,\,\,\,\,\,\,\,\,\,\,\,\,\,\,\,\,\,\,\,\,\,\,\,\,\,\,\,\,\,\,\,\,\,\,\,\,\,\,\,$
.23/.18 & Residential isolation index\\
\texttt{NDI} & %$\,\,\,\,\,\,\,\,\,\,\,\,\,\,\,\,\,\,\,\,\,\,\,\,\,\,\,\,\,\,\,\,\,\,\,\,\,\,\,\,\,\,\,\,\,\,\,\,\,\,$ 
.10/-1.04& Neighborhood deprivation index \\\hline
\end{tabular}
\caption{Variables used in the analysis of the relationship between lead exposure and EoG test scores among fourth grade children in North Carolina. Data are restricted to children with  30--42 weeks of gestation, 0--104 weeks of age-within-cohort, mother's age 15--44, \texttt{Blood\_Lead} $\leq$ 10, birth order $\leq$ 4, no status as an English Language Learner, and residence in North Carolina at the time of birth and the time of the EoG test. For the quantile regressions, numeric covariates are scaled to mean 0 with 0.5 standard deviation.}
\label{tabdata}
\end{table}

We analyze a dataset comprising 
%dataset containing information on 
$\sim$170,000 North Carolina children born between 2003 and 2005.  The dataset is constructed by linking children's data across three databases comprising (i) detailed birth records, which include maternal demographics, maternal and infant health measures, and maternal obstetrics history for all documented live births in North Carolina; (ii) lead exposure surveillance records from a registry maintained by the state,  
%Childhood Lead Poisoning Prevention Program of the Children’s Environmental Health Unit, Department of Health and Human Services in Raleigh, N.C., 
which include integer-valued blood-lead levels; and (iii) test score data from the N.C.\ Education Research Data Center at Duke University, which include EoG reading and mathematics test scores as well as   some 
%from the 1995-1996 school year to the present, 
%student identifying information, and 
%data on 
demographic and socioeconomic information \citep{CEHI}. 
%Our study variables include students' end-of-fourth-grade (EoG) math and reading test scores, child's blood-lead measurements, and demographic and socioeconomic variables for the mother and child. 
Table~\ref{tabdata} summarizes the variables we use, including their sample averages among children with lead exposure (\vtt{Blood_lead}) observed or missing. Notably, 35\% of the lead measurements are missing. Other variables have missing data rates less than 0.02\%.

\begin{figure}[t]%
    \centering
%    \subfloat[\centering CDC published estimates for select quantiles of lead exposure]{{\includegraphics[height=4cm]{Images/CDC.png} }}%
 %   \qquad
 %   \subfloat[\centering  Comparison of CDC estimates with the empirical marginal of lead exposure in the North Carolina Data]{{
\includegraphics[width = 0.75\textwidth,keepaspectratio]{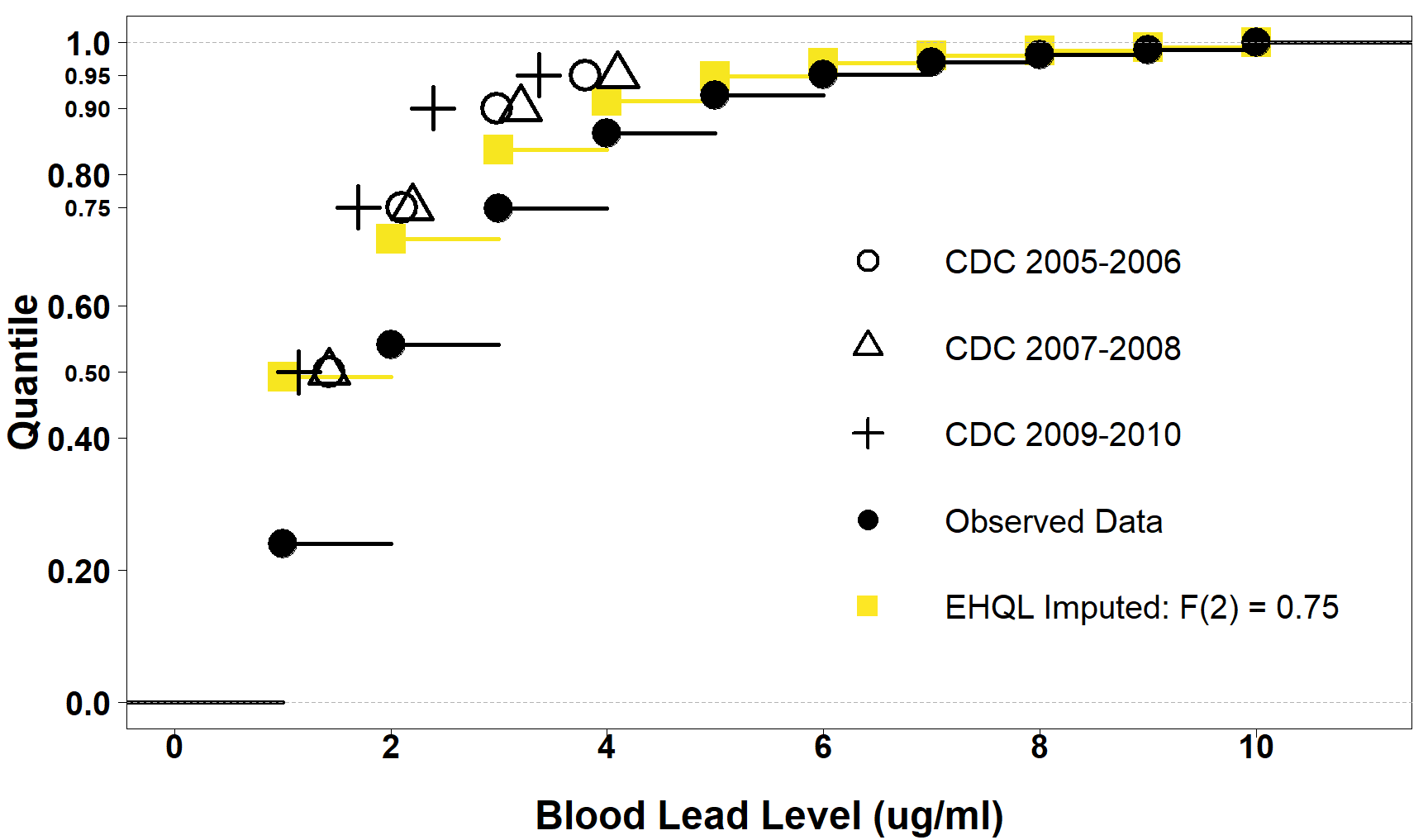}
    \caption{Empirical quantiles of \vtt{Blood_lead} in the North Carolina data and population-level quantiles published by CDC.  Also displayed are average estimated quantiles of \vtt{Blood_lead} in completed datasets after EHQL  imputations. 
    %The agency publishes annual estimates for the 50th, 75th, 90th, and 95th quantiles of lead exposure for different age groups.Since our data set contains children born between 2003 and 2005, who were measured for lead between 2004 and 2008, we include annual estimates for the 1-5 year old age group.  
%We include CDC values consistent with the years most children were measured for lead exposure. 
After imputation, the distribution of \vtt{Blood_lead} more closely matches the population-level distribution.}
    %between the years 2005 and 2009 corresponding to 1--5 years of age. As such, we include yearly population-level values for children of 1--5 years of age during the specified time frame.} 
    %Clear evidence of selection bias is present in the North Carolina data, as the observed quantiles do not correspond to population-level quantiles.}%
    \label{cdccomp}%
\end{figure}

Figure \ref{cdccomp} displays the ECDF of \vtt{Blood_lead} compared to three years of annual point estimates for the 50th, 75th, 90th, and 95th quantiles of lead exposure levels published by the CDC \citep{CDCwebsite}. 
Clearly, the observed distribution 
%of \vtt{Blood_lead} 
does not match the population-level quantile estimates. We therefore consider the missingness in \vtt{Blood_lead} as possibly missing not at random (MNAR). As we discuss later, the EHQL copula correlation estimates suggest this missingness in fact is MNAR. 

%include yearly population-level estimates for children of 1--5 years of age during the specified time frame.

% \end{figure}
%     \caption{}
%     \label{cdccomp}
% \end{figure}

\textcolor{black}{
To adjust for this missingness, we seek to leverage the CDC estimates as auxiliary information for \vtt{Blood_lead}}. 
%However, the  \vtt{Blood_lead} measurements in the North Carolina data are recorded as integers corresponding to the interval in which the child's measurement is contained. For instance, \vtt{Blood_lead}=1 means that the child's measurement was in the interval $(0,1]$ $\mu$g/ml. By contrast, the CDC measurements are continuous.  
%We therefore 
%We \textcolor{black}{illustrate} one potential integration of the CDC estimates into the copula model, which approximately locates the marginal distribution of \vtt{Blood_lead} and incorporates Algorithm \ref{algMA} to infer intermediate quantiles. 
%We consider three specifications of this marginal distribution. 
%for \vtt{Blood_lead}, which accounts for the uncertainty undRLying the CDC estimates coupled with their extrapolation to North Carolina children.
%To do so, we provide to the EHQL a single auxiliary quantile for \vtt{Blood_lead}, which we vary in the sensitivity analysis. 
As evident in Figure \ref{cdccomp}, the 75th quantile estimates from three relevant years of CDC publications are reliably around 2.0. Consequently, we set $\mathcal{A}_{\vtt{Blood_lead}}= \{F_{\vtt{Blood_lead}}^{-1}(0) = 0,F_{\vtt{Blood_lead}}^{-1}(0.75) = 2, F_{\vtt{Blood_lead}}^{-1}(1) = 10\}$. Because we do not know the distribution of lead exposure in North Carolina exactly---the CDC provides national quantiles---we examine results when using $F_{\vtt{Blood_lead}}^{-1}(0.70) = 2$ and $F_{\vtt{Blood_lead}}^{-1}(0.80) = 2$. This sensitivity analysis   
%of the different auxiliary quantile specifications 
is available in Section G of the supplement. In each setting, we estimate distribution of lead exposure by setting $\{y_{\vtt{Blood_lead}}^{q}\} = \{3,4,5,6,7,8,9\}$ and applying Algorithm \ref{algMA}. For the remaining numerical study variables, we use empirical deciles for the auxiliary quantiles and the ECDF for imputation. This is reasonable given the scarce missingness and large sample size, and provides significant computational benefit for estimation of the copula model. These quantities are not varied in the sensitivity analysis.

 We use the EHQL copula with the factor model in \eqref{factor}  to implement multiple imputation of all missing values. We include a missingness indicator for \vtt{Blood_lead} but not the other variables, \textcolor{black}{which are almost completely observed.  Their indicators comprise almost all zeros and thus are not likely to 
 %them to the model would offer scant additional information 
 to inform the imputation but would slow computation.} Consequently, the remaining study variables are treated as MCAR. For each set of auxiliary information, we estimate the EHQL copula by running Algorithm \ref{algEQL} for 10,000 iterations and discarding 5,000 draws as burn-in. We use every 250th posterior sample of model parameters to create $m=20$ multiple imputations. Posterior predictive checks suggest that the model reasonably describes the observed data; these are available in Section G of the supplement. 
 
 Using the completed datasets, we 
 %we target the associations of lead exposure with EoG math and reading test scores. We 
 estimate the 10th, 50th, and 90th
 quantile regressions of math and reading scores on main effects of all the study variables, and derive point estimates and uncertainty quantification using multiple imputation combining rules.  These quantities describe potentially heterogeneous impacts of the covariates on low, middle, and high-achieving students, which is of interest to public health research \citep{miranda:kim:etal}. 

\begin{figure}[t]
    \centering
    \includegraphics[width = .8\textwidth, keepaspectratio]{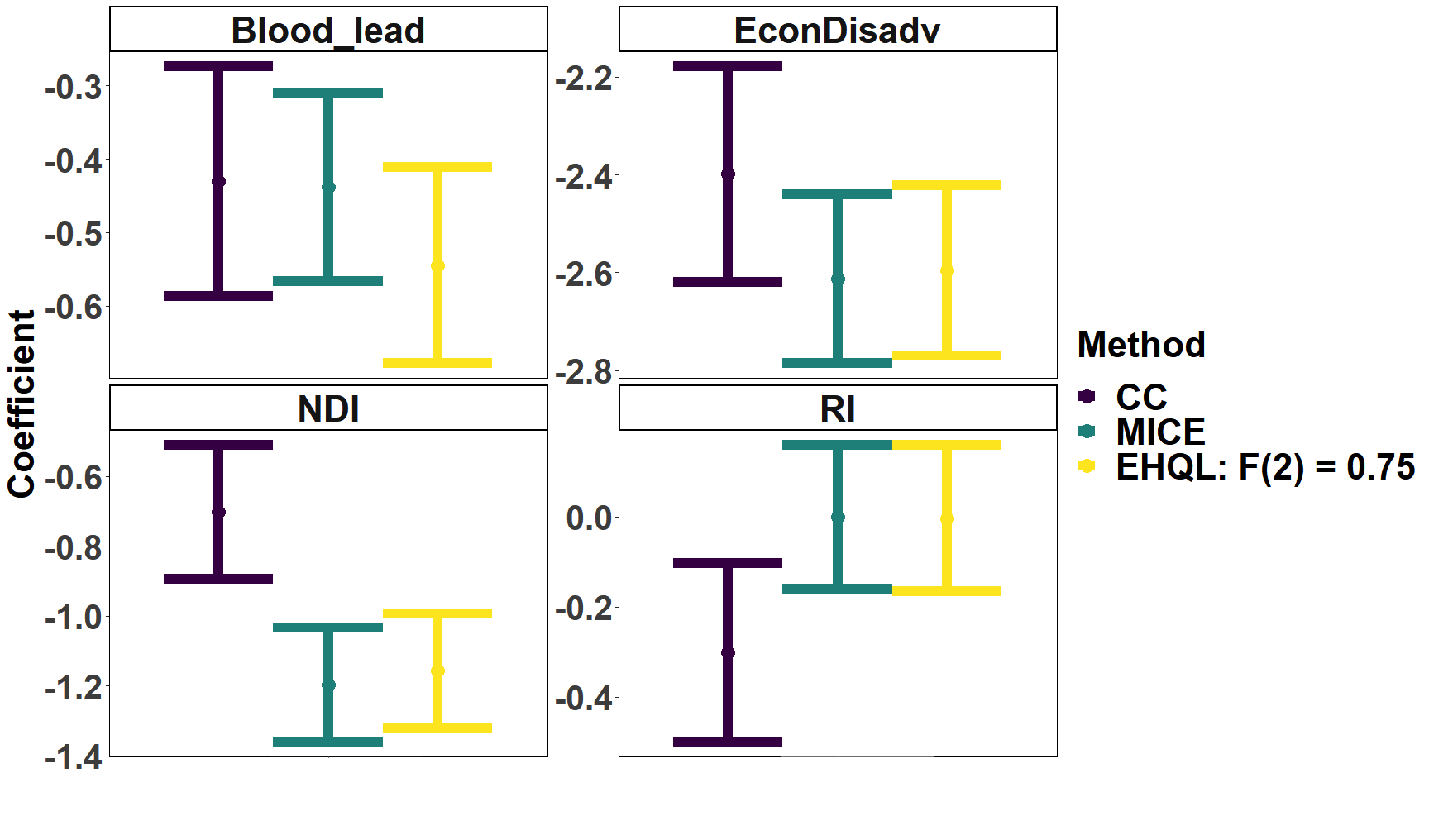}
    \caption{Multiple imputation inferences for coefficients of 10th quantile regression when the 75th quantile of the marginal distribution of \vtt{Blood_lead} = 2. Compared to complete case analysis, 
    %we see sizeable differences especially for the coefficient of 
    \vtt{Blood_lead} is estimated to be more adversely associated with EoG math scores. MICE imputations do not capture this difference, as MICE mimics patterns in the observed data.}
    %The results are insensitive to varying the auxiliary quantile introduced into the model. }
    \label{mathinf}
\end{figure}
Figure \ref{mathinf} summarizes the multiple imputation inferences for the 10th quantile regression coefficients for \vtt{Blood_lead}, \vtt{EconDisadv}, \vtt{NDI}, and \vtt{RI} using \vtt{Math_Score} as the response. The results are presented for $F_{\vtt{Blood_lead}}^{-1}(0.75) = 2$, and are insensitive to the other specifications of this auxiliary quantile. We compare these inferences to those obtained by fitting the quantile regression models to complete case (CC) observations, which exclude any observations with missing study variables, and to results from $m=20$ multiple imputations using a bespoke application of MICE. Results for the other auxiliary quantile specifications, quantile regressions, and using reading scores as the response 
%are qualitatively similar and 
are in the supplement.

After multiple imputation, 
%we see sizeable shifts in the inferences. The estimated 
the associations of \vtt{Blood_lead}, \vtt{EconDisadv}, and \vtt{NDI} with \vtt{Math_Score} are more adverse than in the CC analysis. The association of \vtt{Blood_lead} with \vtt{Math_Score} is estimated to be significantly stronger when using the EHQL imputations rather than the MICE imputations. Specifically, the point estimate for this coefficient under the EHQL is nearly two standard errors more negative than point estimates under CC and MICE, which is a substantively large change.
%This is noteworthy in conjunction with the observation that
The inferences under EHQL and MICE are similar for the other predictors' coefficients, which is not unexpected as these variables have few missing values. Similar shifts are evident in the 50th and 90th quantile regressions as well.

%Evidently, introducing auxiliary quantiles has a sizeable impact on imputation of \vtt{Blood_lead} and creates more plausible completed data sets.  
%By contrast, accounting for the nonignorable missing data in lead measurements results in weaker signal in the quantile regression coefficients for \vtt{RI}. 
%Overall, the results are insensitive to the auxiliary quantiles used in the modeling. 

%\begin{figure}[t]
%    \centering
%    \includegraphics[width = 0.9\textwidth,keepaspectratio]{Images/cdc_comp75.png}
%    \caption{Comparison of quantiles of lead exposure levels in the completed data after EHQL, observed data with no imputation, and the CDC population values. \textcolor{black}{just this plot -- drop other one since it is repetitive.  Or, perhaps we put CDC and green dots in this one.}}
%    \label{cdccomp_imp}
%\end{figure}

 %An explanation for this is available by closer examination of the completed North Carolina datasets. In 
Figure \ref{cdccomp} displays the 
 %To illuminate the effect of the EHQL imputations further, we examine the 
 empirical CDF of \vtt{Blood_lead} obtained by averaging across the $20$ completed datasets.
 %this is displayed in Figure \ref{cdccomp}. 
 %for the setting with $F_{\vtt{Blood_lead}}^{-1}(0.75) = 2$. 
 %and compare them to the empirical quantiles from the observed data and the CDC population values. 
 Because it utilizes $\mathcal{A}$, the EHQL imputes values of missing \vtt{Blood_lead} measurements that are small relative to the observed values.  Simultaneously, the 
 %10th, 50th, and 90th 
 percentiles of EoG math scores are 
 %about four points 
 higher for students missing \vtt{Blood_lead} than for students with recorded values.  Consequently, imputing \vtt{Blood_lead} so that its completed-data distribution accords with $\mathcal{A}$ strengthens its negative association 
 %--- and those for the other exposures ---
 with \vtt{Math_Score}.  We see similar strengthening of inverse associations with the other predictors  %between \vtt{NDI}, \vtt{RI}, the proportion of students with positive indicators for \vtt{EconDisadvantage}, and EoG test scores are similar for individuals with and without a \vtt{Blood_lead} measurement.
 %The exception to this pattern among the selected exposure variables comes 
 except for \vtt{RI}, which has estimates mostly shrunk towards zero. \vtt{RI} is strongly correlated to \vtt{NDI}, which may help explain why its estimates attenuate. We note that a 95\% credible interval for the copula correlation between \vtt{Blood_lead} and its missingness indicator  is (-0.92,-0.91), offering additional evidence that  lead exposure measurements are MNAR.

\section{Concluding Remarks} \label{sec:concl}

Using auxiliary marginal quantiles offers a convenient and flexible way to handle  
%perform sensitivity analysis 
nonignorable missing data in Gaussian copula models. The simulation studies suggest that using reliable $\mathcal{A}$---for example, informed by external sources like national surveys or administrative databases---can result in more accurate inferences than treating nonignorable missing data as MCAR or MAR. In fact, under AN missingness, it is possible to estimate accurately the copula correlation and perform well-calibrated multiple imputation even with just a few auxiliary quantiles on each study variable.  The simulations also suggest that augmenting $\mathcal{A}$ with intermediate quantiles can improve the quality of inferences and imputations. 
%with few discrepancies from using fully specified marginal distributions. 
%This suggest that Theorem \ref{postconsist} may be more broadly applicable to the copula model under the EQL and EHQL. 

%\textcolor{black}{As with other data integration settings, analysts should assess the sensitivity of results to different specifications of auxiliary marginal quantiles.}

 % the model is most reliable when the auxiliary information is acctrue distributions, i.e., absent any missing data, in the populations underlying the auxiliary and study data match}

%There are many topics worthy of future research. For example, the 
%we assume that the marginal information is specified precisely. Sometimes, however,
The marginal quantiles may be known with uncertainty, for example, estimates from a probability sample. 
%One could fit the model with several quantiles, reflecting that uncertainty, and summarize results as part of the sensitivity analysis.  Alternatively,
%In cases where 
When analysts desire a single inference, it may be possible to posit sampling distributions for the true quantiles that can be integrated into the model specification; this is a topic for future research.  Additionally, often data have survey weights. While there are methods for using  auxiliary margins with survey-weighted data for categorical data models  \citep{akande2021multiple, tang2024using}, work is needed to develop methods for the EQL/EHQL copulas.

The North Carolina lead exposure analysis suggests that utilizing auxiliary marginal quantiles to handle nonignorable missing data can impact empirical findings. In particular, inferences drawn from the multiple imputations based on the EHQL copula suggest that lead exposure affects childhood cognitive development more adversely than might be concluded from a complete case analysis.  %This application is representative of broader contexts.  
More broadly, linked data like these 
%like those studied in our application 
are commonly used in public health studies and full of missing values. When 
%there is evidence that 
the missingness may be MNAR,
%systematic in potentially biasing ways, 
analysts can consider imputation strategies that leverage known marginal quantiles 
%distributions 
of the study variables 
%Our techniques should be applied to the study of the impacts of other environmental toxins 
to better inform health policy.
%and intervention strategies.
%NOTE: Deleted to avoid using strategies twice in the same sentence

%\section*{Acknowledgments}
%\noindent To be completed later.
%The authors thank the anonymous reviewers for their valuable suggestions. 
%This work is supported in part by funds from the National Science Foundation (NSF: \# 1636933 and \# 1920920).

%\noindent{\em Conflicts of interest:} None declared.

%\section*{Data availability}
%\noindent ADD A DATA AVAILABILITY STATEMENT
%The data that support the findings in this article were obtained from the Alzheimer’s Disease 
%Neuroimaging Initiative (ADNI) database (https://adni.loni.usc.edu/). The pre-processed data 
%are available from Huang et al. (2010). Restrictions may apply to the availability of these data.

%\section*{Supplementary material}
%\noindent Supplementary material is available online at TBD.
%{\em Journal of the Royal Statistical Society: Series B.}

%\section{Author contributions statement}

%Must include all authors, identified by initials, for example:
%S.R. and D.A. conceived the experiment(s),  S.R. conducted the experiment(s), S.R. and D.A. analysed the results.  S.R. and D.A. wrote and reviewed the manuscript.

\bibliographystyle{abbrvnat}
\bibliography{reference}

\end{document}

\title{Supplement to \\ ``Gaussian Copula Models for Nonignorable Missing Data Using Auxiliary Marginal Quantiles''}
\date{}

\author{Joseph Feldman, Jerome P. Reiter, and Daniel Kowal}

\ifblinded
\author{}
\else

\author{Joseph Feldman, Jerome P. Reiter, Daniel Kowal}

\fi

\maketitle
\large 

\vspace{-10mm}

\appendix

\section{Introduction}

This document includes supplementary material to the main text.  Section \ref{gibbsfullmarginal} outlines the Gibbs sampler for use with full marginal specifications in $\mathcal{A}$. 
%Section \ref{simdesign} presents the data generating model used in Figure 1 in the main text.  
Section \ref{proofs} includes the proofs of the theorems from the main text. Section \ref{cat} presents the extension of the Gaussian copula model to handle unordered categorical variables.  Section \ref{samp} presents the full model specification for the EQL and EHQL including the prior distributions from the factor model described in the main text, as well as the full conditional distributions used in the Gibbs samplers for those models.  Section \ref{sims} presents results of additional simulations.  Section \ref{NCadditional} presents additional results from the analysis of the North Carolina lead data.

\section{Gibbs Sampler for Full Marginal Specifications}\label{gibbsfullmarginal}

Given a set $\{F_j\}_{j=1}^p$ of fully specified marginal distributions for the study variables, we use Algorithm \ref{algwork} to sample from the posterior distribution of $(\boldsymbol C_{\boldsymbol \theta}, \boldsymbol \alpha)$ and impute $\boldsymbol y^{mis}$. We presume the analyst uses the prior distribution described in Section 4 of the main text.   In the algorithm, the subscript $(-j)$ in a vector denotes that vector without the $j$th element; the subscript $(-j)$ in the column (row) index of a matrix indicates %inclusion or 
exclusion of the elements corresponding to the $j$th column (row) of that matrix; and, the subscript $-(jj)$ in a matrix indicates exclusion of all row and column elements for the $j$th variable in that matrix. Algorithm \ref{algwork} can be used generally when full marginal distributions are available.

\addtocounter{algorithm}{2}
\begin{algorithm}[t]
\caption{Bayesian estimation and imputation for the Gaussian copula with fixed marginals.}
\begin{algorithmic}\label{algwork}

     \STATE \textbf{Require:}  prior $p(\boldsymbol C_{\boldsymbol \theta}, \boldsymbol \alpha)$, marginals $\{F_{j}\}_{j=1}^{p}$. Let $\boldsymbol C = \boldsymbol C_{\theta}$.
     \begin{itemize}
         \item  \textbf{Step 1}: Sample $( \boldsymbol z^{mis}, \boldsymbol z_{\boldsymbol r}) \mid \boldsymbol C, \boldsymbol \alpha$
     \end{itemize}
     \bindent
     \FOR{$j \in \{1,\dots,2p\}$}
         \STATE Compute $\mu_{ij} = 
         \alpha_j + \boldsymbol C_{j(-j)} \boldsymbol C_{-(jj)}^{-1}(\boldsymbol z_{i(-j)} - \boldsymbol \alpha_{-j})$ and   $\sigma_{j}^{2} = \boldsymbol C_{jj}- \boldsymbol C_{j(-j)}\boldsymbol  C^{-1}_{-(jj)}\boldsymbol C_{(-j)j}$
            \IF{$ z_{ij} \in \boldsymbol z_{\boldsymbol r}$}
                \STATE $\ell_{ij} = 0 - \infty \mathbbm{1}_{r_{j} =1}, \quad u_{ij} = 0 + \infty \mathbbm{1}_{r_{ij} = 0}$
                \STATE Sample $ z_{ij} \sim \mbox{Normal}(\mu_{ij}, \sigma_{j}^{2}) \mathbbm{1}(\ell_{ij},u_{ij})$
                \ENDIF
            \IF{$ z_{ij} \in \boldsymbol z^{mis}$}
                \STATE Sample $z_{ij} \sim \mbox{Normal}(\mu_{ij}, \sigma_{j}^{2})$
                \ENDIF
             \ENDFOR       

     \eindent

            %         \STATE $$
          \begin{itemize}
         \item  \textbf{Step 2}: Sample $\boldsymbol C, \boldsymbol \alpha \sim p(\boldsymbol C, \boldsymbol \alpha \mid \boldsymbol{z}^{obs}, \boldsymbol z^{mis}, \boldsymbol z_{\boldsymbol r})$
              \end{itemize}
              \bindent
         \STATE where $p(\boldsymbol C, \boldsymbol \alpha \mid \boldsymbol{z}^{obs}, \boldsymbol z^{mis}, \boldsymbol z_{\boldsymbol r}) \propto N_{2p}((\boldsymbol{z}^{obs}, \boldsymbol z^{mis}, \boldsymbol z_{\boldsymbol r}); \boldsymbol \alpha, \boldsymbol C) p (\boldsymbol C, \boldsymbol \alpha) $ 
         \eindent
         \begin{itemize}
             \item \textbf{Step 3:} Impute $y^{mis}_{ij} = F_{j}^{-1}(\Phi(z^{mis}_{ij}))$
         \end{itemize}
\end{algorithmic}
\end{algorithm}

\section{Proofs of Theorems in Main Text}\label{proofs}

In this section we present proofs of Theorem \ref{postconsist}, Lemma \ref{lemma1}, and Theorem \ref{postconsistEQL} from the main text.  In the proofs, equations referenced by number only, e.g., (2)--(3), refer to the corresponding equations in the main text.
\begin{theorem}\label{postconsist}
Suppose $\{(\boldsymbol{y}_i, \boldsymbol r_i) \}_{i=1}^{n}\overset{iid}{\sim} \Pi_{0}$ where $\Pi_{0}$ is the Gaussian copula with
    % SHOULD WE DELETE $\boldsymbol{y}^{mis}$?  \textcolor{black}{comprise $n$} independent and identically distributed samples from \eqref{YRcop}-\eqref{xformYR} with  true
    correlation $\boldsymbol C_{0}$ and  marginals $\{F_{j}\}_{j=1}^{p}$ as in \eqref{YRcop}--\eqref{xformYR}, and 
    %\textcolor{black}{where $\boldsymbol{y}^{mis}$ follows a nonignorable missingness mechanism under the copula model.}  
    $\{\mathcal{A}_{j}\}_{j=1}^{p} = \{F_{j}\}_{j=1}^{p}$. Let $p(\boldsymbol \theta)$ be a prior with respect a measure that induces a prior $\Pi$ over the space of all $2p \times 2p$ correlation matrices 
 $\boldsymbol{\mathbbm{C}}$ with $\Pi(\boldsymbol C_{\boldsymbol \theta})>0$ for all $\boldsymbol C_{\boldsymbol \theta} \in \mathbbm{C}$.  Then, for all $\epsilon > 0$, $\lim_{n\rightarrow \infty}\Pi_{n}\{\mathcal{U}_{\epsilon}(\boldsymbol C_{0})\} \rightarrow 1$ almost surely $[\Pi_{0}]$, where 
    $\mathcal{U}_{\epsilon}(\boldsymbol C_{0}) = \{\boldsymbol C_{\boldsymbol \theta} \in \mathbbm{C}: \lVert \boldsymbol C_{0} - \boldsymbol C_{\boldsymbol \theta}\rVert_{F} < \epsilon\}$ and $\lVert . \rVert_{F}$ is  Frobenius norm.
\end{theorem}
%\begin{theorem}\label{postconsist}
 %Suppose $(\boldsymbol{y}, \boldsymbol r) \overset{iid}{\sim} \Pi_{0}$ where $\Pi_{0}$ is the Gaussian copula with
    % SHOULD WE DELETE $\boldsymbol{y}^{mis}$?  \textcolor{black}{comprise $n$} independent and identically distributed samples from \eqref{YRcop}-\eqref{xformYR} with  true
  %  correlation $\boldsymbol C_{0}$ and univariate marginals $\{F_{j}\}_{j=1}^{p}$ as in \eqref{YRcop}-\eqref{xformYR}, and 
    %\textcolor{black}{where $\boldsymbol{y}^{mis}$ follows a nonignorable missingness mechanism under the copula model.}  
   % assume $\{\mathcal{A}_{j}\}_{j=1}^{p} = \{F_{j}\}_{j=1}^{p}$. Let $p(\boldsymbol \theta)$ be a prior with respect a measure that induces a prior $\Pi$ over the space of all $2p \times 2p$ valid correlation matrices $\boldsymbol{\mathbbm{C}}$ such that $\Pi(\boldsymbol C)>0$ for all $\boldsymbol C \in \mathbbm{C}$.  Then for all $\epsilon > 0$, $\Pi_{n}\{\mathcal{U}_{\epsilon}(\boldsymbol C_{0})\} \rightarrow 1$ almost surely $[\Pi_{0}]$, where 
    %$\mathcal{U}_{\epsilon}(\boldsymbol C_{0}) = \{\boldsymbol C_{\boldsymbol \theta} \in \mathbbm{C}: \lVert \boldsymbol C_{0} - \boldsymbol C_{\boldsymbol \theta}\rVert_{F} < \epsilon\}$ and $\lVert . \rVert_{F}$ is the Frobenius norm.
%\end{theorem}

\begin{proof}
  Without loss of generality, suppose $\boldsymbol \alpha = \boldsymbol 0$. 
  %Consider the sequence of functions 
  Let $f_{n}(\boldsymbol C_{\boldsymbol \theta}) = -n^{-1}\sum_{i=1}^{n} \log p(\boldsymbol z^{obs}_{i}, \boldsymbol z_{\boldsymbol r_i} \in \mathcal{E}(\boldsymbol r_i); \boldsymbol C_{\boldsymbol \theta})$ computed with the $n$ sampled draws of $(\boldsymbol y_i^{obs}, \boldsymbol r_i)$. 
  %from $\Pi_0$. 
  Here, $\mathcal{E}(\boldsymbol r_i)$ is the probit set restriction defined in the main text specific to the values of the nonresponse indicators $\boldsymbol r_i$.  Let $(\boldsymbol Z^{obs}, \boldsymbol Z_{\boldsymbol R})$ represent the latent random variables for a random draw of
  $(\boldsymbol y^{obs}, \boldsymbol r)$ 
%  $(\boldsymbol z^{obs}, \boldsymbol z_{\boldsymbol r})$ 
from $\Pi_0$. Define $f(\boldsymbol C_{\boldsymbol \theta}) = - \mathbbm{E}_{\Pi_{0}} \log p(\boldsymbol Z^{obs}, \boldsymbol Z_{\boldsymbol R} \in \mathcal{E}(\boldsymbol R); \boldsymbol C_{\boldsymbol \theta})$, where the expectation is with respect to $\Pi_0$ with known 
%  $p(\boldsymbol z^{obs}, \boldsymbol z_{\boldsymbol r} \in \mathcal{E}(\boldsymbol r); \boldsymbol C_{\boldsymbol \theta})$ is the Gaussian copula likelihood in \eqref{copwork1}--\eqref{integrand}, fixing 
$\mathcal{A} = \{F_{j}\}_{j=1}^{p}$ and $\mathcal{E}(\boldsymbol R)$ is the probit set restriction corresponding to the random draw of $\boldsymbol r$. 
Let $\phi_{\boldsymbol C_{\boldsymbol \theta}}$ and $\phi_{\boldsymbol C_{0}}$ represent the $2p$-dimensional multivariate Gaussian densities with correlation $\boldsymbol C_{\boldsymbol \theta}$ and $\boldsymbol C_{0}$, respectively. 

The proof establishes  that the following conditions from Theorem 3 in \cite{miller2021asymptotic} hold almost surely $[\Pi_{0}]$:
    \begin{enumerate}
        \item $\Pi\{\mathcal{U}_{\epsilon}(\boldsymbol C_{0})\}>0$;
        \item $f_{n} \rightarrow f$ pointwise on $\mathbbm{C}$;
        \item $f_{n}$ is convex for each $n$;
        \item $\mathbbm{C} \subseteq \mathbbm{R}^{2p(2p-1)}$;
        \item $\boldsymbol C_{0} \in \text{int}(\mathbbm{C})$; and
        \item $f(\boldsymbol C_{\boldsymbol \theta}) > f(\boldsymbol C_{0})$ for all $\boldsymbol C_{\boldsymbol \theta} \in \mathbbm{C}\setminus \boldsymbol C_{0}$.
    \end{enumerate}

Conditions 1, 4, and 5 are satisfied directly by the theorem assumptions on the prior $\Pi$. Condition 3 is satisfied since 
%$f_{n}(\boldsymbol C_{\boldsymbol \theta}) = -n^{-1}\sum_{i=1}^{n} \log p(\boldsymbol z^{obs}_{i}, \boldsymbol z_{\boldsymbol r_i} \in \mathcal{E}(\boldsymbol r_i); \boldsymbol C_{\boldsymbol \theta})$ and 
$p(\boldsymbol z_i^{obs}, \boldsymbol z_{\boldsymbol r_i} \in \mathcal{E}(\boldsymbol r_i); \boldsymbol C_{\boldsymbol \theta})$ is a function of the Gaussian density which is log concave. Condition 2 is satisfied by the strong law of large numbers and since each $(\boldsymbol y_i, \boldsymbol r_i)$, where $i=1, \dots, n$, is an i.i.d.\ sample from $\Pi_{0}$. 

To validate condition 6, we first establish the Kullback-Leibler (KL) divergence between $\Pi^{obs}_{\boldsymbol C_{\boldsymbol \theta}}  = p(\boldsymbol y^{obs}, \boldsymbol r \mid \boldsymbol C_{\boldsymbol \theta}, \{F_{j}\}_{j=1}^{p})$ for $\boldsymbol C_{\boldsymbol \theta} \in \mathbbm{C}$ and the distribution of the observed data given the ground truth copula correlation, $\Pi^{obs}_{0} = p(\boldsymbol y^{obs}, \boldsymbol r \mid\boldsymbol C_{0}, \{F_{j}\}_{j=1}^{p})$.  Here, the KL-divergence between distributions $P$ and $Q$ with density functions $p$ and $q$, respectively, is given by $d_{KL}(P,Q) = \int p(\boldsymbol x) \log\{p(\boldsymbol x)/q(\boldsymbol x)\}d\boldsymbol x$.
% With known marginals and fixed $\boldsymbol C_{\boldsymbol \theta}$, the joint density under the Gaussian copula at any value of $(\boldsymbol y^{obs}, \boldsymbol r)$ is given by
% \begin{equation}\label{cop_dens}
%     \Pi_{\boldsymbol C_{\boldsymbol \theta}}(\boldsymbol y^{obs}, \boldsymbol r) = \int_{\boldsymbol z^{mis}}\int_{\boldsymbol z_{\boldsymbol r} \in \mathcal{E}(\boldsymbol r)}\phi_{\boldsymbol C_{\boldsymbol \theta}}(\boldsymbol z^{obs}, \boldsymbol z^{mis}, \boldsymbol z_{\boldsymbol r}) d \boldsymbol z_{\boldsymbol r} d \boldsymbol z^{mis},
% \end{equation}
% where $\mathcal{E}(\boldsymbol r)$ is the proper probit set restriction on the latent scale given $\boldsymbol R = \boldsymbol r$ and $\phi_{\boldsymbol C_{\boldsymbol \theta}}()$ is the $2p$-dimensional multivariate Gaussian density function with correlation $\boldsymbol C_{\boldsymbol \theta}$. By replacing $\boldsymbol C_{\boldsymbol \theta}$ with $\boldsymbol C_{0}$, the  data generating density $\Pi_{0}$ can be expressed similarly.

%\textcolor{black}{DO we need to define this again here?  I thought we already defined this set restriction in the main text?  If it is the same, we canjust use it without defining again.}

To begin, let $\boldsymbol r^{g}$ be one of the $2^p$ possible combinations of $(R_1, \dots, R_p)$, with $\mathcal{E}(\boldsymbol r^{g})$ the associated probit set restriction for $\boldsymbol z_{\boldsymbol r}$ when  $\boldsymbol r = \boldsymbol r^{g}$. For example, for two study variables consider $\boldsymbol r^{g} = (R_{1} =1, R_{2} = 1)$. For arbitrary $\boldsymbol C_{\boldsymbol \theta}$, we write the marginal probability $p(\boldsymbol R = \boldsymbol r^{g}) = \int_{\boldsymbol z_{\boldsymbol r} \in \mathcal{E}(\boldsymbol r^{g})}\phi_{\boldsymbol C_{\boldsymbol \theta}}(\boldsymbol z_{\boldsymbol r}) d \boldsymbol z_{\boldsymbol r} = \int_{0}^{\infty}\int_{0}^{\infty} \phi_{\boldsymbol C_{\boldsymbol \theta}}(z_{1},z_{2}) d z_{1} d z_{2}$. 
%Writing these integrals for each possible $\boldsymbol r^{g}$ pattern, the
The set restriction $\mathcal{E}(\boldsymbol r^{g})$ implied by the collection of unique missingness patterns in $\boldsymbol r^{g}$ covers a non-overlapping orthant in the $p$-dimensional latent space for $\boldsymbol z_{\boldsymbol r}$.  Thus, we have $\cup_{g =1}^{2^p} \mathcal{E}(\boldsymbol r^{g}) = \sum_{g=1}^{2^{p}} \mathcal{E}(\boldsymbol r^{g}) = \mathbbm{R}^{p}$. Illustrating this with the two-dimensional example above, we observe that $\sum_{g=1}^{2^2} \int_{\boldsymbol z_{\boldsymbol r} \in \mathcal{E}(\boldsymbol r^{g})}\phi_{\boldsymbol C_{\boldsymbol \theta}}(\boldsymbol z_{\boldsymbol r}) d \boldsymbol z_{\boldsymbol r} = \int_{-\infty}^{\infty} \int_{-\infty}^{\infty}  \phi_{\boldsymbol C_{\boldsymbol \theta}}(\boldsymbol z_{\boldsymbol r}) d \boldsymbol z_{\boldsymbol r}$. Thus, for $p$-dimensional $\boldsymbol z_{\boldsymbol r}$, we may express the expectation of $g(\boldsymbol z_{\boldsymbol r})$ for some function $g$ as
\begin{align}
    \sum_{g=1}^{2^p}\int_{\boldsymbol z_{\boldsymbol r} \in \mathcal{E}(\boldsymbol r^g)} g(\boldsymbol z_{\boldsymbol r})\phi_{\boldsymbol C_{\boldsymbol \theta}}(\boldsymbol z_{\boldsymbol r}) d \boldsymbol z_{\boldsymbol r}=\int_{\mathbbm{R}^{p}} g(\boldsymbol z_{\boldsymbol r})\phi_{\boldsymbol C_{\boldsymbol \theta}} d \boldsymbol z_{\boldsymbol r}
\end{align}

% $\mathcal{E}(\boldsymbol r')$ be the set of probit restrictions for a

% $\mathcal{E}(R)$ induced by all 
% %possible combination  $R$. With $p$ study variables, there are
% $2^{p}$ possible combinations of $R_1, \dots, R_p$. 
% In words, $\mathcal{E}(R)$ represents a condition on each $(R_{j}, Z_j)$ so that whenever any $r_{ij}>0$ we have $z_{ij}>0$. 
% \textcolor{black}{JOE: Do we need all the  notation/detail below to define E(R)? It seems like a lot of overhead and we really never use the definition. Also, just to clarify, is this different from what we defined previously as the probit  set restriction?  Is it now over all possible $R_1, \dots, R_p$ whereas before we had the observed $r$?}
% Formally, we define $\mathcal{E}(R) = \{\mathcal{E}(\boldsymbol r'): r'_{ij} =1 \implies z'_{ij} > 0\}$.  THIS FEELS A LITTLE LOOSE TO ME SINCE $i$ IS ONE SPECIFIC INDIVIDUAL.
% Each $\mathcal{E}(\boldsymbol r')$ covers a unique orthant in the $p$-dimensional latent space, and $\cup_{\mathcal{E}(\boldsymbol r') \in \mathbbm{E}(\boldsymbol R)} \mathcal{E}(\boldsymbol r') = \sum_{\mathcal{E}(\boldsymbol r') \in \mathbbm{E}(\boldsymbol R)} \mathcal{E}(\boldsymbol r') =   \mathbbm{R}^{p}$.   

Consequently, with known marginals, we leverage the simplification of the Gaussian copula likelihood in \eqref{copwork1}--\eqref{integrand} to express $d_{KL}(\Pi^{obs}_{0},\Pi^{obs}_{\boldsymbol C_{\boldsymbol \theta}})$ as
\begin{align}
    &d_{KL}(\Pi^{obs}_{0},\Pi^{obs}_{\boldsymbol C_{\boldsymbol \theta}}) = \sum_{g=1}^{2^p}\int p(\boldsymbol y^{obs}, \boldsymbol r^{g} \mid \boldsymbol C_{0}, \{F_{j}\}_{j=1}^{p}) \log \bigg\{\frac{p(\boldsymbol y^{obs}, \boldsymbol r^{g} \mid \boldsymbol C_{0}, \{F_{j}\}_{j=1}^{p})}{p(\boldsymbol y^{obs}, \boldsymbol r^{g} \mid \boldsymbol C_{\boldsymbol \theta}, \{F_{j}\}_{j=1}^{p})}\bigg\} d \boldsymbol y^{obs} \label{obsKL} \\&=\sum_{g=1}^{2^p}\int \int \int_{\boldsymbol z_{\boldsymbol r} \in \mathcal{E}(\boldsymbol r^{g})} \phi_{C_{0}}(\boldsymbol z^{obs}, \boldsymbol z^{mis}, \boldsymbol z_{\boldsymbol r}) \mbox{log}\bigg\{\frac{\phi_{\boldsymbol C_{0}}(\boldsymbol z^{obs}, \boldsymbol z^{mis}, \boldsymbol z_{\boldsymbol r})}{\phi_{\boldsymbol C_{\boldsymbol \theta}}(\boldsymbol z^{obs}, \boldsymbol z^{mis}, \boldsymbol z_{\boldsymbol r})}\bigg\}d \boldsymbol z^{obs} d\boldsymbol z^{mis} d\boldsymbol z_{\boldsymbol r}\label{kl-expand} \\
    &= \int \int \int \phi_{\boldsymbol C_{0}}(\boldsymbol z^{obs}, \boldsymbol z^{mis}, \boldsymbol z_{\boldsymbol r'}) \mbox{log}\bigg\{\frac{\phi_{\boldsymbol C_{0}}(\boldsymbol z^{obs}, \boldsymbol z^{mis}, \boldsymbol z_{\boldsymbol r})}{\phi_{\boldsymbol C_{\boldsymbol \theta}}(\boldsymbol z^{obs}, \boldsymbol z^{mis}, \boldsymbol z_{\boldsymbol r})}\bigg\}d \boldsymbol z^{obs} d\boldsymbol z^{mis} d\boldsymbol z_{\boldsymbol r}.\label{simp-kl}
\end{align}
The equivalence between \eqref{obsKL} and \eqref{kl-expand} arises from \eqref{copwork1}--\eqref{integrand} in the main text: with the margins known, the transformation between $\boldsymbol y^{obs}$ and $\boldsymbol z^{obs}$ is fixed. In addition, since $\cup_{g =1}^{2^p} \mathcal{E}(\boldsymbol r^{g}) = \mathbbm{R}^{p}$, we may collapse the sum into an integral which leads to the equivalence between \eqref{kl-expand} and \eqref{simp-kl}.
% Since $\int_{a}^{c} f(x)dx = \int_{a}^{b} f(x) dx + \int_{b}^{c}f(x) dx$ for $a<b<c$,

Consequently, the KL-divergence between $\Pi^{obs}_{0}$ and $\Pi^{obs}_{\boldsymbol C_{\boldsymbol \theta}}$ when the margins are known reduces to evaluating the KL divergence between $2p$-dimensional multivariate Gaussian distributions. It is well known that 
\begin{align}
    d_{KL}(\Pi^{obs}_{0},\Pi^{obs}_{\boldsymbol C_{\boldsymbol \theta}}) &= 
    \frac{1}{2}\bigg\{tr(\boldsymbol C_{\boldsymbol \theta}^{-1} \boldsymbol C_{0}) - 2p + \log \bigg(\frac{\lvert \boldsymbol C_{\boldsymbol \theta}\rvert}{\lvert \boldsymbol C_{0}\rvert}\bigg)\bigg\}.\end{align}
    % \\
    % &= \frac{1}{2}\bigg\{\sum_{{u\leq v, v = 1}}^{2p}(M_{uv})^2- 2p + \log\bigg(\frac{\lvert \boldsymbol C_{\boldsymbol \theta}\rvert}{\lvert \boldsymbol C_{0}\rvert}\bigg)\bigg\}. \label{simplify} 
This quantity is minimized when $\boldsymbol C_{\boldsymbol \theta} = \boldsymbol C_{0}$, in which case  $d_{KL}(\Pi^{obs}_{0},\Pi^{obs}_{\boldsymbol C_{\boldsymbol \theta}}) = 0$. By construction, $\boldsymbol C_{\boldsymbol \theta} = \boldsymbol C_{0}$ also minimizes $f(\boldsymbol C_{\boldsymbol \theta})$. 
%=  - \mathbbm{E}_{\Pi_{0}} \log p\{\boldsymbol z^{obs}, \boldsymbol z_{\boldsymbol r} \in \mathcal{E}(\boldsymbol r); \boldsymbol C_{\boldsymbol \theta}\}$. 
Therefore, $f(\boldsymbol C_{\boldsymbol \theta}) > f(\boldsymbol{C}_{0})$ for all $\boldsymbol C_{\boldsymbol \theta} \in \mathbbm{C}\setminus \boldsymbol C_{0}$.
\end{proof}

\vspace{12pt}
\begin{lemma}\label{lemma1}
   Suppose $\{(\boldsymbol{y}_i, \boldsymbol r_i) \}_{i=1}^{n}\overset{iid}{\sim} \Pi_{0}$ where $\Pi_{0}$ is the Gaussian copula with
    % SHOULD WE DELETE $\boldsymbol{y}^{mis}$?  \textcolor{black}{comprise $n$} independent and identically distributed samples from \eqref{YRcop}-\eqref{xformYR} with  true
    correlation $\boldsymbol C_{0}$ and  marginals $\{F_{j}\}_{j=1}^{p}$ as in \eqref{YRcop}--\eqref{xformYR}, and 
    %\textcolor{black}{where $\boldsymbol{y}^{mis}$ follows a nonignorable missingness mechanism under the copula model.}  
    $\{\mathcal{A}_{j}\}_{j=1}^{p} = \{F_{j}\}_{j=1}^{p}$. For any value of $(\boldsymbol y^{obs}_{i},\boldsymbol y^{mis}_{i})$,  $p(R_{ij} =1 \mid \boldsymbol y_i^{obs}, \boldsymbol y_i^{mis}, \boldsymbol{C}_{j}^{*},  \alpha_{r_{j}}, \{F_{j}\}_{j=1}^{p})$ satisfies \eqref{AN} with  $x_{ik}=z_{ik}$,
   %= \Phi^{-1}(F_{j}^{-1}(y_{ij}))]_{j=1}^{p}$, 
   $g$ the probit link function in \eqref{copmis}--\eqref{moments}, $\beta_{0} = \alpha_{r_{j}}$, and $\beta_{k}$  the $k$th component of the vector $\boldsymbol{C}_{r_{j} \boldsymbol y}\boldsymbol{C}_{\boldsymbol y}^{-1}$.
\end{lemma}
%\begin{lemma}\label{lemma1}
%   Suppose $(\boldsymbol{y}, \boldsymbol r) \overset{iid}{\sim} \Pi_{0}$ where $\Pi_{0}$ is the Gaussian copula with
    % SHOULD WE DELETE $\boldsymbol{y}^{mis}$?  \textcolor{black}{comprise $n$} independent and identically distributed samples from \eqref{YRcop}-\eqref{xformYR} with  true
 %   correlation $\boldsymbol C_{0}$ and univariate marginals $\{F_{j}\}_{j=1}^{p}$ as in \eqref{YRcop}-\eqref{xformYR}, and 
    %\textcolor{black}{where $\boldsymbol{y}^{mis}$ follows a nonignorable missingness mechanism under the copula model.}  
  %  assume $\{\mathcal{A}_{j}\}_{j=1}^{p} = \{F_{j}\}_{j=1}^{p}$. For any value of $\boldsymbol y^{obs}_{i},\boldsymbol y^{mis}_{i}$, the missingness mechanism $p(R_{ij} =1 \mid \boldsymbol y_i^{obs}, \boldsymbol y_i^{mis}, \{F_{j}\}_{j=1}^{p}, \boldsymbol{C}_{j}^{*}, \boldsymbol \alpha_j^{*})$ satisfies \eqref{AN} with  $x_{ij}=z_{ij}$,
   %= \Phi^{-1}(F_{j}^{-1}(y_{ij}))]_{j=1}^{p}$, 
   %$g$ the probit link function in \eqref{copmis}-\eqref{moments}, $\beta_{0} = \alpha_{r_{j}}$, and $\boldsymbol \beta_{j}$  the $j$th component of the vector $\boldsymbol{C}_{r_{j}y}\boldsymbol{C}_{y}^{-1}$.
%\end{lemma}
\begin{proof}
%The result follows by first noticing the equivalence 
The probit restriction implies that $Z_{R_{ij}}>0$ when $R_{ij}=1$.  Therefore, 
\begin{align}\label{copmis1}
      p(R_{ij} = 1 \mid \boldsymbol y_i^{obs}, \boldsymbol y_i^{mis}, \boldsymbol{C}_j^{*},  \alpha_{r_j}, \{F_{j}\}_{j=1}^{p})&=   p(Z_{R_{ij}} >0 \mid \boldsymbol z_i^{mis}, \boldsymbol z_i^{obs}, \boldsymbol{C}_j^{*}, \alpha_{r_j})\\ &= 1- \Phi_{\alpha_{ij}^{*}, \sigma_{j}^{2*}}(0)\\
      &= 1 - \Phi_{0,\sigma_{j}^{2*}}(\alpha_{ij}^{*}). \label{probit-link1}
\end{align}
%Let $\boldsymbol z_{i} = (\boldsymbol z_{i}^{obs}, \boldsymbol z_{i}^{mis})$. Then, 
Here, we have 
\begin{align}  
\alpha_{ij}^{*} &= \alpha_{r_{j}} + \boldsymbol C_{r_{j}\boldsymbol y}\boldsymbol C_{\boldsymbol y}^{-1}\boldsymbol z_{i}\\
&= \alpha_{r_{j}} + \sum_{k=1}^{p}\sum_{j=1}^{p} \boldsymbol C_{r_{j}y_{j}}\boldsymbol C^{-1}_{y_{k}y_{j}} z_{ij}\\
&= \alpha_{r_{j}} + \sum_{j=1}^{p} \beta_{j} z_{ij}. 
\end{align}
%and $\beta_{j} = \sum_{k=1}^{p} C_{r_{j}y_{k}}C^{-1}_{y_{k},y_{j}}$.
Consequently, the mean of \eqref{probit-link1} is additive in the components of $\boldsymbol z_i$, 
%= [\Phi^{-1}\{F_{j}(y_{ij})\}]_{j=1}^{p}$, 
which implies that $p(R_{ij} = 1 \mid \boldsymbol y_i^{obs}, \boldsymbol y_i^{mis}, \boldsymbol{C}_j^{*}, \alpha_{r_j},\{F_{j}\}_{j=1}^{p})$ satisfies \eqref{AN} with the probit link function.
\end{proof}

\vspace{12pt}

\begin{theorem}\label{postconsistEQL}
   Suppose $\{(\boldsymbol{y}_i, \boldsymbol r_i) \}_{i=1}^{n} \overset{iid}{\sim} \Pi_{0}$ where $\Pi_{0}$ is the Gaussian copula with correlation $\boldsymbol C_{0}$ and marginals $\{F_{j}\}_{j=1}^{p}$ as in \eqref{YRcop}--\eqref{xformYR}. For $j = 1, \dots, p$, suppose $\mathcal{A}_{j}$ comprises $m_{j}\geq 3$ auxiliary quantiles of $F_{j}$, including  $F_{j}^{-1}(0)$ and $F_{j}^{-1}(1)$. Let $p(\boldsymbol \theta)$ be a prior with respect a measure that induces a prior $\Pi$ over the space of all $2p \times 2p$ correlation matrices $\boldsymbol{\mathbbm{C}}$ such that $\Pi(\boldsymbol C_{\boldsymbol \theta})>0$ for all $\boldsymbol C_{\boldsymbol \theta} \in \mathbbm{C}$. Then, for any neighborhood $\mathcal{B}$ of $\boldsymbol C_{0}$, $\lim_{n\rightarrow \infty} \Pi^{*}_{n}(\boldsymbol C_{\boldsymbol \theta} \in \mathcal{B}) \rightarrow 1 \ \textrm{almost surely } [\Pi_{0}]$.
    % MAYBE WE SHOULD ADD THE AN MISSINGNESS MECHANISM AS A CONDITION?
\end{theorem}

%\begin{theorem}\label{postconsistEQL}
%     Suppose $(\boldsymbol{y}, \boldsymbol r) \overset{iid}{\sim} \Pi_{0}$ where $\Pi_{0}$ is the Gaussian copula with
    % SHOULD WE DELETE $\boldsymbol{y}^{mis}$?  \textcolor{black}{comprise $n$} independent and identically distributed samples from \eqref{YRcop}-\eqref{xformYR} with  true
%    correlation $\boldsymbol C_{0}$ and univariate marginals $\{F_{j}\}_{j=1}^{p}$ as in \eqref{YRcop}-\eqref{xformYR}. For $j = 1, \dots, p$, assume $\mathcal{A}_{j}$ is comprised of $m_{j}\geq 3$ auxiliary quantiles of the true marginal $F_{j}$, including  $F_{j}^{-1}(0)$ and $F_{j}^{-1}(1)$. Let $p(\boldsymbol \theta)$ be a prior with respect a measure that induces a prior $\Pi$ over the space of all $2p \times 2p$ valid correlation matrices $\boldsymbol{\mathbbm{C}}$ such that $\Pi(\boldsymbol C)>0$ for all $\boldsymbol C \in \mathbbm{C}$. Then, for any neighborhood $\mathcal{B}$ of $\boldsymbol C_{0}$ $\lim_{n\rightarrow \infty} \Pi^{*}_{n}(\boldsymbol C_{\theta} \in \mathcal{B}) \rightarrow 1 \ a.s. \ [\Pi_{0}]$
%\end{theorem}
\begin{proof}
    
% \textcolor{black}{Need to show that for any two variables (miss indicators or study variables), that we can estimate the copula correlation. We do this with Doob's theorem, and we'll show it for two study variables $(Y_{1}, Y_{2})$, from which the other cases follow easily.
% The key is to recognize that for any value of $b_{1}, b_{2}$, the missingness mechanism $p(R_{1}, R_{2} \mid b_{1}, b_{2})$ can be expressed in terms of latent variables, and can be represented as a sequentially additive non-ignorable missingness mechanism. Also the auxiliary information fully identifies the marginal distribution of each $b$. Therefore, we conclude that $p(b_{1}, b_{2})$ is identified. Since $(y,r)$ come from a copula, we also conclude that with large enough $n$ we can accurately estimate the contingency table}

We prove this result using a variant of Doob's theorem presented in \cite{gu2009bayesian}.

{\bf Doob's Theorem }{\it \ Let $X_{i}$ be observations whose distributions depend on a parameter $\boldsymbol \theta$, both taking values in Polish spaces. Assume $\boldsymbol \theta \sim \Pi$ and $X_{i}\mid \boldsymbol \theta \sim P_{\boldsymbol \theta}$. Let $\mathcal{X}_{N}$ be the $\sigma$-field generated by $X_{1},\dots,X_{N}$, and $\mathcal{X}_{\infty} = \sigma(\bigcup_{i}^{\infty}\mathcal{X}_{i})$. If there exists a $\mathcal{X}_{\infty}$ measurable function $f$ such that for $(\boldsymbol \omega, \boldsymbol \theta) \in \Omega^{\infty} \times \Theta, \ \boldsymbol \theta = f(\boldsymbol \omega) \ a.e. \ [P_{\boldsymbol \theta}^{\infty} \times \Pi]$ then the posterior is strongly consistent at $\boldsymbol \theta$ for almost every $\boldsymbol \theta \ [\Pi]$.}

For  $j = 1, \dots, p$, let $B_{j}$ be the ordinal random variable with $m_{j}-1$ levels resulting from the coarsening of $Y_{j}$ from the intervals in \eqref{bins}. Let $R_j$ be the random variable corresponding to the process of nonresponse for $Y_j$. Because of Doob's theorem, it suffices to show that, for any pair of $(B_{j}, B_{j'})$, $(B_{j}, R_{j'})$, or $(R_{j}, R_{j'})$, we can recover the true copula correlation $\rho_{0,jj'}$ between the pair of variables by a function that is measurable with respect to the $\sigma$-field generated by the sequence of $\{\boldsymbol z^{obs} \in \mathcal{D}(\boldsymbol b^{obs}), \boldsymbol z_{\boldsymbol r} \in \mathcal{E}(\boldsymbol r) \}$ as $n \rightarrow \infty$.

We first note that the marginal distribution of any $B_{j}$ is fully specified by $\mathcal{A}_{j}$. To see this, for any level $q = 1,\dots, m_{j}-1$ with $B_{j} = q$ defined as in \eqref{bins}, the marginal probability is exactly $p(B_{j} = q) = p(Z_{j} \in (\Phi^{-1}(\tau_{j}^{q}),\Phi^{-1}(\tau_{j}^{q+1})]) = \tau_{j}^{q+1}- \tau_{j}^{q}$. Furthermore, $\sum_{q=1}^{m_{j} -1} p(B_{j} = q) = \sum_{q=1}^{m_{j} -1} \tau_{j}^{q+1}- \tau_{j}^{q} =1$, since $\tau_{j}^{m_{j}} = 1$ and $\sum_{q=1}^{m_{j} -1} \tau_{j}^{q+1}- \tau_{j}^{q}$ is a telescoping series. Thus, $\mathcal{A}_{j}$ provides the entire marginal distribution function of  $B_{j}$. 

We next consider the joint distributions arising from the three combinations of binned study variables and missingness indicators for any $j,j' =\in \{1,\dots,p\}$.
%\begin{enumerate}
%    \item $p(B_{j}, R_{j'})$
%%    \item $p(B_{j}, B_{j'})$
%\end{enumerate}
The joint distribution for $(B_j, B_{j'})$, where $j \neq j'$, can be represented by a contingency table concatenated to a  $m_{j}m_{j'} \times 1$ vector $\boldsymbol \pi_{B_j,B_{j'}} = \{\pi_{qk}: q = 1, \dots, m_j; k = 1, \dots, m_{j'}\}$, where any $\pi_{qk} = p(B_{j} = q, B_{j'} = k)$.  Similarly, we can define $\boldsymbol \pi_{B_j, R_{j'}} = \{\pi_{qk}: q = 1 \dots, m_j; k = 0,1\}$ where $\pi_{qk} = p(B_{j} = q, R_{j'} = k)$.  And, for $j \neq j'$,  we can define $\boldsymbol \pi_{R_j, R_{j'}} = \{\pi_{qk}: q = 0,1; k = 0,1\}$ where $\pi_{qk} = p(R_{j} = q, R_{j'} = k)$. 
%$m_{j} \times 2$, $2 \times 2$, or
By construction, we have that $p(B_{j} = q) = p(Y_{j} \in \mathcal{I}_{j}^{q})$, and so any of the joint probabilities above involving $B_{j}$ may  be expressed in terms of $Y_{j}$. Since $\{(\boldsymbol y_i, \boldsymbol r_i)\}_{i=1}^{n}$ come from a Gaussian copula with correlation $\boldsymbol C_{0}$, and any pair of variables also follows a Gaussian copula with sub-correlation $\rho_{0,jj'}$, we may express the joint distributions in terms of the data generating Gaussian copula parameters. For example,
\begin{align}
    p(B_{j} = q, B_{j'} = k) &= p(Y_{j} \in \mathcal{I}_{j}^{q}, Y_{j'} \in \mathcal{I}_{j'}^{k})\label{bins2obs} \\
    &= \int_{\Phi^{-1}(\tau_{j}^{q})}^{,\Phi^{-1}(\tau_{j}^{q+1})}\int_{\Phi^{-1}(\tau_{j'}^{k})}^{\Phi^{-1}(\tau_{j'}^{k+1})}\phi(z_{j}, z_{j'}; \rho_{0,jj'})d z_j dz_{j'}.
\end{align}
For notational convenience, we drop the subscripts from $\boldsymbol \pi_{B_j,B_{j'}}$, $\boldsymbol \pi_{B_j,R_{j'}}$, and $\boldsymbol \pi_{R_j,R_{j'}}$ and let $\boldsymbol \pi$ be defined in context. 

%For each combination of variables, c
We consider three types of sequences of empirical contingency tables.  For pairs $(B_j, B_{j'})$, the sequence is $\boldsymbol \pi^{n}$ with entries $\pi^{n}_{qk} = n^{-1}\sum_{i=1}^{n} \mathbbm{1}(B_{ij}= q, B_{ij'} = k)$.  Re-using $\pi^n_{qk}$ for economy of notation, for pairs $(R_j, R_{j'})$, the sequence has entries $\pi^{n}_{qk} = n^{-1}\sum_{i=1}^{n} \mathbbm{1}(R_{ij}= q, R_{ij'} = k)$.  Finally, for pairs $(B_j, R_{j'})$, the sequence has entries $\pi^{n}_{qk} = n^{-1}\sum_{i=1}^{n} \mathbbm{1}(B_{ij}= q, R_{ij'} = k)$.
%As a slight abuse of notation, we use $./.$ to denote the empirical contingency table resulting from each of the possible combinations of $B_{j}$ and $R_{j}$ above.
%Note that for each combination, the 
%The information contained in $\boldsymbol \pi^{n}$ is also contained in the EQL. 
Note that we may construct $\boldsymbol \pi^{n}$ from the information contained in any realization of $\{\boldsymbol z^{obs} \in \mathcal{D}(\boldsymbol b^{obs}), \boldsymbol z_{\boldsymbol r} \in \mathcal{E}(\boldsymbol r)\}$  by simply observing the counts falling into corresponding intervals specified by $\mathcal{A}$ or the probit restrictions on the $R_j$. 
Therefore, the $\sigma$-field generated by the sequence of $\boldsymbol \pi^{n}$ is a sub $\sigma$-field of that generated by the sequence of $\{\boldsymbol z^{obs} \in \mathcal{D}(\boldsymbol b^{obs}), \boldsymbol z_{\boldsymbol r} \in \mathcal{E}(\boldsymbol r) \}$. Thus, any function that is measurable with respect to the $\sigma$-field generated by the sequence of $\boldsymbol \pi^{n}$ is also measurable with respect to that generated by the sequence of $\{\boldsymbol z^{obs} \in \mathcal{D}(\boldsymbol b^{obs}), \boldsymbol z_{\boldsymbol r} \in \mathcal{E}(\boldsymbol r) \}$.  We will work exclusively with the former. 

We proceed to show that for any pair of variables among $\{B_{1}, \dots, B_p, R_1, \dots,  R_p$\}, 
each entry in $\boldsymbol \pi^{n}$ is identified by the observed (coarsened) data. As a consequence of the model for $\{(\boldsymbol y_i, \boldsymbol r_i)\}_{i=1}^{n}$, we show that $\boldsymbol \pi^{n} \overset{a.s.}{\rightarrow} \boldsymbol \pi$. We then show that there exists an estimator $\hat{\rho}_{jj'}$ which may be obtained by a function that is measurable with respect to the $\sigma$-field generated by the sequence of $\boldsymbol \pi_{n}$. This estimator must converge to a limit, $\rho^{*}_{jj'}$, which is therefore measurable with respect to the $\sigma$-field generated by the infinite sequence of $\boldsymbol \pi_{n}$. Finally, we show that $\rho^{*}_{jj'}=\rho_{0,jj'}$ which concludes the proof.

We first consider  $\boldsymbol \pi^{n}$ constructed from $(R_{j}, R_{j'})$. Since each $R_{ij}$ is always observed, it is immediately apparent through the strong law of large numbers (S.L.L.N.) that $\boldsymbol \pi^{n} \overset{a.s.}{\rightarrow} \boldsymbol \pi$. 

For $\boldsymbol \pi^{n}$ constructed using $(B_{j}, R_{j'})$, we first utilize theory on the additive nonignorable missingness mechanism  along with the identifying information for $B_{j}$ to demonstrate that $\boldsymbol \pi^{n}$ is identified by the observed data.  We have the following equivalences under the data generating Gaussian copula.
\begin{align}
    &p(B_{j} = q, R_{j'} = 1) =p(B_{j} = q)p(R_{j'} = 1\mid B_{j} = q)\label{binned_scale}\\
    &= p(Y_{j} \in \mathcal{I}_{j}^{q})p(R_{j'} = 1 \mid Y_{j} \in \mathcal{I}_{j}^{q})\label{obs_scale}\\
    &= p(Z_{j} \in (\Phi^{-1}(\tau_{j}^{q}), \Phi^{-1}(\tau_{j}^{q+1})])p(Z_{R_{j'}} >0 \mid Z_{j} \in (\Phi^{-1}(\tau_{j}^{q}), \Phi^{-1}(\tau_{j}^{q+1})]).\label{latent_scale}
\end{align}
The first term in \eqref{latent_scale} is known since the marginal distribution of $B_{j}$ is fully specified from $\mathcal{A}_j$. To characterize the missingness mechanism implied by the second term, we require theory on the selection normal distribution \citep{arellano2006unification}, given in Lemma \eqref{SLCT}.

\begin{lemma}\label{SLCT}
    Suppose $(\boldsymbol x_{0}, \boldsymbol x_1)$ are jointly Gaussian such that $\boldsymbol X_{0} \sim N_{p}(\boldsymbol \mu_0, \boldsymbol \Sigma_{0})$ and $\boldsymbol X_{1} \sim N_{p^*}(\boldsymbol \mu_1,\boldsymbol\Sigma_{1})$  with joint  covariance matrix  $\boldsymbol \Sigma = \begin{bmatrix}
        \boldsymbol \Sigma_{0} & \boldsymbol \Sigma_{01} \\
        \boldsymbol \Sigma_{10} &\boldsymbol \Sigma_{1}
    \end{bmatrix}$.
    Define $\boldsymbol X_{0}^{(\mathcal{C})} \overset{d}{=} [\boldsymbol X_{0} \mid \boldsymbol X_{0} \in \mathcal{C}]$ for $\mathcal{C}$ some $p$-dimensional hypercube. Then, $[\boldsymbol X_{1} \mid \boldsymbol X_0 \in \mathcal{C}] \overset{d}{=} \mu_0 + \boldsymbol X_{1} + \boldsymbol \Sigma_{10}^{T}\boldsymbol \Sigma_{0}^{-1} \boldsymbol X_{0}^{(\mathcal{C})} $, and we say that $[\boldsymbol X_{1} \mid \boldsymbol X_0 \in \mathcal{C}] \sim \mbox{SLCT-N}_{p,p^*}(\boldsymbol \mu_0, \boldsymbol \mu_1,\boldsymbol \Sigma_{0}, \boldsymbol \Sigma_{1}, \boldsymbol \Sigma_{01}, \mathcal{C})$.
\end{lemma}

Setting $\mathcal{C} = (\Phi^{-1}(\tau_{j}^{q}), \Phi^{-1}(\tau_{j}^{q+1})]$ we observe from \eqref{latent_scale} that $[Z_{R_{j'}} \mid Z_{j} \in \mathcal{C}]\sim \\\mbox{SLCT-N}_{1,1}(\alpha_{R_j}, 0, 1, 1, \rho_{0,jj})$. 
%Thus, $Z_{R_{ij}}$ in \eqref{slct-mis} follows a SLCT-N distribution.  
%#It is straightforward to verify that t
This distribution satisfies the requisite properties of a link function outlined in Lemma \ref{lemma1}. Furthermore, the construction $[\boldsymbol X_{1} \mid \boldsymbol X_0 \in \mathcal{C}] \overset{d}{=} \mu_0 + \boldsymbol X_{1} + \boldsymbol \Sigma_{10}'\boldsymbol \Sigma_{0}^{-1} \boldsymbol X_{0}^{(\mathcal{C})}$ of SLCT-N random variables demonstrates that this missingness mechanism is additive in  $Z_{j} \in \mathcal{C}$, with $\beta = \rho_{0,jj'}$. Using the equivalence between \eqref{binned_scale} and \eqref{latent_scale}, together with the fact that the marginal distribution of $B_{j}$ is completely specified, by Theorem 1 of \cite{sadinle2019sequentially} this establishes that each component of $\boldsymbol \pi^{n}$ is identified by the observed data.  Furthermore, due to the correspondence between $(B_{j}, R_{j'})$ and $(Y_{j}, R_{j'})$, and since $(Y_{j}, R_{j'})$ are distributed according to a Gaussian copula with correlation $\rho_{0,jj'}$, the identified probability $\pi^{n}_{qk}$ must converge to $\pi_{qk}$ generated by the true copula by the S.L.L.N.
% \ we have 
% \begin{align}
%     \pi^{n}_{qk} &= n^{-1}\sum_{i=1}^{n} \mathbbm{1}(b_{ij} =q, r_{ij'} = k)\label{emp}\\ &= n^{-1}\sum_{i=1}^{n} \mathbbm{1}(y_{ij} \in \mathcal{I}_{j}^{q}, r_{ij'} = k) \\
%     &\overset{a.s}{\rightarrow} p(y_{ij} \in \mathcal{I}_{j}^{q}, R_{j'} = k) \\
%     &= p(B_{j} = q, R_{j'} = k) = \boldsymbol \pi_{qk} \label{limit}
% \end{align}
Applying this to each entry of the vector we have $\boldsymbol \pi^{n} \overset{a.s.}{\rightarrow} \boldsymbol \pi$ for any pair $(B_j, R_{j'})$. 

For $\boldsymbol \pi^{n}$ constructed using $(B_{j}, B_{j'})$, we use similar logic as used for $(B_{j}, R_{j'})$. We first consider the joint probability, $p(B_{j} = q, B_{j'} = k, R_{j} = 1, R_{j'} = 1)$.  This may be written as 
\begin{align}
     &p(B_{j} = q, B_{j'} = k, R_{j} = 1, R_{j'} = 1)  \nonumber\\ &= p(B_{j} = q, B_{j'} = k)p(R_{j} = 1, R_{j'} = 1 \mid B_{j}=q, B_{j'} = k). \label{jointexpand}
\end{align}
Of course, this decomposition holds for any combination in the sample space of $(B_{j}, B_{j'}, R_{j},R_{j'})$. We can expand the joint missingness mechanism in \eqref{jointexpand} as
\begin{align}
    &p(R_{j} = 1, R_{j'}=1 \mid B_{j}=q, B_{j'} = k) \nonumber\\ &= p(R_{j} = 1\mid B_{j}=q, B_{j'} = k)p(R_{j'} = 1 \mid B_{j'} = q, B_{j} = k, R_{j} = 1) \\
    &= p(Z_{R_{j}}>0 \mid Z_{j} \in \mathcal{C}, Z_{j} \in \mathcal{C'})p(Z_{R_{j'}} >0 \mid Z_{j} \in \mathcal{C}, Z_{j} \in \mathcal{C'}, Z_{R_{j}}>0), \label{sqnAN}
\end{align}
where $\mathcal{C} = (\Phi^{-1}(\tau_{j}^{q}), \Phi^{-1}(\tau_{j}^{q+1})]$ and $\mathcal{C'} = (\Phi^{-1}(\tau_{j'}^{k}), \Phi^{-1}(\tau_{j'}^{k+1})]$. Thus, each term in \eqref{sqnAN} reveals an additive nonignorable missingness mechanism via SLCT-N random variables per Lemma \ref{SLCT}. This form defines a sequential additive nonignorable missingness mechanism for multivariate nonignorable missing data (\cite{sadinle2019sequentially}, Definition 6). Since the marginal distributions of both $B_{j}$ and $B_{j'}$ are fully specified by $\mathcal{A}_{j}$ and $\mathcal{A}_{j'}$, and the joint missingness mechanism is sequentially additive nonignorable under a SLCT-N link, the conditions in Theorem 3 of \cite{sadinle2019sequentially} are satisfied. Therefore, the joint probabilities $p(B_{j}, B_{j'} , R_{j}, R_{j'})$, and thus the $\boldsymbol \pi^{n}$, are identified from the observed data. As in the $(B_{j}, R_{j'})$ case, the correspondence between $(B_{j}, B_{j'})$ and $(Y_{j}, Y_{j'})$ coupled with the true data generating model enables application of the S.L.L.N.\  to conclude that $\boldsymbol \pi^{n} \overset{a.s.}{\rightarrow}\boldsymbol \pi$ for any pair $(B_j, B_{j'})$.

% \begin{align}
%     \pi_{qk}^{n} &= n^{-1} \bigg(\sum_{i \in I^{obs}}\mathbbm{1}(b_{ij} = q, r_{ij'} = k) +  \sum_{i \in I^{mis}}\mathbbm{1}(b_{ij} = q, r_{ij'} = k)\bigg)\label{expandxtab}
%     % \\
%     % & =(\frac{\lvert I^{obs} \rvert}{n}) \lvert I^{obs} \rvert^{-1} \sum_{i \in I^{obs}}\mathbbm{1}(b_{ij} = q, r_{ij'} = k)+ (\frac{\lvert I^{mis} \rvert}{n}) \lvert I^{mis} \rvert^{-1} \sum_{i \in I^{mis}}\mathbbm{1}(b_{ij} = q, r_{ij'} = k) \label{expandxtab}
% \end{align}

% The right quantity in \eqref{expandxtab} is not directly computable from the sample. As such, we work with its expectation under the data generating model.

% expectation under the copula model is estimable using the theory of the 
% Here, $I^{obs}, I^{mis}$ are the indices for which $B_{j}$ are observed and missing, respectively in the sample. The right quantiIn the limit, this quantity converges almost surely to
%  \begin{align}
%      p(R_{j} = 0) p(B_{j} =q , R_{j'} = k \mid R_{j} = 0) + p(R_{j} = 1) p(B_{j} =q , R_{j'} = k \mid R_{j} = 1) 
%  \end{align}
% Though right side of \eqref{expandxtab} is not observed in the sample, it is estimable in a finite sample 

%\textcolor{black}{JOE: I feel like this paragraph is a bit redundant with my edits saying that the identified probability is in fact the probability implied by the true copula correlation.  If so, perhaps you might edit my edits at the end of each paragraph, or edit the start of this paragraph somehow.}
Finally, any $\boldsymbol \pi^{n}$ arises from discretizing latent Gaussian variables at fixed cut-points given by $\mathcal{A}$ and the probit restrictions on $R$. Thus, the problem of estimating $\rho_{jj'}$ from $\boldsymbol \pi^{n}$ reduces to estimating the polychoric correlation coefficient \citep{olsson1979maximum}. The resulting likelihood is a regular parametric family admitting a consistent estimator through maximum likelihood estimation (MLE). Therefore, the MLE $\hat{\rho}_{jj'}$ of the polychoric correlation coefficient estimated from $\boldsymbol \pi^{n}$ is measurable with respect the $\sigma$-field generated by the sequence of  $\boldsymbol \pi^{n}$, and its limit as $n \rightarrow \infty$, denoted $\rho^{*}_{jj'}$, is measurable with respect to the $\sigma$-field generated by the infinite sequence of $\boldsymbol \pi^{n}$.

It remains to show that $\rho^{*}_{jj'} = \rho_{0,jj'}$.
% Based on the results presented above, the AN missingness mechanism and auxiliary marginal information in $\mathcal{A}$ uniquely identifies the full data joint distribution of the ordinal variables constructed from the EQL and the missingness indicators. Therefore, it follows that any marginal distribution comprising a combination of study variables and missingness indicators is also identified under the AN missingness. By the strong law of large numbers (SLLN), this implies that
% %  \begin{align}
% %      \pi^{n}_{qq'} \overset{a.s}{\rightarrow} \pi^{0}_{qq'}
% %  \end{align}
% % where $\pi^{0}_{qq'}$ is the ground-truth joint probability $p(b_{ij} = q, b_{ij'} = q' \mid \rho^{*}_{jj'}, \mathcal{A})$, since $\boldsymbol y, r \overset{i.i.d}{\sim} \Pi_{0}$. 
Suppose towards a contradiction that $\rho^{*}_{jj'} \neq \rho_{0,jj'}$. For $(B_{j}, B_{j'})$, by construction, the limiting polychoric correlation satisfies
\begin{align}
    \pi_{qk} &= p(B_{j} = q, B_{j'} = k \mid \rho^{*}_{jj'})\\
    &= p(Y_{j} \in \mathcal{I}_{j}^{q}, Y_{j'} \in \mathcal{I}_{j'}^{k}\mid \rho^{*}_{jj'})\\ &= p(z_{j} \in (\Phi^{-1}(\tau_{j}^{q}),\Phi^{-1}(\tau_{j}^{q+1})], z_{j'} \in (\Phi^{-1}(\tau_{j'}^{k}),\Phi^{-1}(\tau_{j'}^{k+1})];\rho^{*}_{jj'})\\
    &= \int_{\Phi^{-1}(\tau_{j}^{q})}^{\Phi^{-1}(\tau_{j}^{q+1})}\int_{\Phi^{-1}(\tau_{j'}^{k})}^{\Phi^{-1}(\tau_{j'}^{k+1})} \phi(z_{j}, z_{j'}; \rho^{*}_{jj'}) d z_{j} d z_{j'}\\
    & \neq \int_{\Phi^{-1}(\tau_{j}^{q})}^{\Phi^{-1}(\tau_{j}^{q+1})}\int_{\Phi^{-1}(\tau_{j'}^{k})}^{\Phi^{-1}(\tau_{j'}^{k+1})} \phi(z_{j}, z_{j'}; \rho_{0,jj'}) d z_{j} d z_{j'}.
\end{align}
This is a contradiction, since under $\Pi_{0}$, $\pi_{qk} = \int_{\Phi^{-1}(\tau_{j}^{q})}^{\Phi^{-1}(\tau_{j}^{q+1})}\int_{\Phi^{-1}(\tau_{j'}^{q'})}^{\Phi^{-1}(\tau_{j'}^{q'+1})} \phi(z_{j}, z_{j'}; \rho_{0,jj'}) d z_{j} d z_{j'} $. We have shown the contradiction for joint probabilities involving $(B_{j}, B_{j'})$, but this construction holds for combinations of $(B_{j}, R_{j'})$ as well. Therefore, we conclude that $\rho^{*}_{jj} = \rho_{0,jj'}$. Consequently, we have shown the existence of a consistent estimator of $\rho_{0,jj'}$ which is measurable with respect to the $\sigma$-field generated by the sequence  of $\{\boldsymbol \pi^{n}\}_{n=1}^{\infty}$. 

\end{proof}

\section{Extensions for Unordered Categorical Variables}\label{cat}

In the main text, we present the EQL and EHQL likelihoods for mixed count and continuous variables. %subject to nonignorable missing data. 
We now describe how to incorporate unordered categorical variables with no missing or MCAR values. In this case, we need not include nonresponse indicators for the unordered categorical variables in the copula model, and we do not need auxiliary information about these variables in $\mathcal{A}$.  We took this modeling  approach for the analysis of the North Carolina lead exposure data.  We leave to future research handling nonignorable missing data and incorporating known marginal information about unordered categorical variables.
%as these variables are completely observed.  Hence, we present the extension with that model in mind.  After this presentation, we suggest a way to incorporate nonresponse in the unordered categorical variables. 

We use
%which may be MAR or MCAR, 
a diagonal orthant \citep{johndrow2013diagonal} representation for the unordered categorical study variables. The basic idea is to model each unordered categorical $Y_j$ with a set of binary indicators for the levels of $Y_j$, adding a restriction that for any individual only one of these indicators can equal one.  Following the Gaussian copula, we include a latent variable for each binary indicator, carrying the restriction on the set of indicators to the set of latent variables. 

Suppose we have $t < p$ unordered categorical  variables among the $p$ study variables. Without loss of generality, let $Y_1, \dots, Y_{t}$ represent these $t$ unordered categorical study variables, and let $Y_{t+1}, \dots, Y_p$ represent the remaining study variables, which may be continuous or ordinal. 
%Let $\boldsymbol y= \boldsymbol{y}^{w} \cup \boldsymbol{y}^{x}$, where $w$ indexes the numeric variables and $x$ indexes the unordered categorical variables.
%and $p = w+ z$.
 For $j=1, \dots, t$, each $Y_{j}$ takes one of $c_j$ levels, which we write as $\{1,\dots, c_{j}\}$. For any $y_{ij} = c \in \{{1},\dots, {c_{j}}\}$, we define a diagonal orthant representation that encodes a vector of $c_j$ binary variables,  $\boldsymbol \gamma_{j} = (\gamma_{j1}, \dots, \gamma_{jc_j})$.  For any individual's observed data, only one element in $(\gamma_{1}, \dots, \gamma_{c_j})$ equals one.  For example, if $y_{ij}=2$ and $c_{j}=4$, then $\boldsymbol \gamma_{j} = (0, 1, 0, 0)$ for individual $i$. We refer to each individual's $\boldsymbol \gamma_j$ as 
 $\boldsymbol \gamma_{ij} = (\gamma_{ij1}, \dots, \gamma_{ijc_j})$.

 In lieu of a single latent variable $Z_j$ for unordered categorical $Y_j$, we add $c_j$ latent variables, $(Z_{j1}, \dots, Z_{jc_j})$, to the copula model corresponding to each $\{\gamma_{j1}, \dots, \gamma_{jc_j}\}$. Thus, the copula model with $t$ unordered categorical study variables includes $\sum_{j=1}^t c_j + (p-t)$ latent variables for the study variables.  We refer to each individual's vector of latent values for $Y_j$ as $\boldsymbol z_{ij}= (z_{ij1}, \dots, z_{ijc_j})$.
 Since these latent variables model indicators, we use the probit representation so that $z_{ijc}>0$ when $\gamma_{ijc}=1$ and  $z_{ijc}<0$ when $\gamma_{ijc}=0$, for any $j$ and $c$. We also add a restriction that only one of $(z_{ij1}, \dots, z_{ijc_j})$ for any individual can be positive.  That is, for all individuals $i$ and unordered categorical variables $Y_j$, we require for any $c$ that  
 %whenever $y_{ij} = c$ implies that only the $c$th component is positive and the others are all negative.
\begin{equation}
\{\gamma_{ijc} = 1, \gamma_{ijc'} = 0\,\,\, \forall c' \neq c\} \implies  \{z_{ijc} >0,  z_{ijc'} < 0\,\,\, \forall c' \neq c\}.
\end{equation}
 This representation avoids the need to select one level of $Y_j$ as a reference group. Aggregating this representation across $(Y_1, \dots, Y_t)$, the observed categorical variables must satisfy the event 
 \begin{equation}\label{orthantevent}
\mathcal{D}'(\boldsymbol y^{obs}) = 
     \{\boldsymbol{z}^{obs}: \gamma_{ijc} = 1 \implies z_{ijc} >0, z_{ijc'} < 0 \, \forall c' \neq c, \,\, j=1, \dots, t\}.
     \end{equation}
 This representation is also recommended for ordinal variables with few levels \citep{feldman2022bayesian}.

 To estimate the copula model when the data are comprised of mixed continuous, ordinal, and unordered categorical variables, we combine the EQL/EHQL event from the main text defined over $(Y_{t+1}, \dots, Y_{p})$ with the diagonal orthant probit event in \eqref{orthantevent}.  This combined event, which we write as  
 %We have 
 $\mathcal{D}^{*}(\boldsymbol b^{obs}) \cup \mathcal{D}'(\boldsymbol y^{obs})$,  is used in  
 \eqref{marg-work}--\eqref{marg-work-expand} in the main text. Posterior inference for $\boldsymbol C_{\boldsymbol \theta}$ under the factor model of Section 4 in the main text
 %\eqref{factor},
 %conditioning on the event $\mathcal{D}^{*}(\boldsymbol b^{x}) \cup \mathcal{D}'(\boldsymbol y^{w})$, 
 requires simple modifications to Algorithm 1 of the main text, which are outlined in Section  \ref{samp}. As with the specification of the model for $\boldsymbol z_{\boldsymbol r}$ corresponding to the indicators for $R$, we add a non-zero $\alpha_j$ term to the model for any indicator variable introduced into the model by the diagonal orthant probit representation.
 %we add a non-zero $\alpha_j$ term to the model 
 %and estimate the components of $\boldsymbol \alpha$
 %corresponding to the augmented latent variables for those indicators.
 %encode category probabilities.

%When an unordered categorical study variable $Y_j$ has potentially nonignorable missing values---which was not the case in our application---we suggest the following approach.  We add the nonresponse indicator $R_j$ to the Gaussian copula, using a probit model with a single latent variable $Z_j$ corresponding to $R_j$.  To incorporate known auxiliary marginal probabilities $P(Y_{j}=c)$, where $c=1, \dots, c_j$, 
 %for this $Y_j$, we propose to fix the components of the augmented $\boldsymbol{\alpha}$ corresponding to the latent variables $(Z_{j1}, \dots, Z_{jc_j})$ to make the induced probabilities for $Y_j$ match the auxiliary marginal probabilities.   
 %This would be done 
 %Within the Gibbs sampler (see Section \ref{samp}), for the current draw of $\boldsymbol{C}_{\boldsymbol \theta}$,  we solve for the components of $\boldsymbol \alpha_{j}$ so that marginally, $p(z_{igc} > 0, z_{ijc'} <0) = P(Y_{j}=c)$. 
 %In our analyses of the North Carolina data, no unordered categorical variables exhibited evidence of non-ignorable missingness.
 
 \section{Model Specification and Gibbs Samplers for EQL and EHQL}\label{samp}

\subsection{Hierarchical Specification of the Factor Model}\label{sec:prior}

We provide a hierarchical specification of the latent factor model in Section 4 in the main text, which is used to estimate the copula correlation matrix $\boldsymbol C_{\boldsymbol \theta}$. For ease of notation and to match the setting of the simulations in Section 4 of the main text,  we presume each $Y_j$ is continuous or ordinal. Let $\lambda_{jh}$ be the element in the $j$th row and $h$th column of the factor loadings matrix $\boldsymbol \Lambda$. When we include latent variables for missingness indicators for all $p$ study variables, so that we have $2p$ variables in total, the model is given by
\begin{align*}
\delta_{1} \sim \mbox{Gamma}(a_{1},1)&, \delta_{l} \sim(a_{2},1), l \geq 2\\
    \xi_{h} = \prod_{l=1}^{h}\delta_{l}, \quad \phi_{jh} &\sim \mbox{Gamma}(\nu/2,\nu/2)\\
    [\lambda_{jh} \mid \phi_{jh},\xi_{h}] \sim N(0,\phi_{jh}^{-1},\xi_{h}^{-1}), \boldsymbol \eta_{i} \sim N_{k}(\boldsymbol 0, \boldsymbol I_{k}), \alpha_{j} &\sim N(0,1), \sigma_{j}^{-2} \sim \mbox{Inverse Gamma}(a_{\sigma}, b_{\sigma})\\
        \boldsymbol z_i  = \boldsymbol \alpha + \boldsymbol \Lambda \boldsymbol \eta_i + \boldsymbol \epsilon_i, &\quad \boldsymbol \epsilon_i \sim N_{2p}(\boldsymbol 0, \boldsymbol \Sigma).
    \end{align*}
%\textcolor{red}{JOE: We should make clearer that only $\alpha_j$ for binary $Y_j$ and $R_j$ need a prior, and that we set $\alpha_j=0$ if $Y_j$ is continuous.  The model as written makes it look like we have a prior for all $\alpha_j$.}
We place a prior on each non-zero component $\alpha_{j}$ of $\boldsymbol\alpha$.  These correspond to latent variables for each $R_j$. The prior for $\boldsymbol \Lambda$ adopts the global-local shrinkage structure from \cite{bhattacharya2011sparse}, which encourages column-wise shrinkage for rank selection. By design, this ordered shrinkage prior reduces sensitivity to the choice of the rank of $\boldsymbol \Lambda$, which we label $k$, provided the rank is sufficiently large.  
In the simulation studies, we set $k= 2p$ to be full-rank, although our results were not sensitive to $k < 2p$. We set $a_1 = 2, a_2 = 3, \nu = 3, a_{\sigma} =1,  b_{\sigma} = 0.3$ for the simulation studies, as well as for the North Carolina data analysis.

When the observed data include binary variables, for example, a binary $Y_j$ or a set of $\boldsymbol \gamma_j$ created by expressing multinomial variables using the diagonal orthant representation from Section \ref{cat}, the model also should include a prior distribution on the non-zero $\alpha_j$ for the latent $Z_j$ corresponding to each binary variable.  The $\alpha_j$ for any continuous or ordered categorical $Y_j$ remains set to zero.
%This includes $\alpha_j$ for the indicators created by expressing multinomial variables using the diagonal orthant representation from Section \ref{cat}.}

In the North Carolina data analysis, the Gaussian copula model does not have $2p$ variables. Rather, we use $18$ latent variables corresponding to the study variables used in the modeling, which include latent variables for the binary indicators for the unordered categorical variables per Section \ref{cat}.  We also use one nonresponse indicator for blood-lead measurements.  Thus, we have $19$ latent variables in the copula model, with dimensions adjusted accordingly. We set $k=19$ for the rank of $\boldsymbol \Lambda$. 

\subsection{Gibbs Sampling for the EHQL Gaussian Copula}

Bayesian estimation of the EHQL Gaussian copula involves sampling from the conditional distributions for the model parameters, latent variables, and marginal distributions per Algorithm 2 of the main text.
%with the margin adjustment in the presence of missing data 
%alternates sampling model parameters conditional on the latent variables, and then sampling latent variables corresponding to $\boldsymbol{Y}^{mis}$ given $\boldsymbol{Z}^{obs}$ and model parameters. 
Here, we assume that the auxiliary information set $\mathcal{A}^{*}$ includes  intermediate quantiles.  We present the sampler for study variables that include binary variables, including the diagonal orthant representation of unordered categorical variables. We do not include an $R_j$ for those variables; we do include $R_j$ for continuous and ordinal variables.  As discussed previously, this requires adding a non-zero $\alpha_j$ for each indicator variable with prior distribution as described in Section \eqref{sec:prior}. Modeling unordered categorical variables as nonignorable under the copula model is an area for future research.  

As such, let  $p^{*}$ be the dimension of the combined set of study variables and missingness indicators, with unordered categorical variables augmented with their diagonal orthant representation per Section \ref{cat}. We index the $p^{*}$ variables with $j=1, \dots, p^*$. Let $k$ be the rank of $\boldsymbol \Lambda$ and dimension of latent factors $\boldsymbol \eta$, which we index with $h$. Finally, we index the observations with $i=1, \dots, n$.
%Under the EQL, only one step changes 

We present the steps in the Gibbs sampler for the EHQL. Under the EQL, only two steps change. We describe these changes after presenting the EHQL sampler.

%The algorithm is broken down into five blocks. In each, $\boldsymbol{z_{i}}$ is assumed complete, meaning that components corresponding to missing values in $\boldsymbol{y}_{i}$ ($\boldsymbol{z}^{mis}_{i}$) have been sampled.

\begin{enumerate}
    \item \textbf{Sample the factor model parameters}:
    \begin{itemize}
        \item $\boldsymbol \lambda_{j-} \mid - \sim N_{k}((\boldsymbol{D_{j}}^{-1} + \sigma_{j}^{-2}\boldsymbol{\eta}^{T}\boldsymbol{\eta})^{-1}\boldsymbol{\eta}^{T}\sigma_{j}^{-2}(\boldsymbol{z_{j}} - \alpha_{j}), (\boldsymbol{D_{j}}^{-1} + \sigma_{j}^{-2}\boldsymbol{\eta}^{T}\boldsymbol{\eta})^{-1})$, 
        where $\boldsymbol{D_{j}}^{-1} = diag(\phi_{j1}\xi_{1}, \dots, \phi_{jk}\xi_{k})$, $\boldsymbol{z_{j}} = (z_{1j}, \dots z_{nj})^{T}$, and $\boldsymbol{\eta} = (\boldsymbol \eta_{1}, \dots \boldsymbol \eta_{n})^{T}$, for $j = 1, \dots, p^{*}$.
        \item $\sigma_{j}^{-2} \mid - \sim Gamma(a_{\sigma} + \frac{n}{2}, b_{\sigma} + \frac{1}{2}\sum_{i = 1}^{n}\sum_{h=1}^{k}(z_{ij} - (\alpha_{j} + \lambda_{jh}\eta_{ih}))^{2})$, for $j = 1, \dots, p^{*}$.
        \item $\boldsymbol \eta_{i}\mid - \sim N_{k}(\boldsymbol{\boldsymbol{I_{k}}} + (\boldsymbol{\Lambda^{T}\Sigma^{-1}\Lambda)^{-1}}\boldsymbol{\Lambda^{T}\Sigma^{-1}}(\boldsymbol{z_{i} - \alpha}), (\boldsymbol{I_{k}} + \boldsymbol{\Lambda \Sigma^{-1}\Lambda})^{-1})$, where $\boldsymbol{z_{i}} = (z_{i1}, \dots, z_{ip^{*}})$, for $i = 1, \dots, n$.
        \item $\phi_{jh} \mid - \sim Gamma( \frac{\nu + 1}{2}, \frac{\nu + \xi_{h}\lambda_{jh}^{2}}{2})$, for $j = 1, \dots, p^{*}, \ h = 1, \dots, k$. 
        \item $\delta_{1} \mid - \sim Gamma(a_{1} + \frac{p^{*}k}{2}, 1 + \frac{1}{2}\sum_{h = 1}^{k}\xi_{h}^{(1)}\sum_{j =1}^{p}\phi_{jh}\lambda_{jh}^{2})$, and for $h \geq 2$. \\
        $\delta_{h} \mid - \sim Gamma(a_{1} + \frac{p^{*}(k -h + 1)}{2}, 1 + \frac{1}{2}\sum_{h = 2}^{k}\xi_{h}^{(h)}\sum_{j =1}^{p}\phi_{jl}\lambda_{jh}^{2})$, where $\xi_{h}^{(h)} = \prod_{w = h, w \neq h}^{h} \delta_{w}$, for $h = 1, \dots, k$.
    \end{itemize}
% \textcolor{red}{Joe: should the following be deleted?}
%     $\delta_{1} \mid - \sim \mbox{Gamma}(a_{1} + \frac{pk}{2}, 1 + \frac{1}{2}\sum_{t = 1}^{k}\tau_{t}^{(1)}\sum_{j =1}^{p}\phi_{jt}\lambda_{jt}^{2})$, and for $v \geq 2$ \\
%         $\delta_{t} \mid - \sim \mbox{Gamma}(a_{1} + \frac{p(k -t + 1)}{2}, 1 + \frac{1}{2}\sum_{t = v}^{k}\tau_{t}^{(v)}\sum_{j =1}^{p}\phi_{jt}\lambda_{jt}^{2})$, where $\tau_{t}^{(v)} = \prod_{w = 1, w \neq v}^{t} \delta_{w}$, for $v = 1, \dots, k$
    \item \textbf{Sample all $\alpha_{j}$}\\ 
    For all $j$ corresponding to binary study variables, missingness indicators, or diagonal orthant expanded unordered categorical variables, we sample $\alpha_j$ from 
 %    \textcolor{red}{JOE: should we just say each binary variable? 
 % We don't have any levels}:
    \begin{itemize}
        \item $\alpha_{j}\mid - \sim N((n\sigma_{j}^{-2} + 1)^{-1}\sigma_{j}^{-2}\sum_{i = 1}^{n}\sum_{h =1}^{k}(z_{ij} - \lambda_{jh}\eta_{ih}), (n\sigma_{j}^{-2} + 1)^{-1}).$
    \end{itemize}
    \item \textbf{Sample $\boldsymbol z^{obs}, \boldsymbol z^{mis}, \boldsymbol z_{\boldsymbol r}$} \\
    Given the conditional independence among the components of $\boldsymbol{z}_{i}$ given $\boldsymbol{\eta}_{i}$, we sample components of $\boldsymbol{z}$ corresponding to observed data points column-by-column, consistent with the ordering induced by the EHQL. For components of $\boldsymbol{z}_{i}$ associated with missing values, no ordering is imposed, and only the diagonal orthant restriction for categorical variables is enforced.
    
    \begin{itemize}
        \item \textit{Missing unordered categorical/binary data}: For $z^{mis}_{ijc}$  corresponding to the $c$th level of categorical variable $Y_{j}$ with $c_{j}$ levels, we first calculate the predictive probability that $y_{ij}^{mis} = c$. To so, we compute the categorical probabilities for each level in $Y_{j}$, using the diagonal orthant set restriction of Section \ref{cat}. That is, we calculate the probability that $z^{mis}_{ijc}>0$ while the components of $z^{mis}_{ijc'}<0$ for $c \neq c'$ corresponding to the remaining levels. Explicitly, this is written
        \begin{align}\label{imputecat}
            &P(z^{mis}_{ijc} >0, \{z^{mis}_{ijc'}<0 : c' \neq c, c' = 1, \dots, c_j\}  \mid -) \propto\\& 1- \Phi(0; \sum_{h=1}^{k}\lambda_{ct}\eta_{ih}, \sigma_{j}^{2})\prod_{c' \in \{c_{1},\dots,c_{j}\}, c' \neq c} \Phi(0; \sum_{h=1}^{k}\lambda_{c'h}\eta_{ih}, \sigma_{c'}^{2}). \nonumber  
            %c_{m} \in \{c_{1},\dots,c_{j}\}. \nonumber
        \end{align}
        We sample $y^{mis}_{ij}$ using these probabilities, with the resulting imputation used in the sampling of $z^{mis}_{ijc}$ under the diagonal orthant set restriction in Section \ref{cat}. Let $\mbox{TN}(\mu, \sigma^{2},a,b)$ denote a truncated univariate normal with  mean $ \mu$, variance $\sigma^{2}$, lower truncation $a$, and upper truncation $b$. The re-sampling step for any $z^{mis}_{ijc}$ is given by
    \begin{equation}\label{trunc}
        z^{mis}_{ijc}\sim \begin{cases} \mbox{TN}(\sum_{h=1}^{k}\lambda_{ct}\eta_{ih},\sigma_{j}^{2},0, \infty), & y^{mis}_{ij} = c\\ \mbox{TN}(\sum_{h=1}^{k}\lambda_{ct}\eta_{ih},\sigma_{j}^{2},-\infty,0), & y^{mis}_{ij} \neq c. \end{cases}
    \end{equation}
    If $Y_j$ is binary, the probability is instead given by $P(z^{mis}_{ij} >0\mid -) = 1- \Phi(0; \sum_{h=1}^{k}\lambda_{jh}\eta_{ih}, \sigma_{j}^{2})$, but the re-sampling step~\ref{trunc} remains the same
    \item \textit{Missing numeric data:}
    In this case, $z^{mis}_{ij}$ is sampled from the unrestricted univariate Gaussian, 
    \begin{equation}
        z^{mis}_{ij} \mid - \sim N(\sum_{h=1}^{k}\lambda_{jt}\eta_{ih},\sigma_{j}^{2}).
    \end{equation}
    \item \textit{Observed data:} For each $j$, sample $z^{obs}_{ij}$ from a truncated normal, with lower and upper bounds for each observation specified by the EHQL/diagonal orthant probit restriction: 
    \begin{equation}
    z^{obs}_{ij} \mid - \sim \mbox{TN}(\sum_{h=1}^{k}\lambda_{jt}\eta_{ih},\ell_{ij}, u_{ij}).
    \end{equation}
For ordinal, count, and continuous variables, the truncation limits are 
\begin{eqnarray}
\label{low}\ell_{ij} &=& \max\{\Phi^{-1}(\tau_{ij}^{\ell}), \max(z_{vj}^{obs}: z^{obs}_{vj} \in \mathcal{I}^{q-1}_{j}, v = 1,\dots,n)\}\\ 
\label{up}u_{ij} &=& \min\{\Phi^{-1}(\tau_{ij}^{u}), \min(z_{vj}^{obs}: z^{obs}_{vj} \in \mathcal{I}^{q+1}_{j},  v = 1,\dots,n)\}.
\end{eqnarray}
with $\tau_{ij}^{\ell}, \tau_{ij}^{u}$ defined as in Step 1 of Algorithm 1. For binary study variables,  the upper and lower truncation limits are 
    \begin{equation}\label{trunc2}
        \ell_{ij} = \begin{cases} 0, & y^{obs}_{ij} = 1\\ -\infty, &y^{obs}_{ij} = 0 \end{cases}, \quad    \quad          
        u_{ij} = \begin{cases} \infty, & y^{obs}_{ij} = 1\\ 0, & y^{obs}_{ij} = 0.\end{cases}
    \end{equation}
Similarly, for missingness indicators, the upper and lower truncation limits are
    \begin{equation}\label{trunc2}
        \ell_{ij} = \begin{cases} 0, & r_{ij} = 1\\ -\infty, &r_{ij} = 0 \end{cases}, \quad    \quad          
        u_{ij} = \begin{cases} \infty, & r_{ij}= 1\\ 0, & r_{ij} = 0.\end{cases}
    \end{equation}
Finally, for unordered categorical variables augmented with the diagonal orthant representation of Section \ref{cat}, the upper and lower truncation limits for each component of $\boldsymbol z_{ij}$ are

    \begin{equation}\label{trunc2}
        \ell_{ijc} = \begin{cases} 0, & \gamma_{ijc} = 1\\ -\infty, &\gamma_{ijc} = 0 \end{cases}, \quad    \quad          
        u_{ijc} = \begin{cases} \infty, & \gamma_{ijc}= 1\\ 0, & \gamma_{ijc} = 0.\end{cases}
    \end{equation}
    \end{itemize}

    \item \textbf{Sample} $\tilde{F}_{j}$\\
    For each unique $\{y_{j}^{q}\}_{j=1}^{s_{j}}$ , we first 
    find $Z_{j}(y_{j}^{q})$ as defined in the main text 
    %= \max\{Z_{ij}^{obs}: y_{ij}^{obs} \in \mathcal{I}^{q}_{j}\}$, 
    and compute 
    \begin{equation}\label{margadjsupp}
        \tilde{F}_{j}(y^{q}_{j}) = \Phi_{j}\{Z_{j}(y^{q}_{j})\}.
    \end{equation}
Here, $\Phi_{j}$ is the Gaussian CDF for $Z_{j}$ under the  current draw of copula parameters.  To estimate $\tilde{F}_{j}$ across unobserved values, we fit a monotone interpolating spline to $[\{y_{j}^{q}\}_{q=1}^{s_{j}}\cup \mathcal{A}_{j},\{\tilde{F}_{j}(y_{j}^{q})\}_{q=1}^{s_{j}}\cup \{\tau^{q}_{j}\}_{q=1}^{\ell_j}]$ as described in Section~\ref{MA-EHQL} of the main text, and use this estimate to approximate $\tilde{F}_{j}(x')$ for $x' \notin \{y_{j}^{q} \cup \mathcal{A}_{j}\}$.
\end{enumerate}
The smoothing step in the sampling of $\tilde{F}_{j}$ is needed for multiple imputation, as the transformation $y^{mis}_{ij} = \tilde{F}^{-1}_{j}(z^{mis}_{ij})$ provides realizations across the entire support of $Y_j$.

We now describe how to modify these steps for the EQL. The sampling of latent variables corresponding to observed numeric (ordered discrete and continuous) variables, and specifically the lower and upper bounds in \eqref{low} and \eqref{up}, are now given by $\tau^{\ell} = \max\{ F_{j}^{-1}(\tau^{q}_{j}) \in \mathcal{A}_{j}: y_{ij} > F_{j}^{-1}(\tau^{q}_{j})\}$ and $\tau^{u} = \min\{ F_{j}^{-1}(\tau^{q}_{j}) \in \mathcal{A}_{j}: y_{ij} < F_{j}^{-1}(\tau^{q}_{j})\}$, since we no longer have intermediate quantile points. In addition, the interpolation step for $F_{j}$ smooths between pairs $[\mathcal{A}_{j}, \{\tau^{q}_{j}\}_{q=1}^{\ell_{j}}]$ since we no longer compute \eqref{margadjsupp}.

%\end{document}

\section{Additional Simulation Results} \label{sims}
This section includes supporting results for the some of the statements made in Section 4.1 of the main text.

%\subsection{Efficiency with Sparse Auxiliary Information and Accuracy of the Margin Adjustment}

%#We proceed to provide additional simulation results, akin to Figure \ref{figconsist} in the main text, which complete the analysis.

\subsection{Accuracy with Sparse Auxiliary Information}

\begin{figure}[t]
    \centering
    \includegraphics[width = 0.99\textwidth, keepaspectratio]{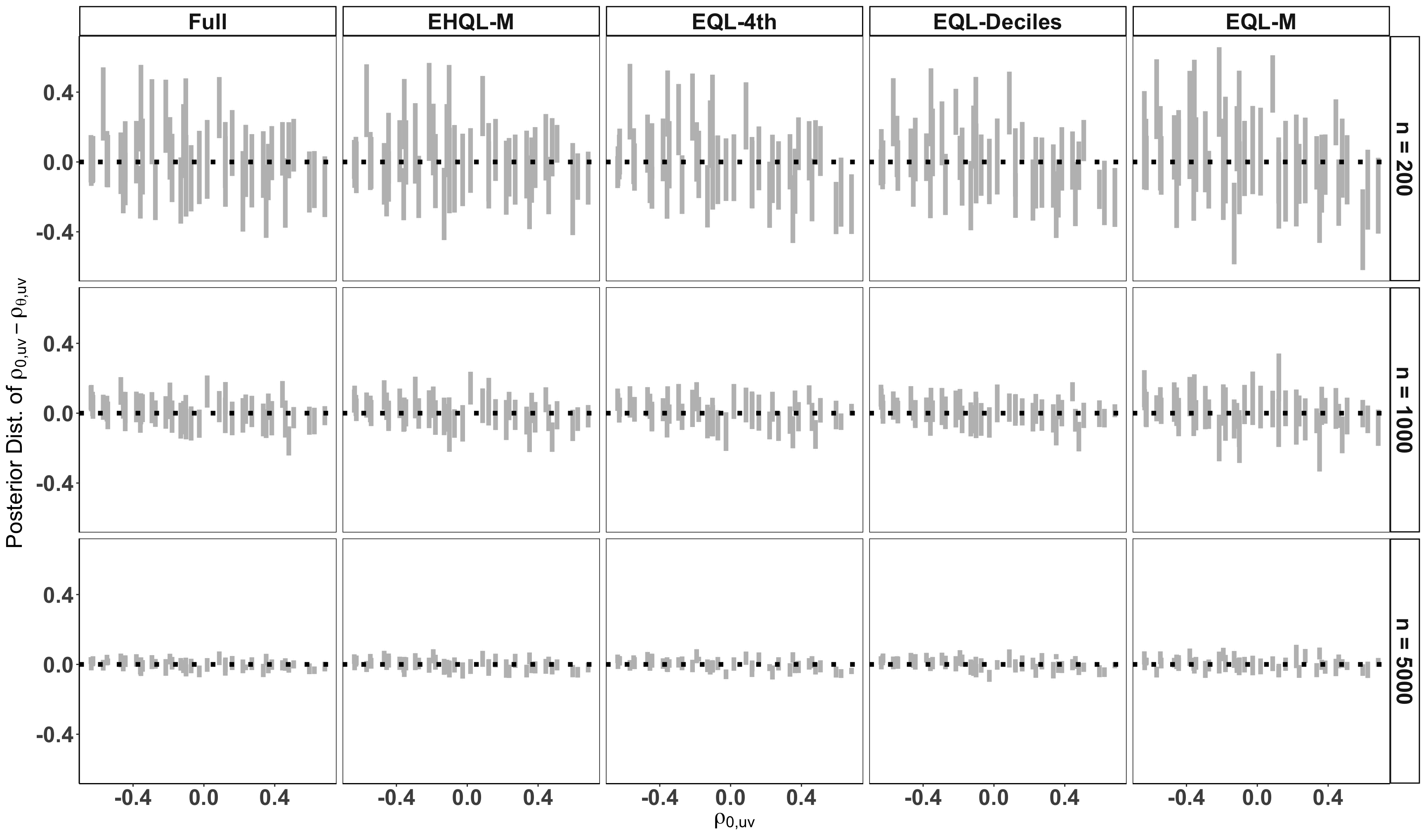}
    \caption{Expanding  Figure \ref{figconsist} in the main text to all levels of auxiliary information considered in the simulation study. Minor gains for smaller sample sizes are observed by incorporating more auxiliary quantiles into the model relative to using just the median. Even with every fourth quantile, the inferences are virtually indistinguishable from the MA + median method.}
    \label{figconsist-allaux}
\end{figure}

Fixing $p = 10$ study variables and the proportion of marginal missingness at 50\%, we now provide  comparisons of the efficiency of posterior inference for the copula correlation by gradually increasing the set of auxiliary quantiles incorporated into the model. In Figure \ref{figconsist-allaux}, we include the same plots as the main text, augmented with inference under the auxiliary sets comprising every fourth quantile (i.e., $\mathcal{A}_{j} = \{0, 0.04, 0.08,\dots,1\}$) and every decile (i.e., $\mathcal{A}_{j} = \{0,0.1,0.2,\dots,1\}$).  We observe small gains when increasing the number of true auxiliary quantiles introduced for each margin. However, as the sample size increases, the gains become practically negligible relative to using  Median or MA + Median.

%\textcolor{black}{JOE: I suggest not including 75\% missingness.  This is just too much to be realistic. So, we can delete this paragraph and corresponding figure.}
%In Figure \ref{figconsist-allaux_75}, we increase the proportion of missingness to $75\%$, holding $p = 10$. \textcolor{blue}{double check: 10 study variables or 5?} There is evidence that the efficiency of posterior inference is affected by the number of auxiliary quantiles introduced into the EQL, as the Median case is generally inferior to denser auxiliary settings. However, incorporating the margin adjustment in addition to the median largely eliminates these deficiencies.

%\begin{figure}[t]
%    \centering
%    \includegraphics[width = 0.8\textwidth, keepaspectratio]{Images/Consist_75.png}
 %   \caption{The same comparisons as Figure \ref{figconsist-allaux}, but now increasing the marginal proportion of missingness to 75\%. In this setting, we do see moderate gains in inference by including more auxiliary quantiles into the model, but the MA + median is competitive with the dense auxiliary information models. }
 %   \label{figconsist-allaux_75}
%\end{figure}

\begin{figure}[t]
    \centering
    \includegraphics[width = 0.99\textwidth, keepaspectratio]{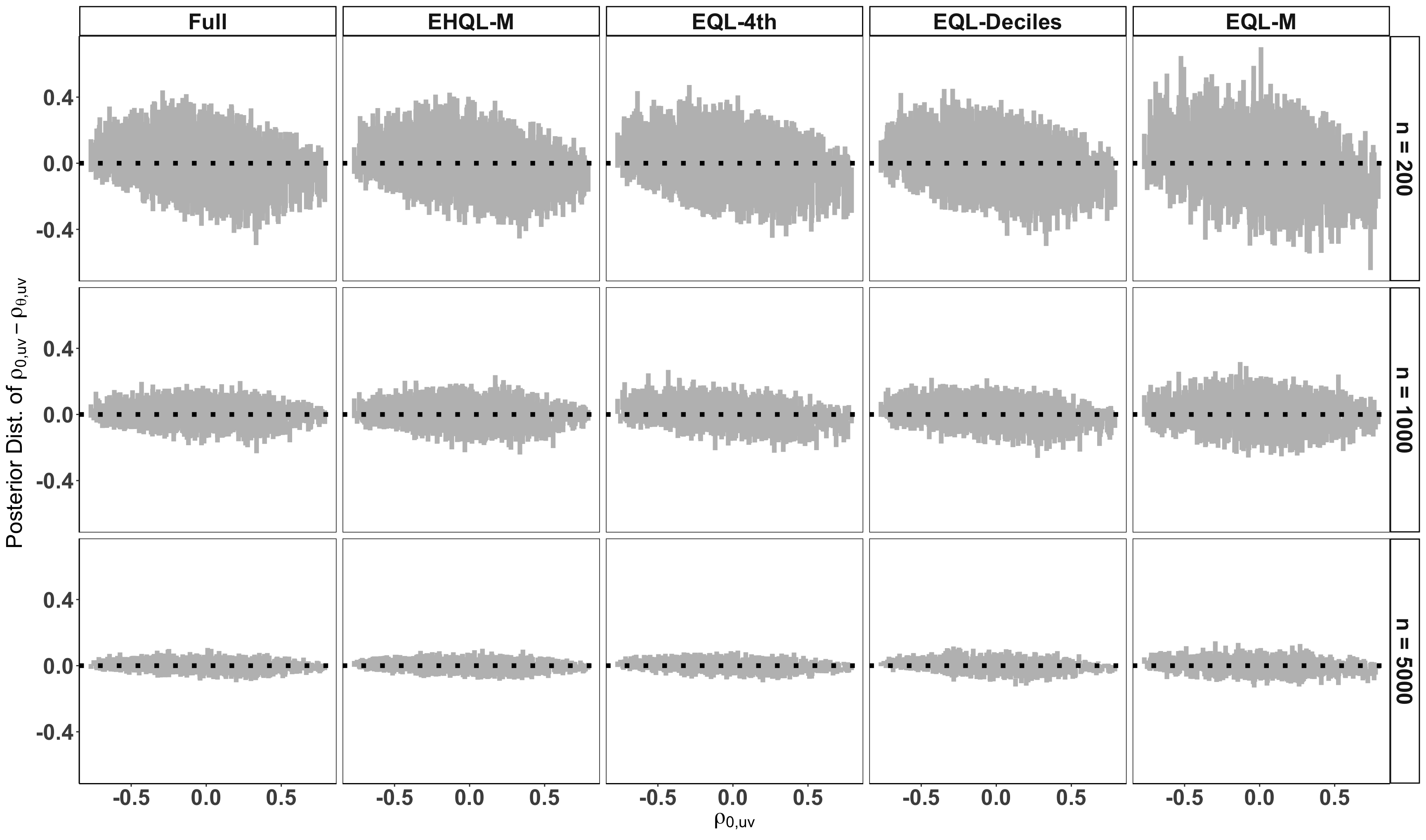}
    \caption{Results of the simulation using $p = 20$ study variables. As in the main paper, the contraction of the posterior is moderately impacted by the level of auxiliary information introduced in the model, particularly for smaller sample sizes. This is evidenced by the interval widths for EQL-M, which are wider than those for the other auxiliary specifications.  }
    \label{figconsist-p20}
\end{figure}

We also check whether the results are sensitive to the dimension of the study variables. Figure \ref{figconsist-p20}  increases the dimension of the study variables to $p = 20$,  holding the missingness at 50\%. The results show similar patterns as Figure \ref{figconsist} in the main text, owing to the scalability of the factor model used to model the latent variables.  We do not include the RL of \citep{hoff2007extending} in these comparisons because the computation is infeasible with our setup when $p = 20$ and $n > 1000$. For $p = 5$, the RL copula sampler took nearly 4 hours to complete for $n = 5000$. By contrast, with $n = 5000$ and $p = 20$ study variables, the EQL/EHQL completed 10000 iterations in around 15 minutes on average using a 2023 Macbook Pro. The computation could be sped up with parallel computing for the sampling of $\boldsymbol z$, since the columns are conditionally independent under the factor model. We leave this to future research.

\begin{figure}[t]
    \centering
    \includegraphics[width = 0.99\textwidth, keepaspectratio]{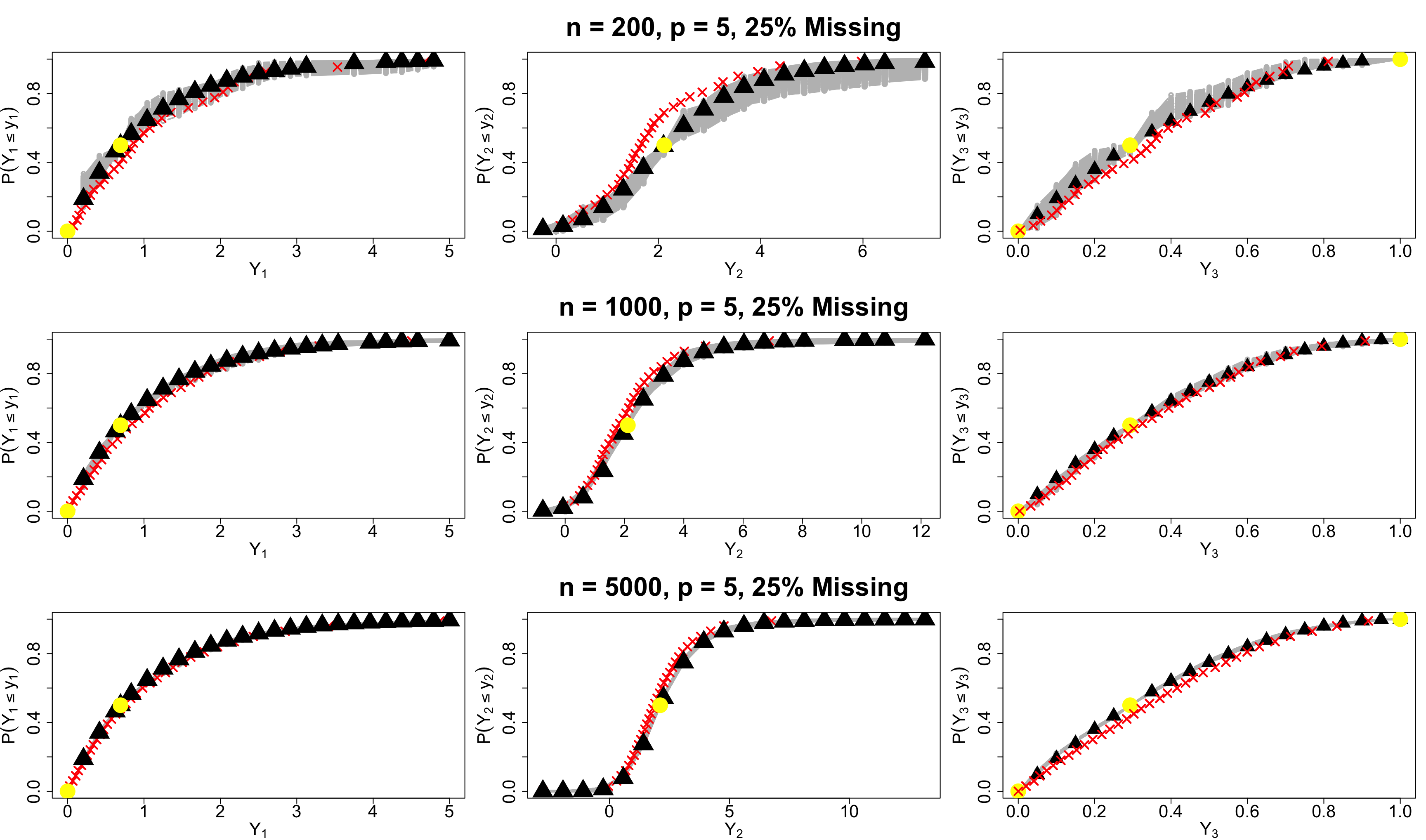}
    \caption{The same comparisons as Figure \ref{figconsistMA} in the main text, but now with 25\% marginal missingness. Since there is only 25\% missingness, the ECDF is less biased. Again, the interpolation strategy at intermediate quantile points enables accurate estimation of each marginal, which holds for each study variable. }
    \label{figconsistMA-mis25}
\end{figure}

Finally, we complete the information in  Figure \ref{figconsistMA} in the main text by lowering the missingness to 25\% and plotting posterior samples of the interpolated marginals obtained using Algorithm 2 in the main text. As shown in Figure \ref{figconsistMA-mis25}, the interpolation accurately captures notable features of each distribution function, with minor gains under less severe missingness, particularly for smaller sample sizes. These findings are not sensitive to the type of $F_j$ or $p$.

\subsection{Simulation of Repeated Sampling Performance}

\begin{figure}[t]
     \centering
     \includegraphics[width = .49\textwidth, keepaspectratio]{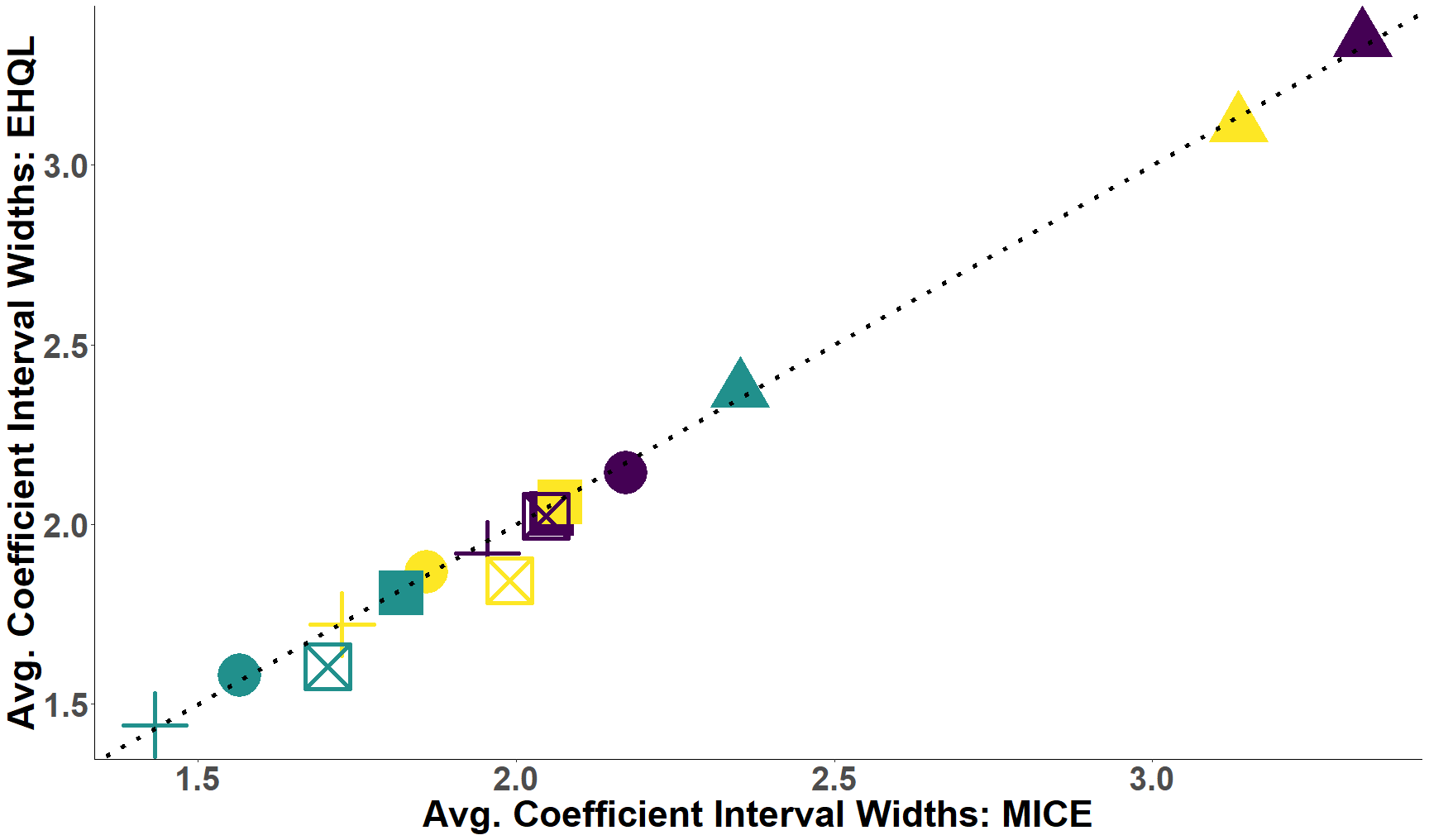}
     \caption{Average interval widths of multiple imputation confidence intervals for the EHQL copula and MICE. Results are presented for the auxiliary specification presented in the main text (i.e., the median for \vtt{NDI} is incorporated into the copula model). The gains in empirical coverage rates under the proposed approach are not simply due to wide confidence intervals}
     \label{intwidths}
     \end{figure}

In the main text, we summarize the improvement of the EHQL copula over MICE in empirical coverage rates under the repeated sampling experiment. We mention that the average interval widths are similar, and so the gains are not merely due to wider uncertainty under the proposed approach. We support this claim in Figure \ref{intwidths}, which plots the average confidence interval widths for the quantile regression coefficients across the experiment. The results for the other settings considered are consistent.

\begin{figure}[t]
     \centering
     \includegraphics[width = .49\textwidth, keepaspectratio]{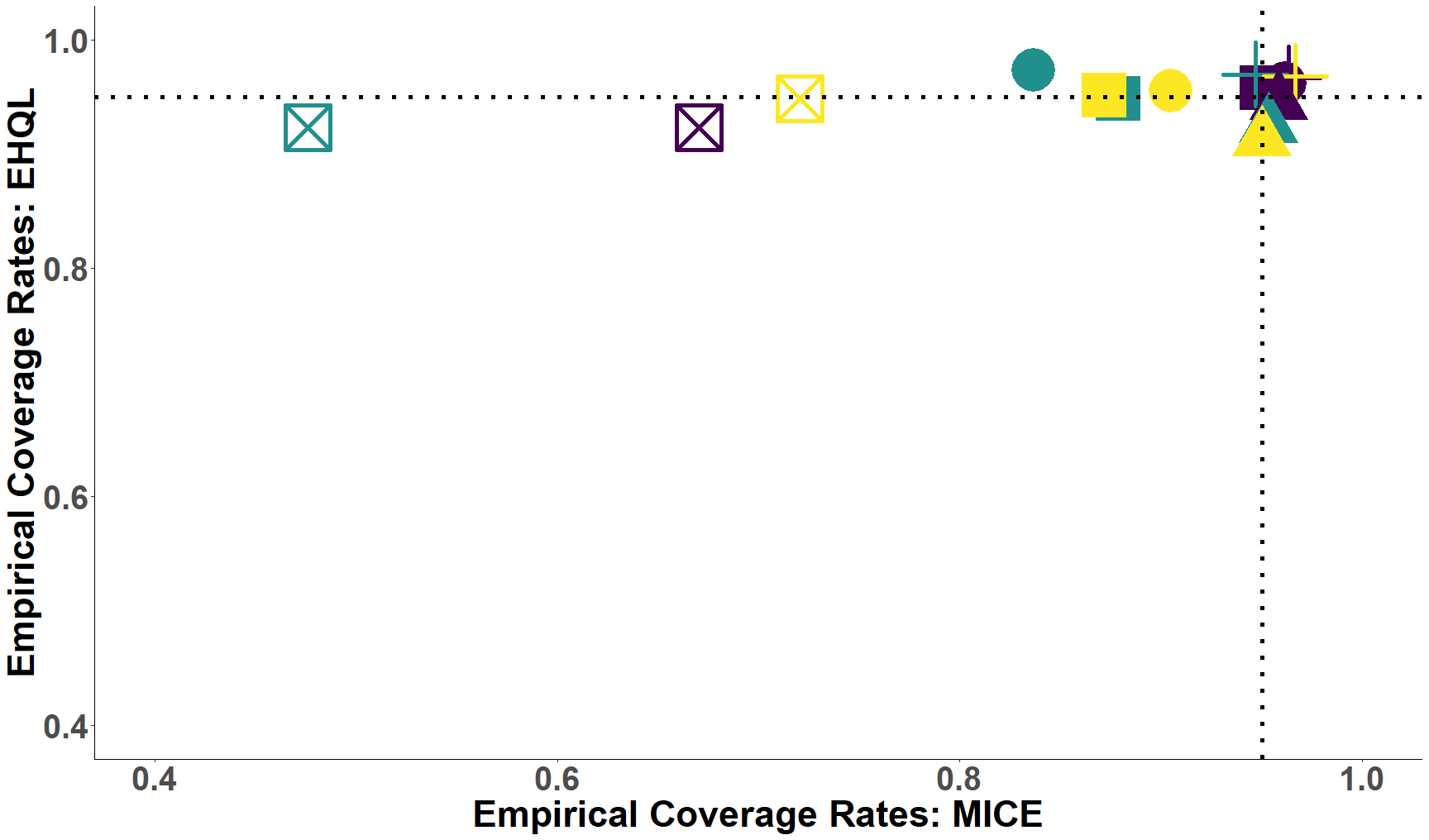}
            \includegraphics[width = .49\textwidth,keepaspectratio]{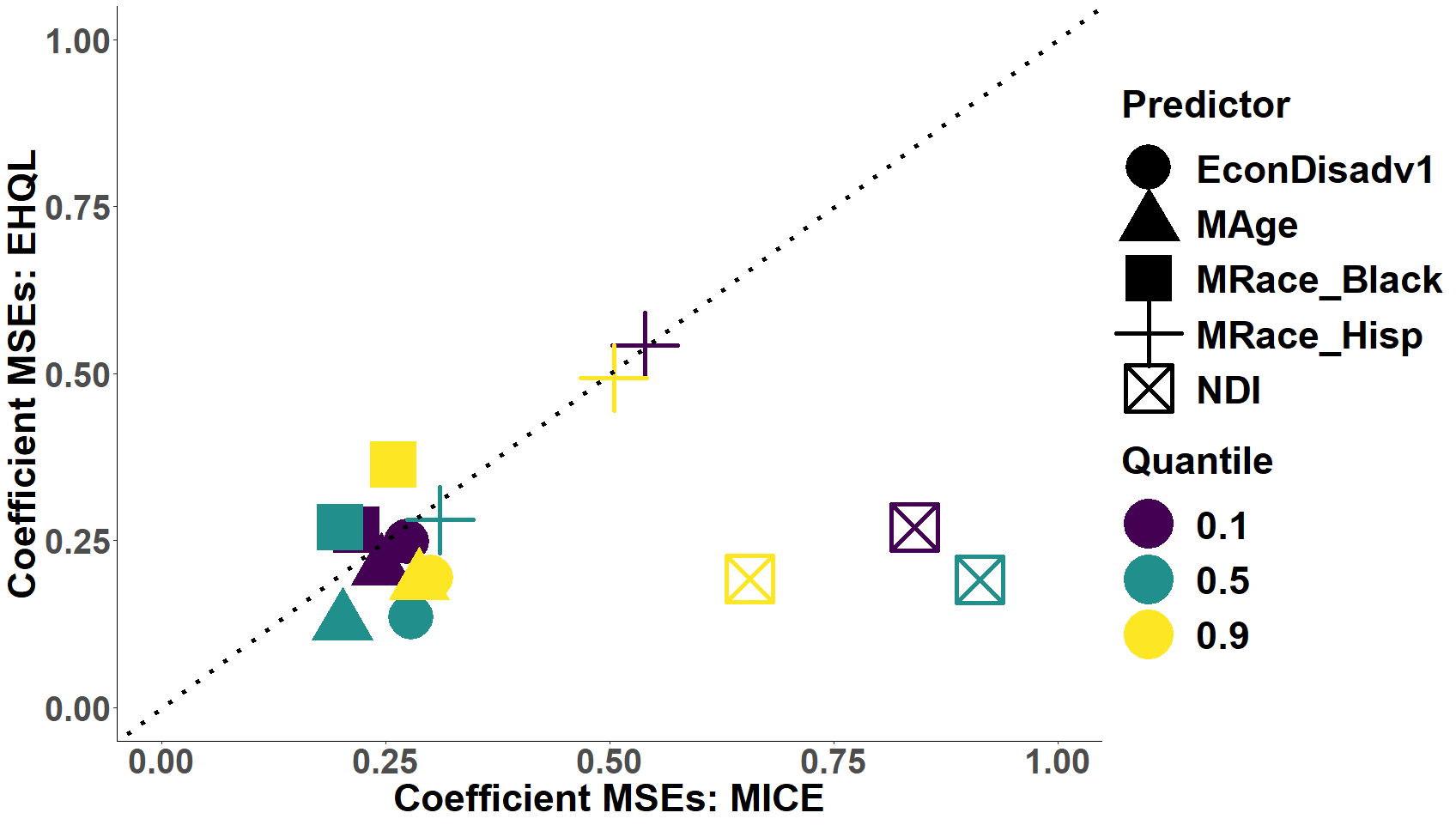}
        \caption{Empirical coverage rates ( left) and average mean squared error (MSE) of multiple imputation point estimates for the EHQL copula and MICE imputations in the repeated simulation study. Instead of specifying the median for \vtt{NDI}, we use the 75th population quantile (in addition to the upper and lower bounds) as the auxiliary information set. Both approaches perform similarly for coefficients other than \vtt{NDI}, which is subject to nonignorable missingness. For this coefficient, the EHQL provides lower bias and higher coverage rates.}
     % \caption{Imputation under the proposed approach (EHQL Copula) compared to MICE. Across quantiles, the inference is similar between the two methods besides the \vtt{NDI_Birth} variable, which is subject to nonignorable missingness. Clearly, the copula model is capable of capturing the dependence in the data, while incorporation of auxiliary quantiles helps to identify the extrapolation distribution and improve imputation.  }
     \label{impinferencesim}
 \end{figure}

In  Section 4.2 of the main text, we specified the median as the additional auxiliary quantile for \vtt{NDI} to the lower and upper bounds.  Here, we repeat the simulation in Section 4.2 of the main text using the auxiliary 75th quantile for \vtt{NDI} in addition to lower and upper bounds. As shown in Figure \ref{impinferencesim}, the results are consistent with what is presented in the main text. The EHQL copula outperforms MICE on multiple imputation inferences for all quantile regression coefficients corresponding to \vtt{NDI}.

\begin{figure}[t]
     \centering
     \includegraphics[width = .49\textwidth, keepaspectratio]{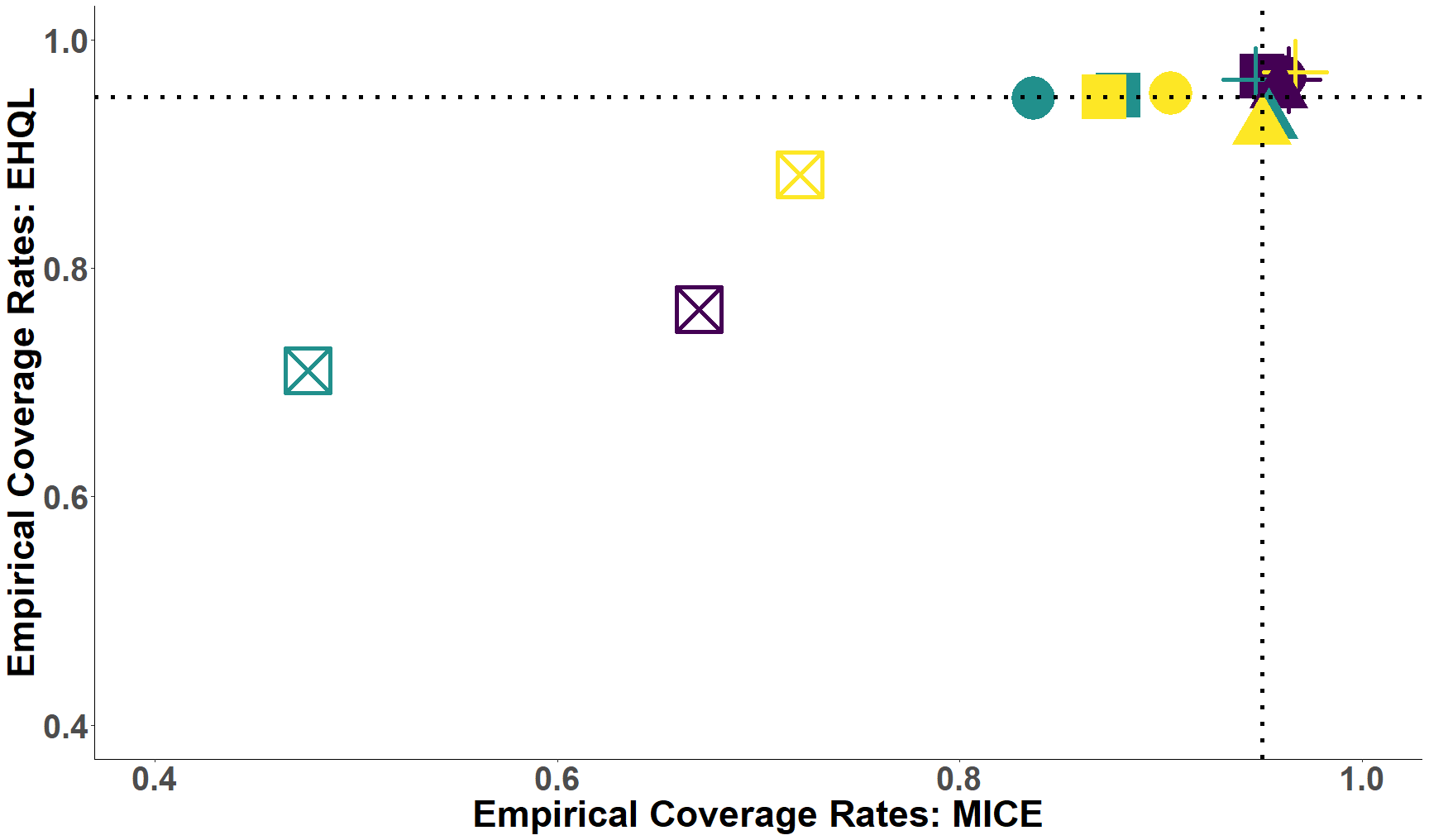}
            \includegraphics[width = .49\textwidth,keepaspectratio]{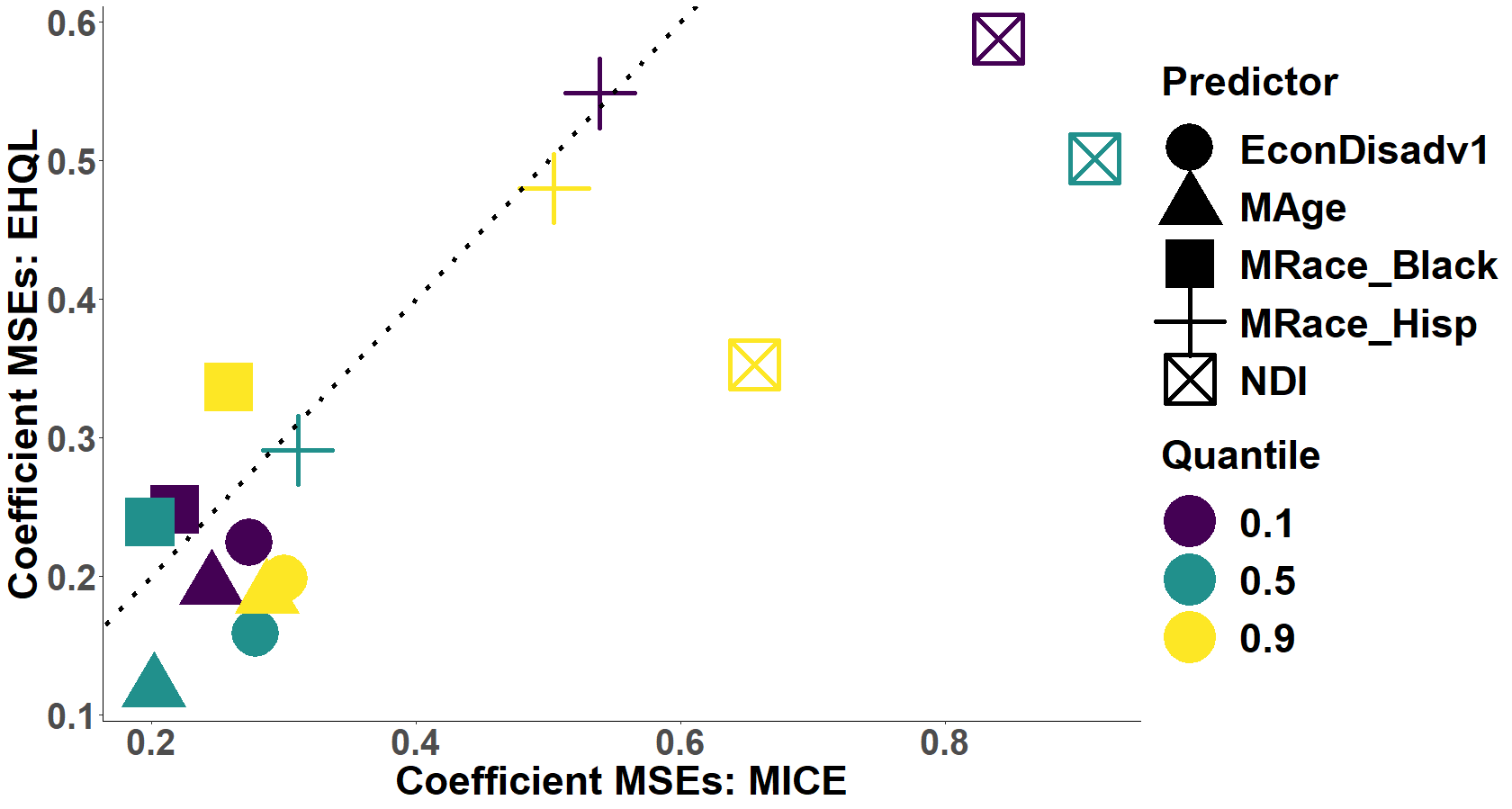}
            \includegraphics[width = .49\textwidth, keepaspectratio]{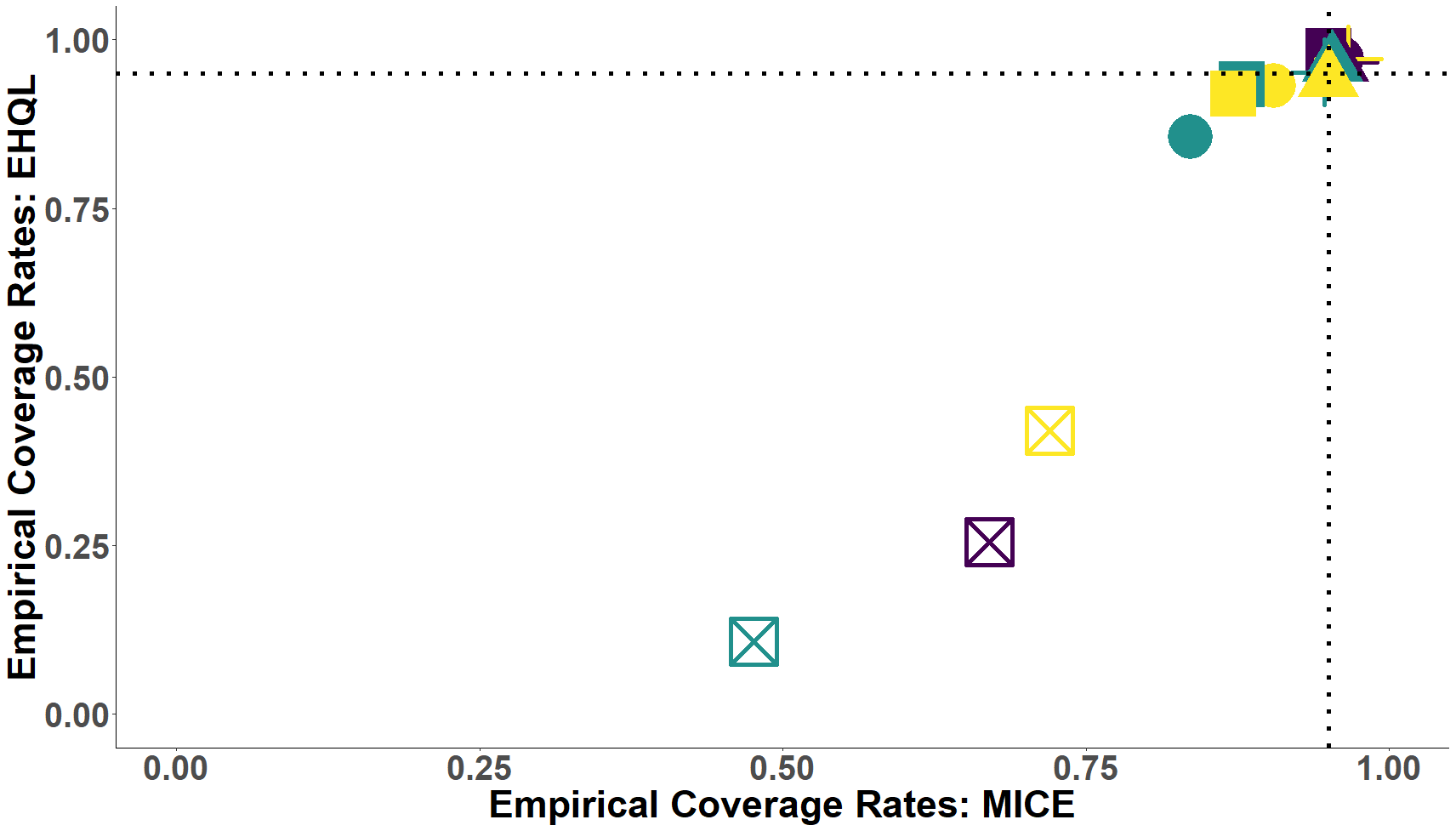}
            \includegraphics[width = .49\textwidth,keepaspectratio]{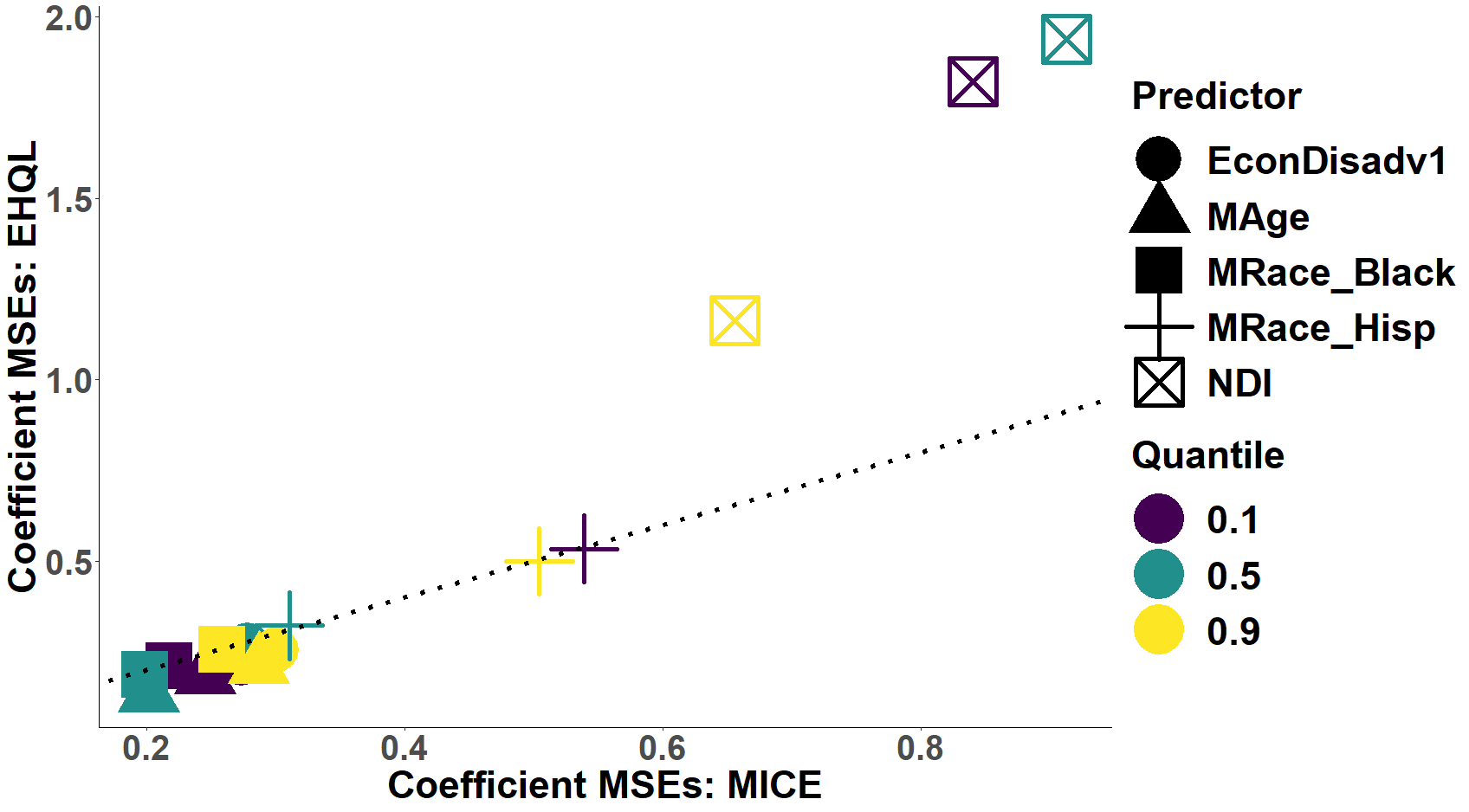}
        \caption{Empirical coverage rates (left column) and average mean squared error of multiple imputation point estimates (right column) for the EHQL copula and MICE imputations in the repeated simulation study. Here, we misspecify the auxiliary median for \vtt{NDI} by $\epsilon = 0.5$ (top row) and $\epsilon = 1$ (bottom row). For $\epsilon = 0.5$, the EHQL still performs better than MICE in inference for the quantile regression coefficients of \vtt{NDI}. However, with extreme misspecification, the EHQL copula does not offer reliable inferences under nonignorable missing data. }
     % \caption{Imputation under the proposed approach (EHQL Copula) compared to MICE. Across quantiles, the inference is similar between the two methods besides the \vtt{NDI_Birth} variable, which is subject to nonignorable missingness. Clearly, the copula model is capable of capturing the dependence in the data, while incorporation of auxiliary quantiles helps to identify the extrapolation distribution and improve imputation.  }
     \label{badmargin}
 \end{figure}

We also include results from misspecifying the auxiliary quantile introduced into the model. To do so, we set $\mathcal{A}_{\vtt{NDI}} = \{F_{}^{-1}(0), F_{j}^{-1}(0.5) + \epsilon, F_{j}^{-1}(1)\}, \ \epsilon \in \{0.5,1\}$. That is, we increasingly perturb the true median for NDI. Figure \ref{badmargin} displays the results.  As expected, the performance of the EHQL copula model deteriorates as the auxiliary quantiles are increasingly biased. However, the performance still improves upon MICE for $\epsilon = 0.5$.

% \begin{figure}[t]
%     \centering
%     \includegraphics[width = .65\textwidth, keepaspectratio]{Images/Margcomp_75.png}
%     \caption{Marginal distribution of \vtt{NDI} in one set of completed datasets for EHQL and MICE. The EHQL imputations capture the bi-modality and skewness of the population marginal, whereas the MICE imputations  match the distribution in the observed sample.}
%     \label{NDImarg_75}
% \end{figure}

\section{Analysis of North Carolina Data: Additional Results}\label{NCadditional}
In this section, we provide details on the auxiliary information from the CDC used in $\mathcal{A}$, posterior predictive checks of the Gaussian copula model on the observed data, and additional quantile regression inferences not presented in the main text.

\subsection{Determining Auxiliary Information from CDC Estimates}

\begin{figure}[t]%
    \centering
    \includegraphics[width = \textwidth, keepaspectratio]{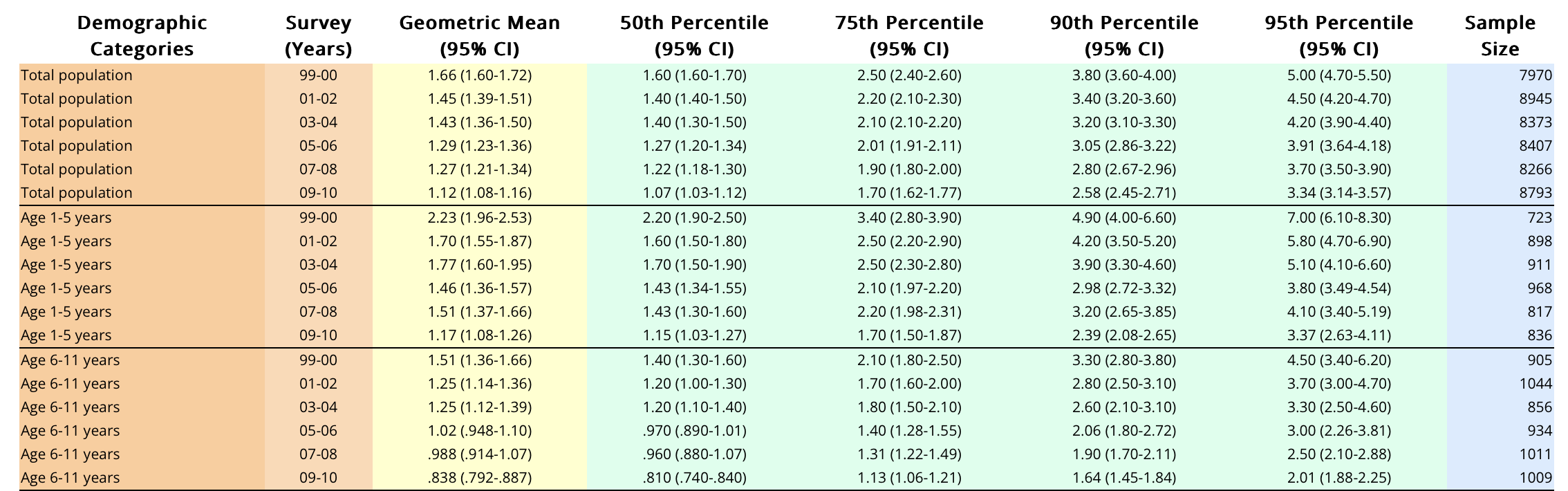}
    \caption{CDC published estimates for select quantiles of lead exposure.}\label{tab:cdc}
\end{figure}

As mentioned in Section 5 of the main text, we leverage published quantile estimates of lead exposure from the CDC \citep{CDCwebsite} to specify auxiliary quantiles on blood-lead levels.  Table \ref{tab:cdc} displays the population-level estimates provided by the CDC. Among the children in the North Carolina data who were measured for lead, 90\% were measured between 2005 and 2009. Since they were born between 2003 and 2005, we base the auxiliary information $\mathcal{A}$ on the published estimates between 2005 and 2010 for the 1-5 years old group. 

We incorporate a single auxiliary quantile (besides the lower and upper bounds) for two reasons. First, the CDC estimates are  continuous, whereas the North Carolina data are binned into intervals of exposure. For instance, observing $\vtt{Blood_lead} = 1$ implies that an individual has a measure between $(0,1]$ $\mu$g/ml. Second, the reported quantiles are national estimates, whereas we use data from North Carolina. 
%As such, we anchor the marginal of \vtt{Blood_lead} to a single quantile formulated from the CDC published estimates, and leverage the margin adjustment to infer the intermediate points of this distribution. 

\subsection{Posterior Predictive Checks}
We verify that the EHQL copula model generates predictive samples that resemble plausible realizations of the observed values in the North Carolina lead exposure data. Specifically, we examine the joint predictive distribution of \vtt{Blood_lead} and the other study variables given that \vtt{Blood_lead} is observed. 

Given a posterior draw of $\{\boldsymbol C_{\theta}, \boldsymbol \alpha, \{\tilde{F}_{j}\}\}$,
we can generate replicate values of $\boldsymbol y^{obs}$ in three steps. Let $j=1$ when $Y_j$ is \vtt{Blood_lead}.  
%For each posterior sample of model parameters, we first  simulate ten 
%i.i.d.\  
We generate the predictive latent variables corresponding to lead exposure being observed, i.e., $\tilde{z}_{r_1}
%\textcolor{black}{What is the 10 for?  Is it 10 per each draw of the parameters?  Or are we using ten parameter draws? Good to clarify explicitly.} 
\sim N(\alpha_{r_1}, 1)\mathbbm{1}_{(-\infty,0)}$. 
%for $i = 1,\dots, 10$. 
We then generate predictive latent variables for the study variables conditional on the sampled $\tilde{z}_{r_1}$ using the distribution for $\tilde{\boldsymbol z}_{\boldsymbol y} \mid \tilde{z}_{r_1}$. This is conditionally multivariate Gaussian with the mean varying as a function of the realized $\tilde{z}_{r_1}$ and covariance $\boldsymbol \Sigma^{*}$ derived from $\boldsymbol C_{\boldsymbol \theta}$. For each numeric $Y_j$, for a hypothetical individual $i$ we obtain $\tilde{y}_{ij} = \tilde{F}^{-1}_{j}(\Phi(\tilde{z}_{ij}))$. For binary variables, sampling $\tilde{z}_{ij} >0$ indicates for our hypothetical individual that $\tilde{y}_{ij}=1$.  For $Y_{j}$ that is unordered categorical, we generate  the vector $\tilde{\boldsymbol{z}}_{ij}$, and set $\tilde{Y}_{ij} = c \iff \tilde{z}_{ijc} = \max\{\boldsymbol{\tilde{z}}_{ij}\}$. 
We repeat this process ten times per parameter draw, resulting in 50,000 posterior predictive samples. 

We note here that our posterior predictive sampling of the categorical variables does not enforce the diagonal orthant restriction of Section \ref{cat} as part of the latent variable generation; rather, for any $y^{mis}_{ij}$, we generate its value by selecting the $c$ corresponding to the maximum value in $\{z_{ijc}: c = 1, \dots, c_j\}$. This is done for computational convenience.  To enforce it, 
%Empirically, we maintain satisfactory performance (see Figure \ref{catpred}) using the max among $\boldsymbol z_{ij}$ and alleviate a substantial computational burden. Specifically, 
we would have to decompose the joint distribution of the latent study variables into a sequence of conditionals for each $\boldsymbol z_{ij}$ corresponding to an unordered categorical variable. Then, we would compute the mean and variance, derive the categorical probabilities, and sample a multinomial. The resulting categorical membership would specify the diagonal orthant set restriction for those variables, after which predictive latent variables could be sampled.

To compute $\boldsymbol \Sigma^{*}$, we first partition each posterior sample of 
\begin{equation}
\boldsymbol C_{\boldsymbol \theta} = \begin{bmatrix}
    \boldsymbol C_{\boldsymbol y} & \boldsymbol C_{\boldsymbol{y}r}\\ \boldsymbol{C}_{r\boldsymbol{y}} & \boldsymbol C_{r}
\end{bmatrix}.
\end{equation}
Then, 
$\boldsymbol \Sigma^{*} = \boldsymbol C_{\boldsymbol y}- \boldsymbol C_{\boldsymbol yr}\boldsymbol C_r^{-1}\boldsymbol C_{r\boldsymbol{y}}$.

\begin{figure}[t]
    \centering
        \includegraphics[width = 0.48\textwidth,keepaspectratio]{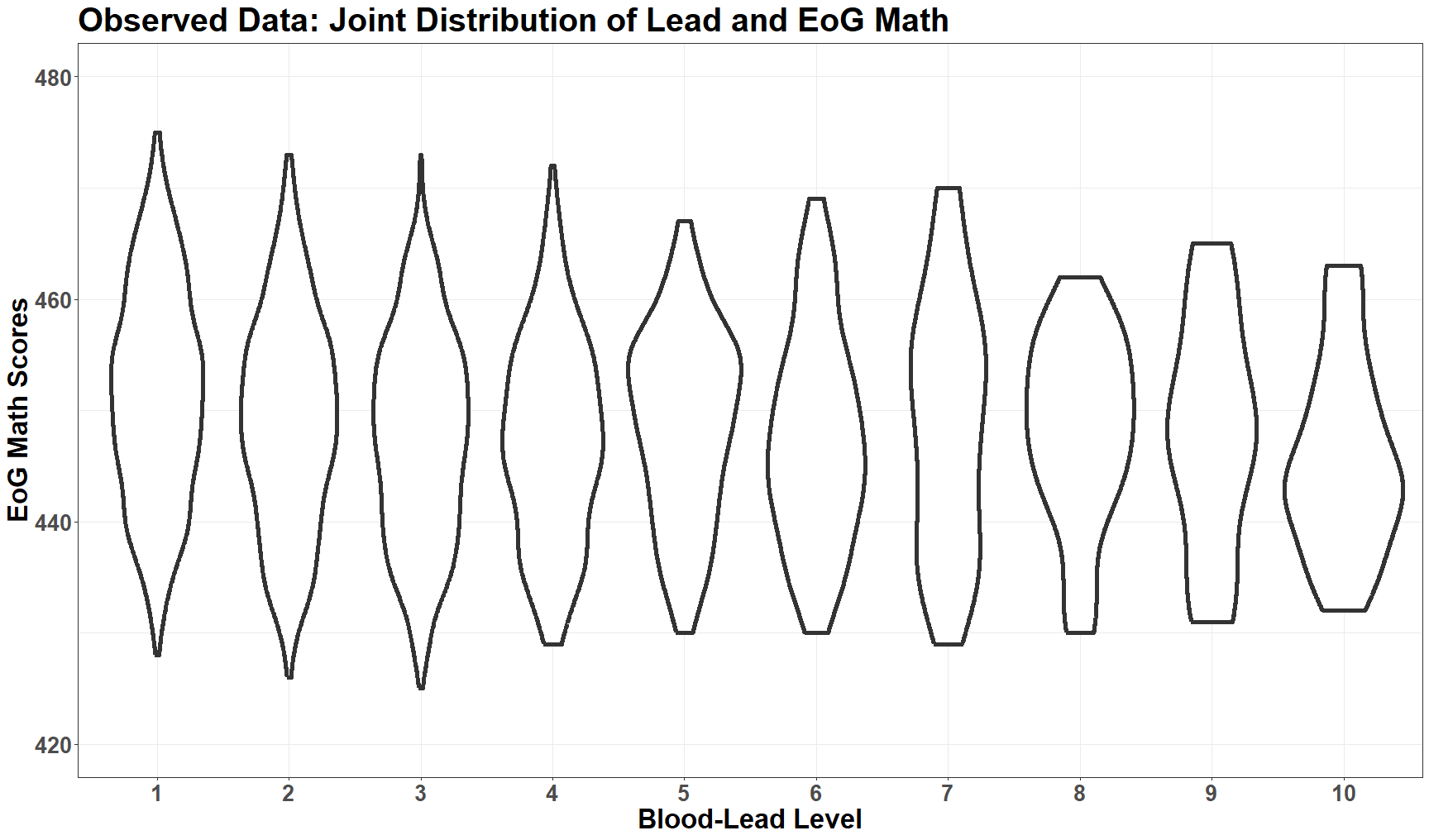}
                    \includegraphics[width = 0.49\textwidth,keepaspectratio]{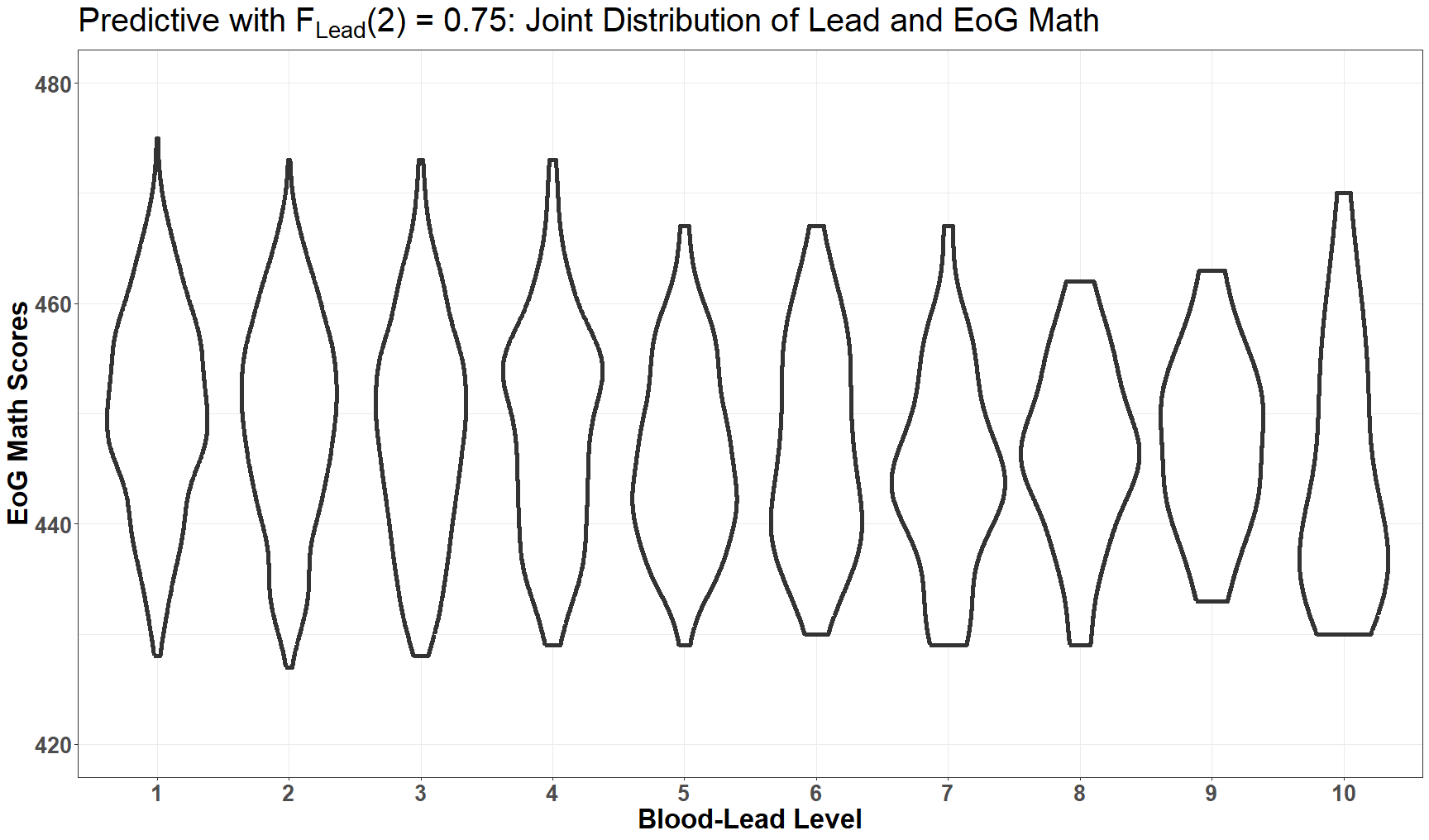}

            \includegraphics[width = 0.49\textwidth,keepaspectratio]{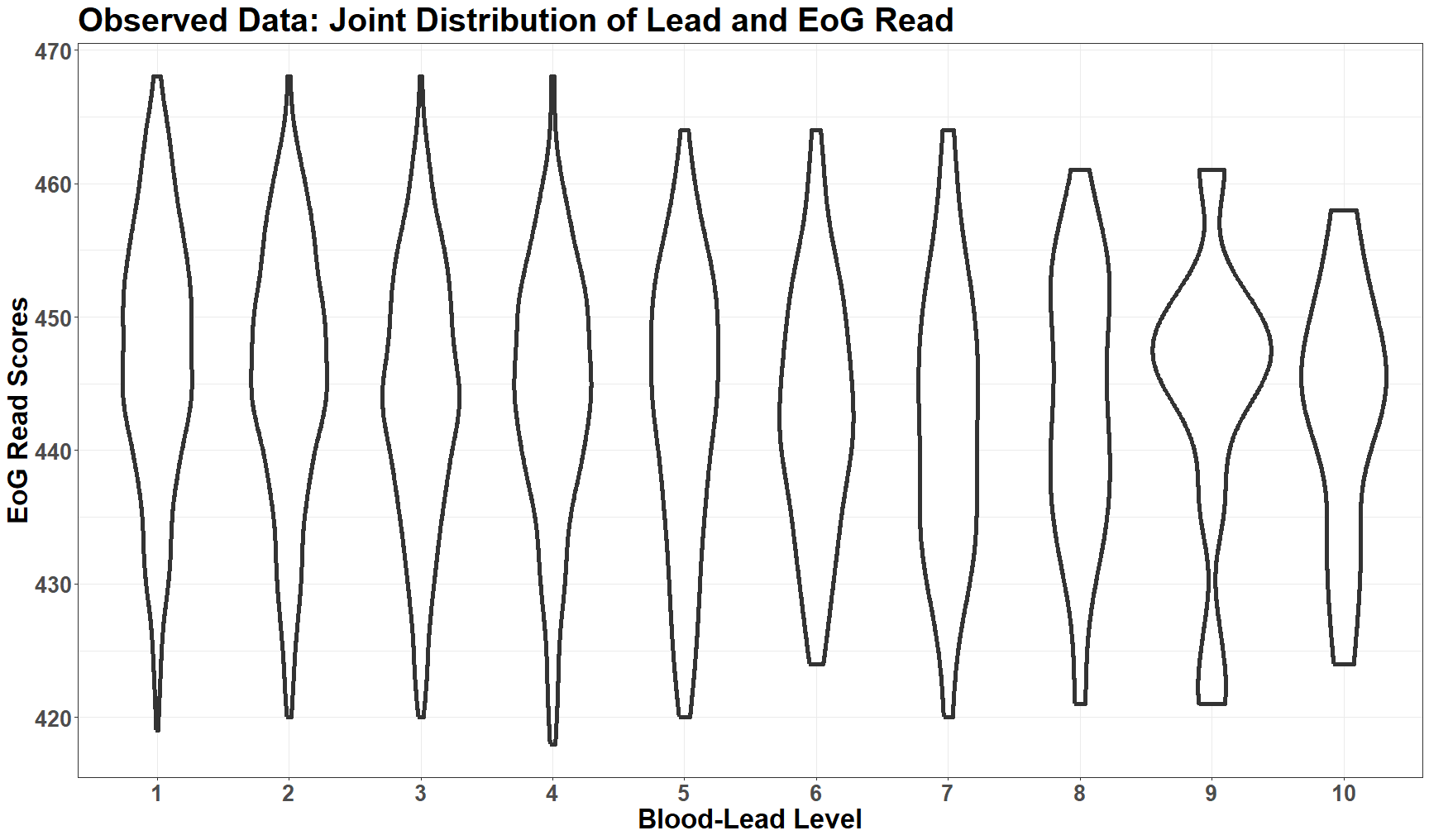}
                \includegraphics[width = 0.49\textwidth,keepaspectratio]{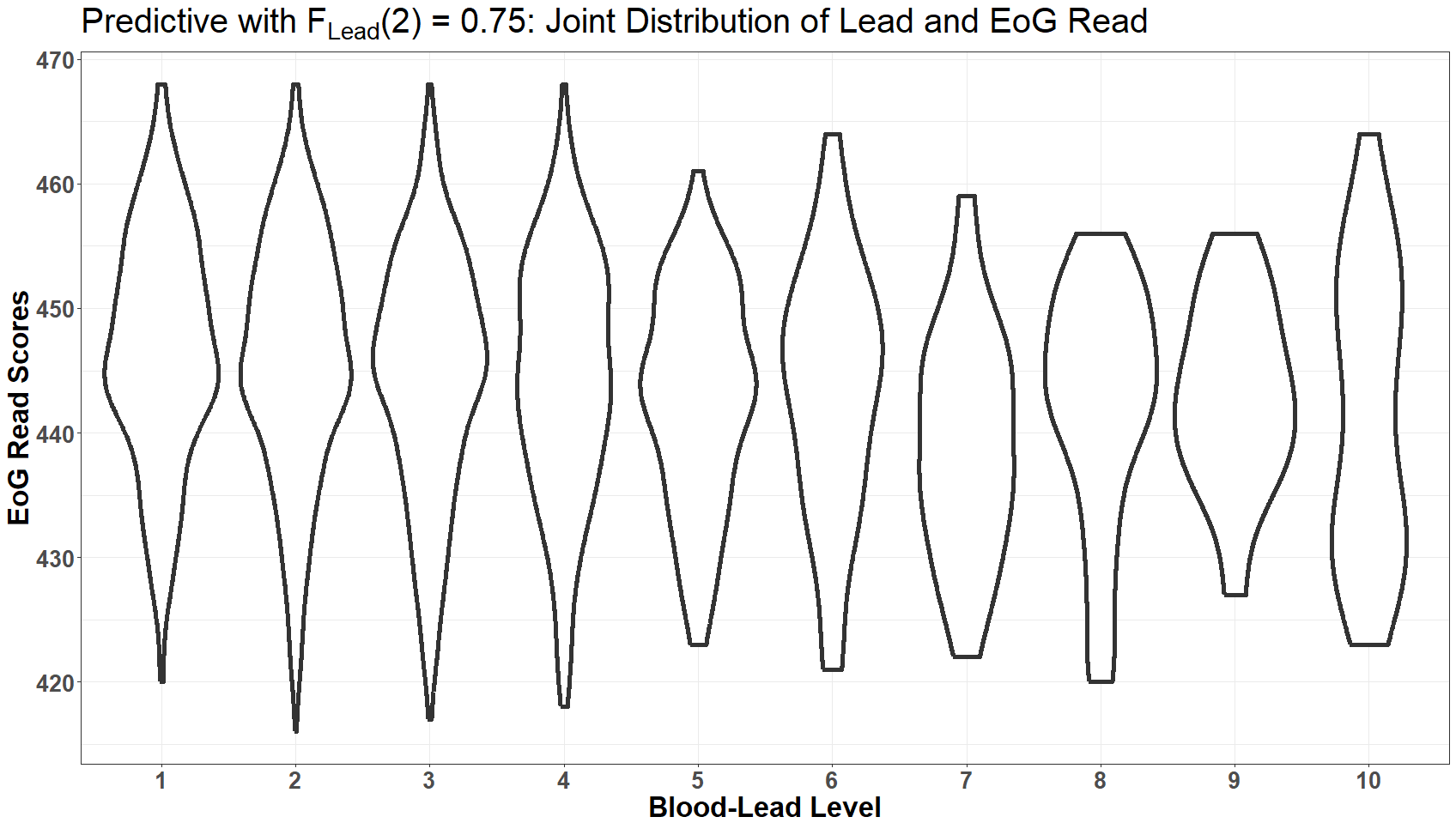}
    
    \caption{Observed (left column) vs posterior predictive (right column) distributions of math/reading EoG test scores and blood-lead levels. The EHQL copula model adequately captures the bivariate associations between the EoG test scores and lead exposure.}
    \label{postpredcheck}
\end{figure}

%We which are then compared to the observed data.
Figure \ref{postpredcheck} displays the joint predictive distribution of reading/math scores and blood-lead levels.  Table \ref{predsampmean} compares the posterior predictive means of $\vtt{Blood_lead}$ ($\overline{Y}^{pred}_{\vtt{blood_lead}}$) for each level of the binary/categorical variables in the model with their counterparts in the observed data. Table \ref{predcor} compares the correlations between $\tilde{Y}_{\vtt{Blood_lead}}$ of the remaining numeric variables  in the predictive samples and observed data. We see some evidence of lack of fit for average lead levels for Hispanic children, who represent less than 10\% of the observations.  Nonetheless, overall, the copula model adequately captures the associations in the observed data. All results are for the EHQL copula with auxiliary \vtt{Blood_lead} quantile $F(2) = 0.75$.

\begin{table}[h]
\centering 
\begin{tabular}{ccccccc}
& \vtt{mRace} & \vtt{mEduc} &\vtt{EconDisadv} & \vtt{Male} & \vtt{NotMarried} &\vtt{Smoker} \\ &(Wh./Bl./Hisp.) & (No H.S., H.S., Coll.) & (No/Yes) & (No/Yes) & (No/Yes) & (No/Yes)\\ \hline
$\overline{Y}^{pred}_{\vtt{Blood_lead}}$ & 2.68/3.28/2.92 &3.20/2.98/2.38 & 2.49/3.13 & 2.82/2.92 &2.58/3.25& 2.80/3.23\\
$\overline{Y}^{obs}_{\vtt{Blood_lead}}$&2.65/3.16/2.59 & 3.07/2.84/2.31 & 2.46/3.02 & 2.74/2.85 &2.58/3.05&2.75/3.07
\end{tabular}\label{catpred}
\caption{Posterior predictive means  of \vtt{Blood_lead} compared to observed means by each level of the categorical and binary study variables. The EHQL copula model accurately captures these multivariate associations in the observed data.}
\label{predsampmean}
\end{table}

\begin{table}[h]
\centering 
\begin{tabular}{cccccc}
 Y{j} & \vtt{mAge} & \vtt{NDI} & \vtt{RI} & \vtt{BWTpct} & \vtt{Gestation}\\\hline
 $\text{cor}(Y_{j}, \vtt{Blood_lead})^{pred}$ & -0.12 &0.22 & 0.11 & -0.03 & 0.03 \\
  $\text{cor}(Y_{j}, \vtt{Blood_lead})^{obs}$&-0.14 &0.15 &0.12 & -0.07 & -0.001
\end{tabular}
\caption{Predictive correlations compared to observed correlations between \vtt{Blood_lead} and the other numeric study variables, excluding reading and math scores. The EHQL copula approximately captures these pairwise associations in the observed data.}
\label{predcor}
\end{table}

% \subsection{Quantile Regression Inference}
\subsection{Complete Results for the 10th Quantile Regression}
In the main paper, we present results for the 10th quantile regression coefficients for selected exposure variables using EoG math scores as the response. In this section we provide the multiple imputation coefficient estimates and 95\%  confidence intervals for the remaining coefficients, and also include parallel results using EoG reading scores as the response. Here, $F_{\vtt{Blood_lead}}(2) = 0.75$. 

\begin{figure}[t]
    \centering
    \includegraphics[width = .8\textwidth, keepaspectratio]{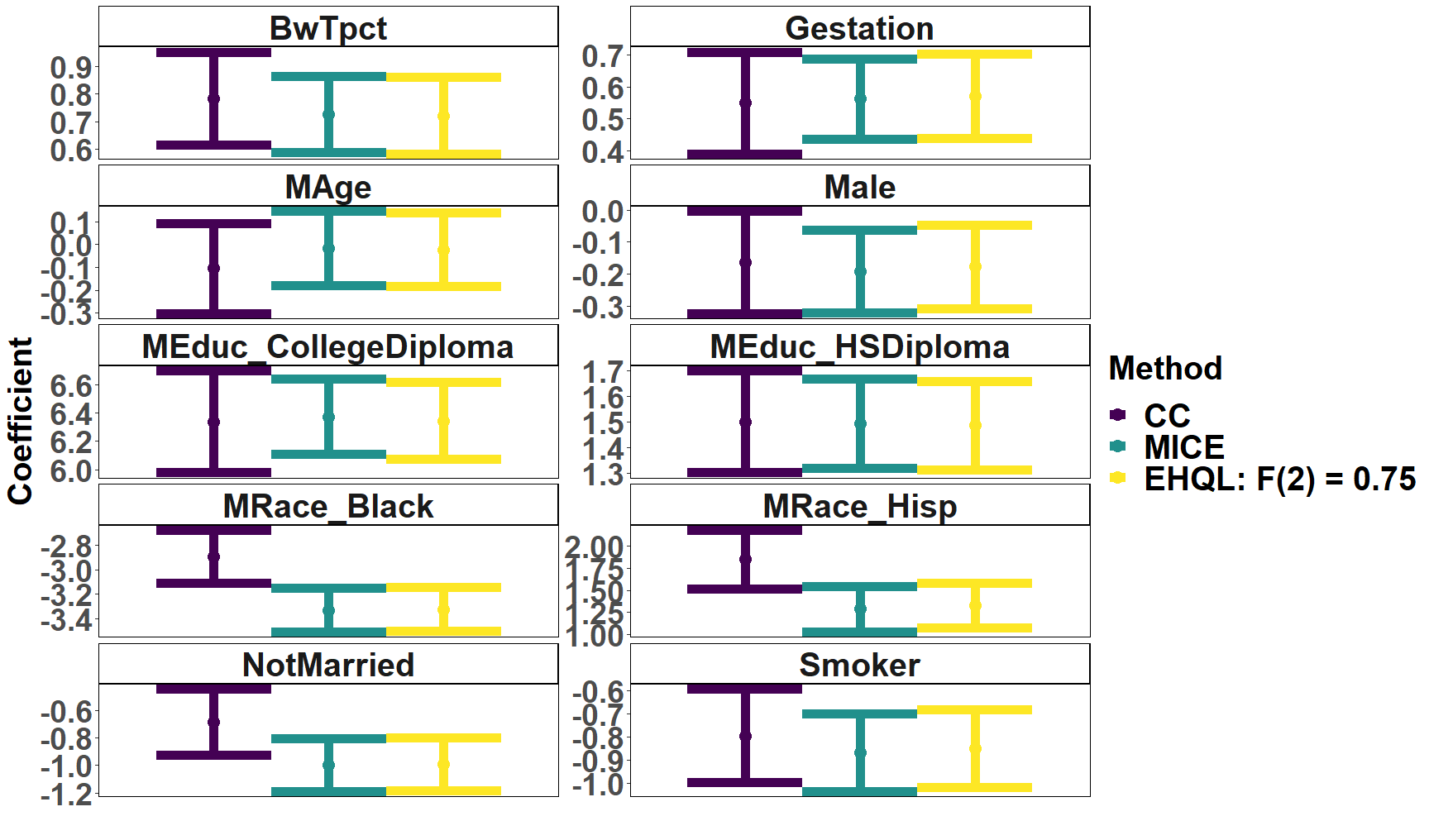}
    \caption{Multiple imputation inferences for the remaining coefficients under the 10th quantile regression. Here, $F_{\vtt{Blood_lead}}(2) = 0.75$ and the response variable is EoG math scores. For both the EHQL and MICE, multiple imputation offers more precision and shifts certain coefficients relative to CC analysis. The inferences between the two methods are virtually identical, owing the fact that the variables are almost completely observed and we treat their missingness as MCAR. }
    \label{q10_math_rest}
\end{figure}

Figure \ref{q10_math_rest} provides the multiple imputation inferences for the remaining coefficients with EoG math scores as the response variable. Both the EHQL and MICE provide inferences that are notably different than complete cases (CC) analysis. Unsurprisingly, the inferences from both approaches are very similar, as these variables are almost completely observed and both the EHQL and MICE treat their missingness as MCAR.

\begin{figure}[h]
    \centering
    \includegraphics[width = .8\textwidth, keepaspectratio]{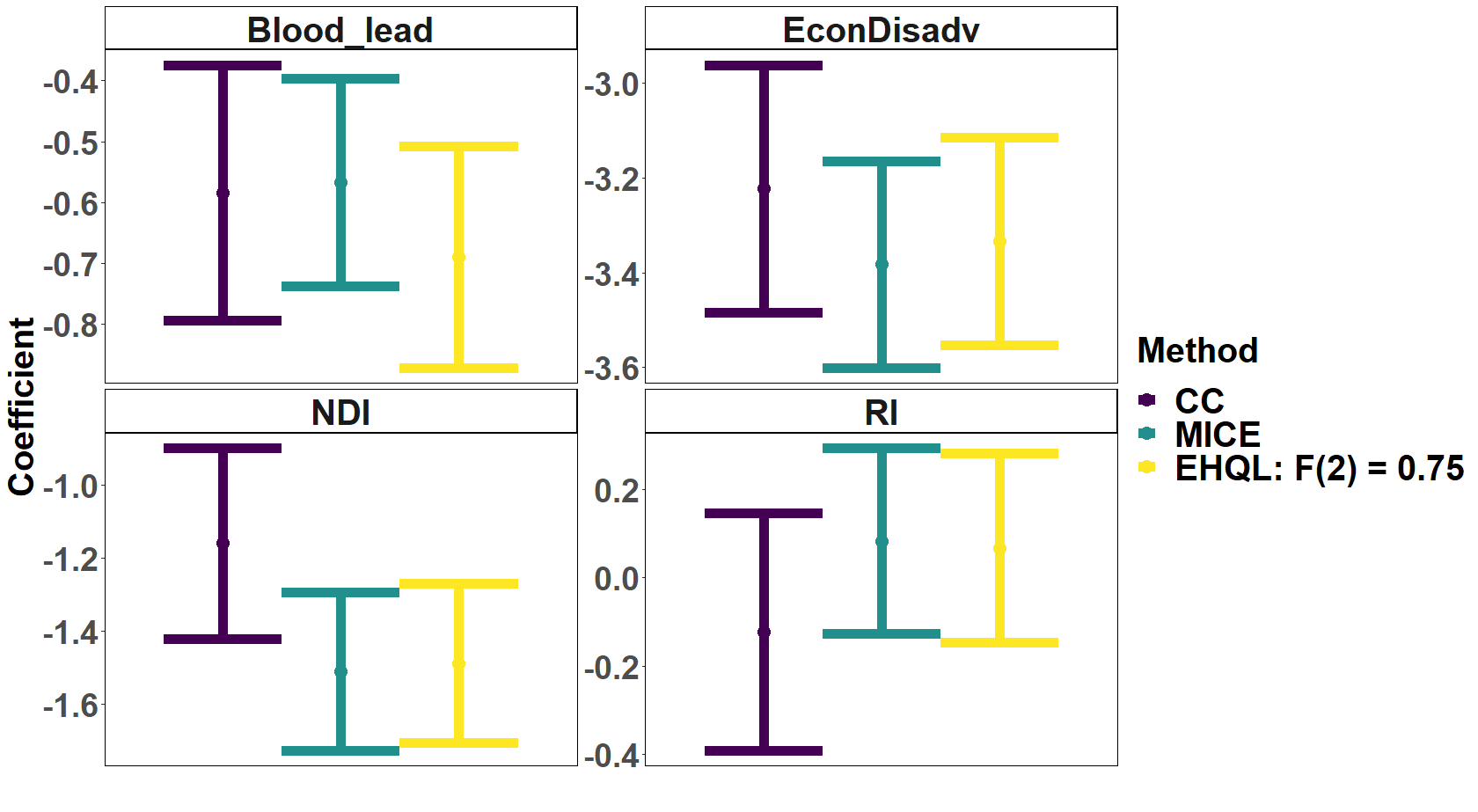}
    \caption{Multiple imputation inference for the remaining coefficients under the 10th quantile regression. Here, $F_{\vtt{Blood_lead}}(2) = 0.75$ and the response variable is EoG reading scores. Similar to the model with EoG math scores as the response, the EHQL estimates a more strong, adverse association between lead exposure and reading scores than both CC and MICE.}
    \label{q10_read_select}
\end{figure}

Figure \ref{q10_read_select} presents multiple imputation inferences for the 10th quantile regression for the four selected exposure variables presented in  the main text with EoG reading test scores as the response.  Accounting for the nonignorable missingness in lead exposure measurements still results in stronger, more adverse associations for reading test scores and lead exposure, although the shift is not as pronounced as for math test scores.  As with the quantile regressions using math test scores as the dependent variable, we see little practical difference between the EHQL and MICE results for the three coefficients that do not correspond to \vtt{Blood_lead}.

\begin{figure}[h]
    \centering
    \includegraphics[width = .8\textwidth, keepaspectratio]{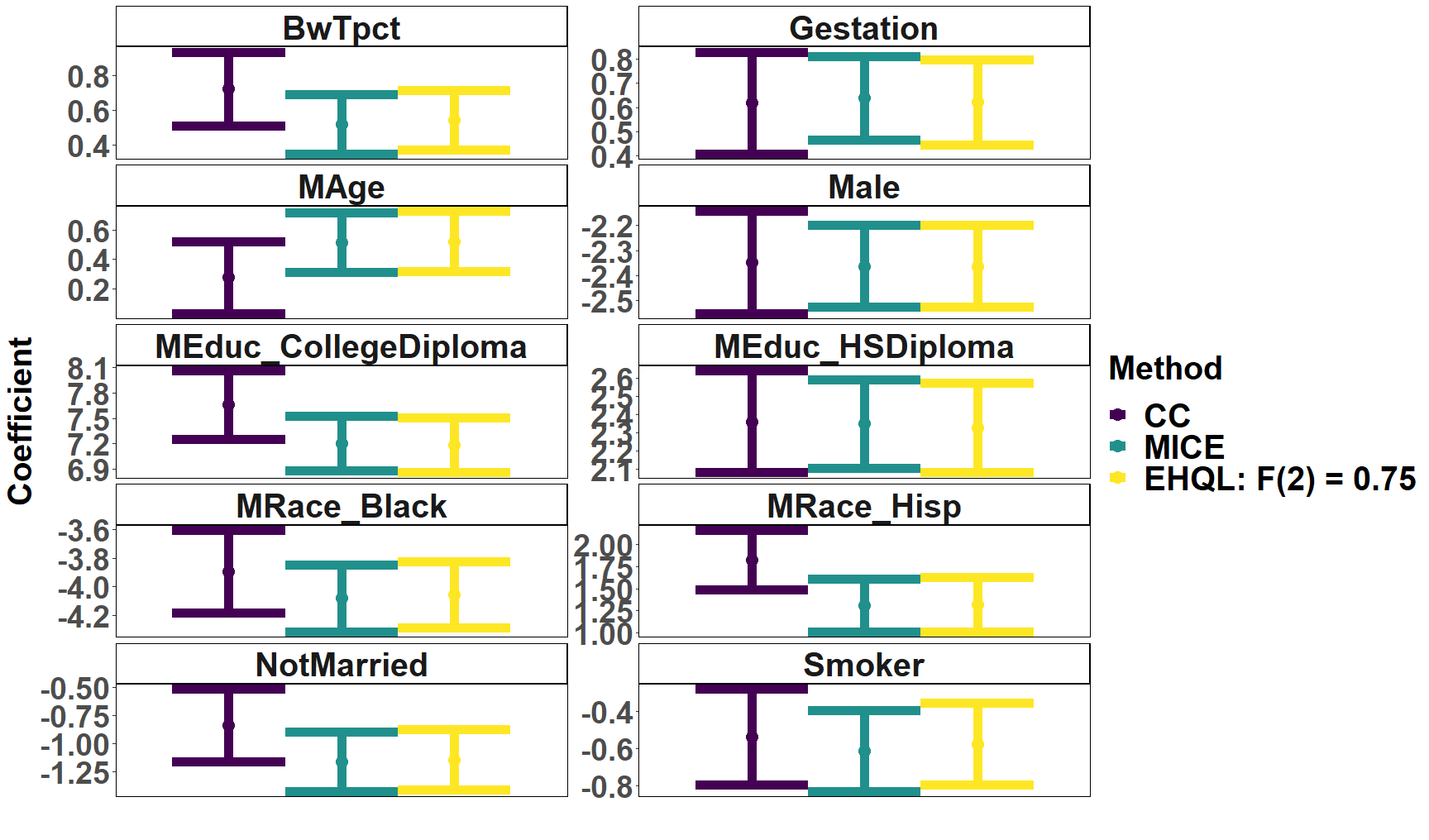}
    \caption{Multiple imputation inference for the remaining coefficients under the 10th quantile regression. Here, $F_{\vtt{Blood_lead}}(2) = 0.75$ and the response variable is EoG reading scores. For both the EHQL and MICE, imputation offers more precision and shifts certain coefficients relative to CC analysis. The inferences between the two methods are virtually identical, owing the fact that the variables are almost completely observed and we treat their missingness as MCAR.}
    \label{q10_read_rest}
\end{figure}

 Finally, Figure \ref{q10_read_rest} displays multiple imputation inferences for the remaining coefficients with EoG reading test scores as the response.  The multiple imputation inferences result in greater precision.  The EHQL and MICE offer very similar inferences---which are sometimes quite different than the CC inferences---due to the dearth of missing values for these other variables.

\subsection{Results from Additional Quantile Regressions and Alternative Auxiliary Quantile Specifications}
In this section, we demonstrate that the quantile regression inferences are not sensitive to the alternative auxiliary specifications considered ($F_{\vtt{Blood_lead}}(2) = 0.80, F_{\vtt{Blood_lead}}(2) = 0.70$). 
%The next pages  present the multiple imputation inferences for the quantile regressions from Section \ref{realdat} of the main text. 
We combine these with results for the quantile regressions at the 50th and 90th quantiles, when either EoG math or reading test scores is the response variable. To economize on the number of figures, we present the results for all quantile regressions and auxiliary specifications in the same plots. We again distinguish between the four selected exposure variables and the remaining covariates.  

\begin{figure}[h]
    \centering
    \includegraphics[width = \textwidth, keepaspectratio]{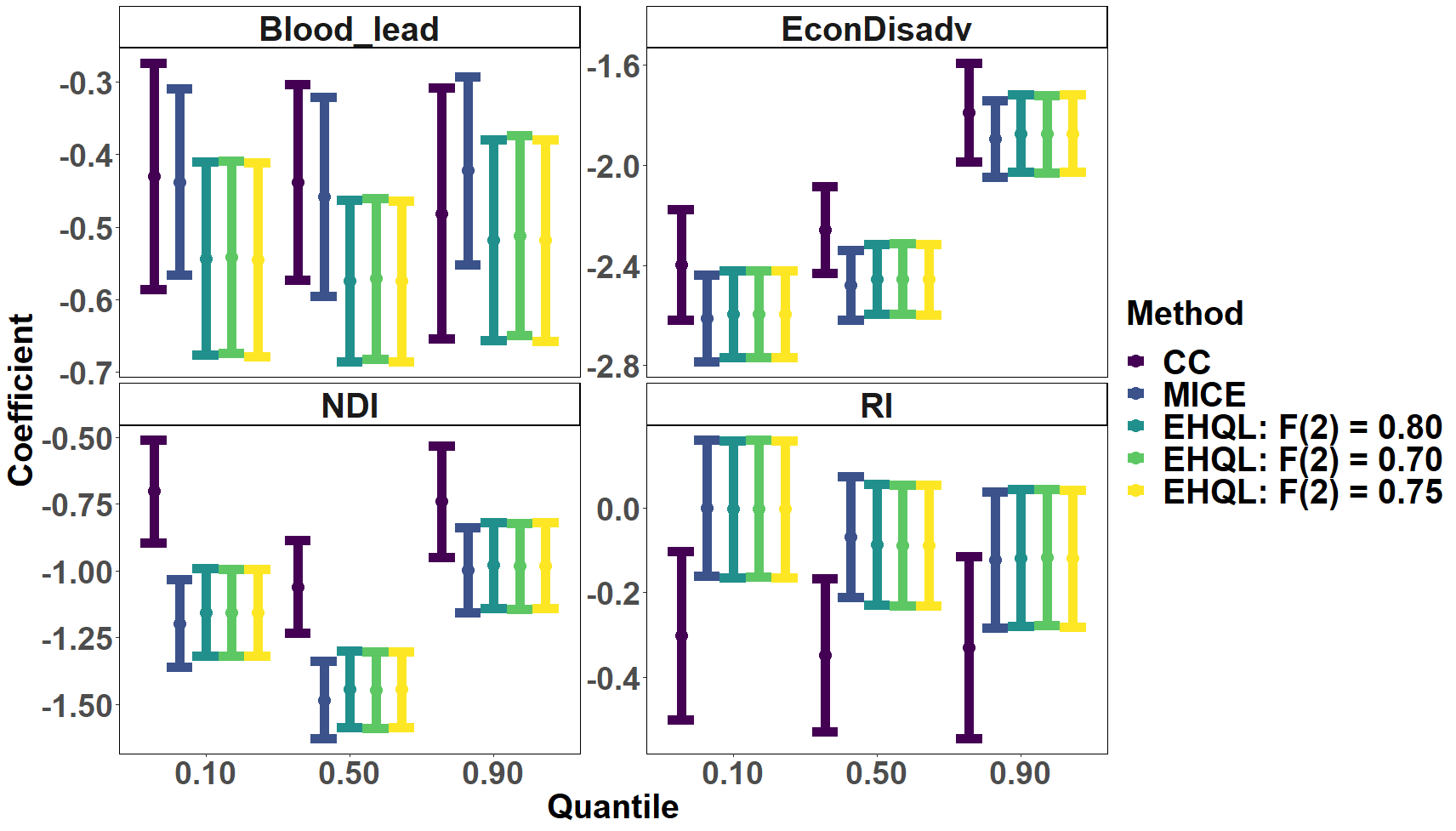 }
    \caption{Multiple imputation inferences for quantile regression coefficients for the four predictors highlighted in the main text for all quantile regressions and auxiliary quantile settings. EoG math test score is the response variable. We observe that inference is not sensitive to auxiliary quantile specifications for lead exposure under the proposed approach, with sizeable shifts between complete case (CC) inference for all coefficients across quantiles. Furthermore, for each auxiliary quantile specification and quantile regression, \vtt{Blood_lead} is more adversely associated with EoG math scores than it is in both the CC and MICE inferences.}
    \label{mathselect}
\end{figure}

\begin{figure}[h]
    \centering
    \includegraphics[width = \textwidth, keepaspectratio]{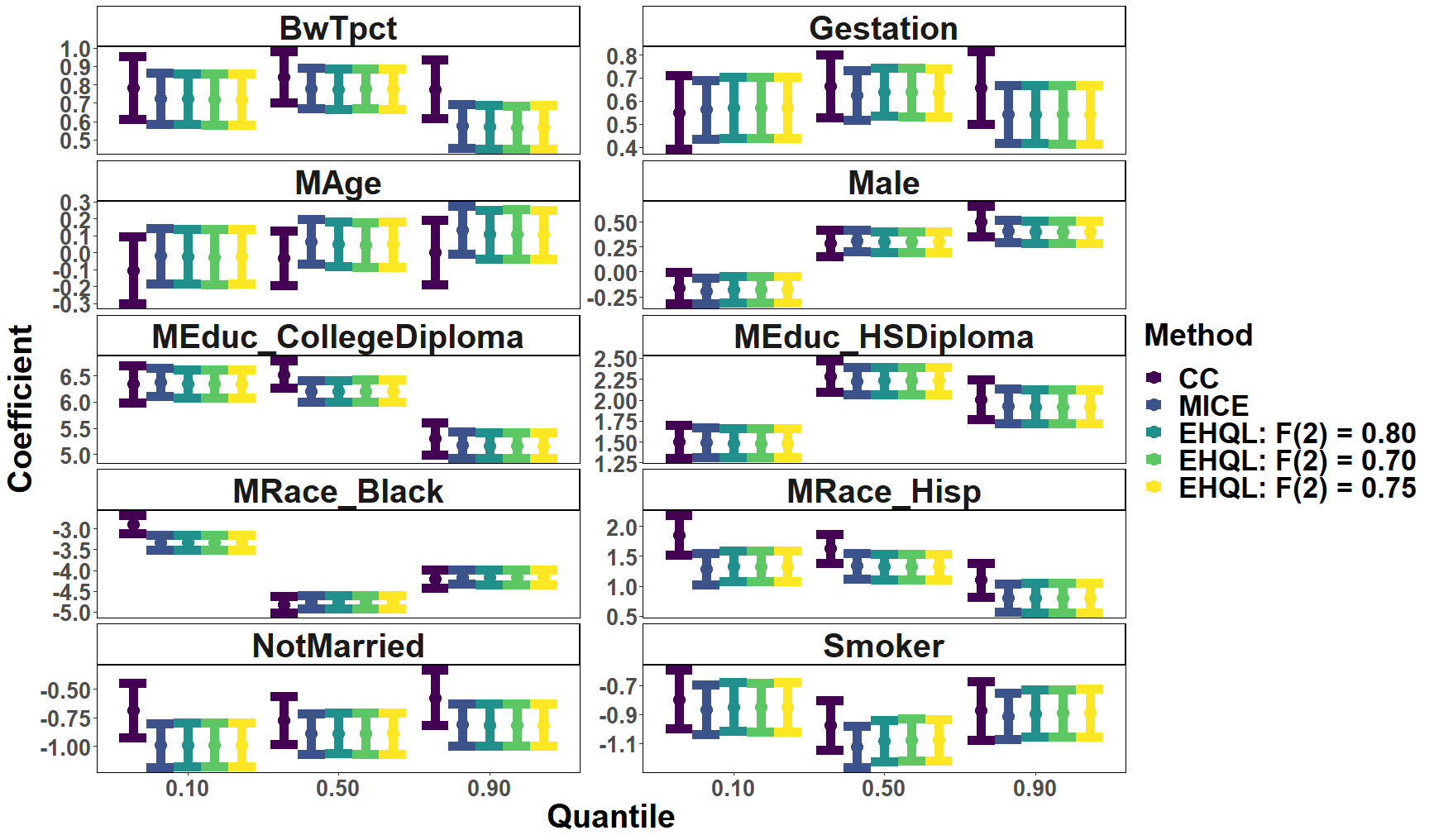}
    \caption{Multiple imputation inferences for quantile regression coefficients for the remaining covariates in the analysis of the North Carolina lead exposure data presented in the main text. EoG math test score is the response variable.}
    \label{mathrest}
\end{figure}

\begin{figure}[h]
    \centering
    \includegraphics[width = \textwidth, keepaspectratio]{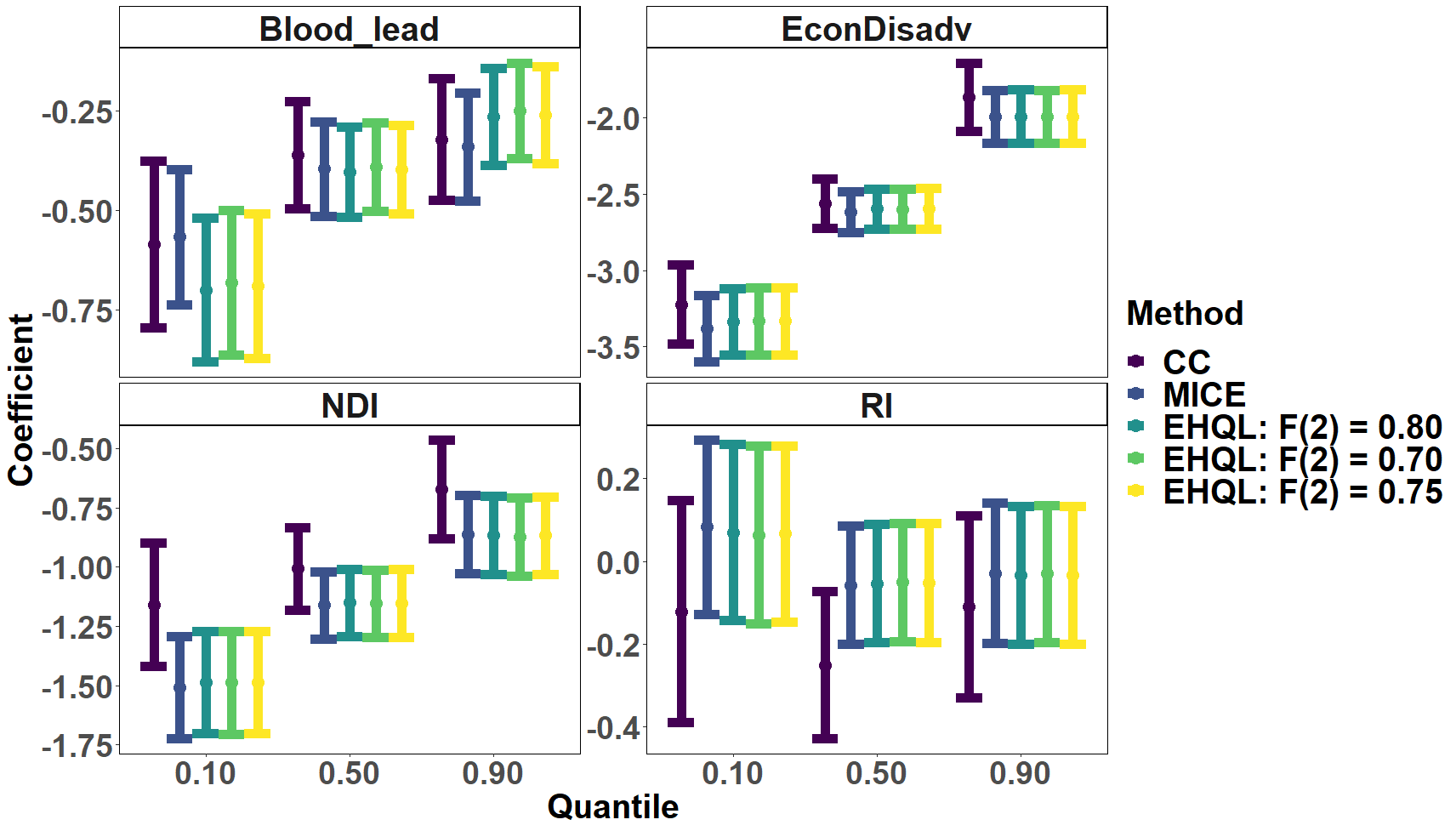}
    \caption{Multiple imputation inferences with EoG reading test score as the response variable. Results are for the quantile regression coefficients for the four predictors highlighted in the main text for all quantile regressions and auxiliary quantile settings. For EoG reading scores, the imputations under the EHQL copula suggest more heterogeneous impacts of \vtt{Blood_lead} across the distribution of reading scores. This includes estimating a more adverse impact of lead for lower scoring children. By contrast, the MICE imputations estimate the  quantile regression coefficients as closer to one another.}
    \label{readselect}
\end{figure}

\begin{figure}[h]
    \centering
    \includegraphics[width = \textwidth, keepaspectratio]{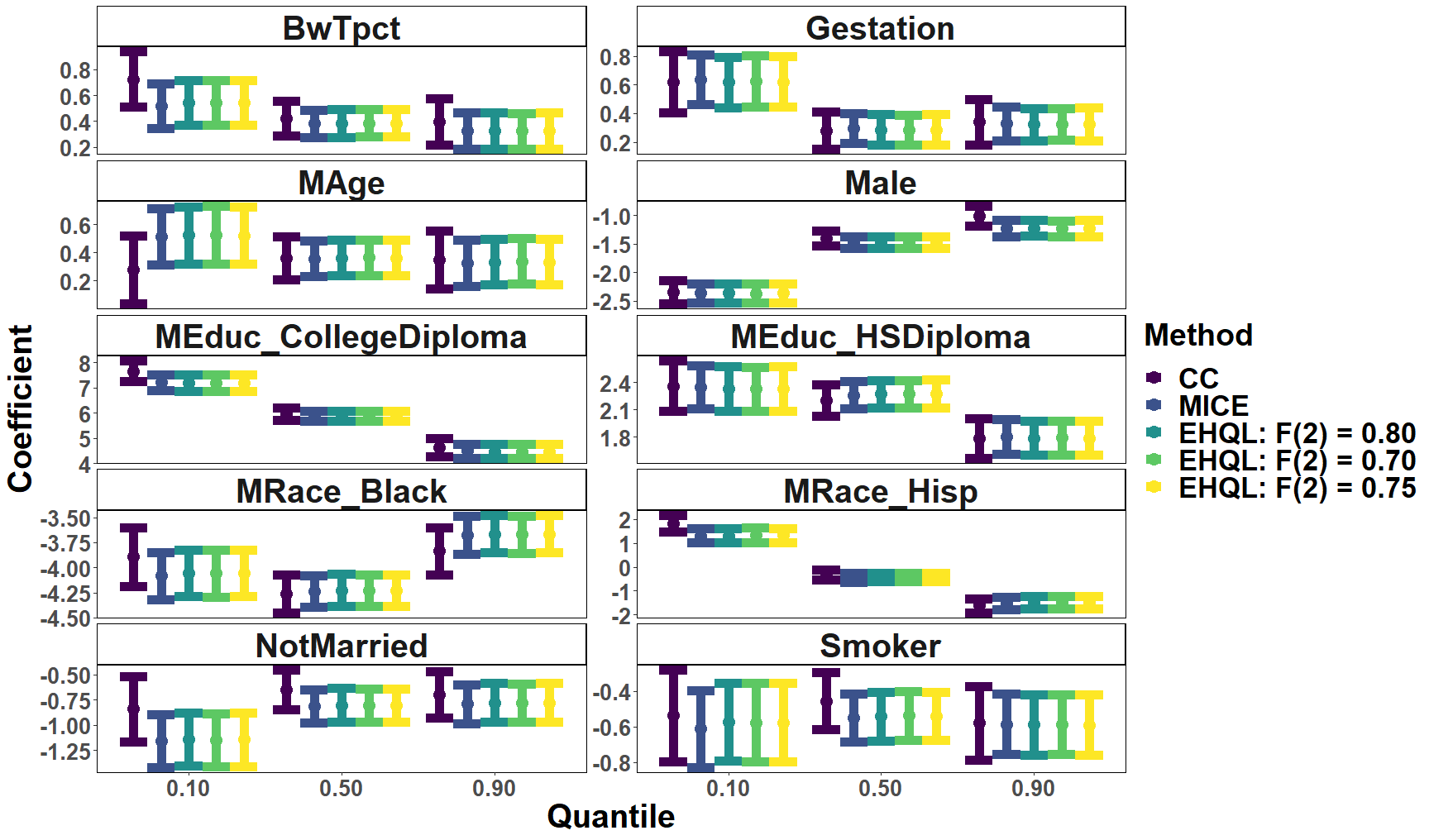}
    \caption{Multiple imputation inferences with EoG reading test score as the response variable for the remaining coefficients.}
    \label{readrest}
\end{figure}

Figures \ref{mathselect} and \ref{mathrest} provide results for the models with EoG math scores as the response, and 
Figures \ref{readselect} and \ref{readrest} provide results for the models with EoG reading scores as the response.  Across all analyses, we see little differences in the inferences for these different specifications of the auxiliary quantiles, suggesting the results are not sensitive to modest perturbations of the  auxiliary quantile specification. Accounting for the nonignorable missingness in  \vtt{Blood_lead} seems to have greater effects on the inferences for the math regressions than for the reading regressions. The effects of lead exposure appear relatively constant across quantiles for the math regressions, whereas they are more adverse at lower quantiles for the reading regressions. 

%Across quantiles and coefficients, the inferences between auxiliary specifications are virtually indistinguishable 
 %The conclusions are largley consistent with what is presented in the paper: imputing lead scores has sizeable impact on the associations between select covariates and the response variable.

%\subsection{Inference for the Remaining Covariates with EoG Math Scores as the Response}

%\subsection{Inference in the Models for EoG Reading Scores}
%We now change the response variable to EoG reading scores and perform the same inference on the target quantile regression models.

% This can easily be verified empirifcally given $\boldsymbol C_{0}$ and $\epsilon >0$, in which case one would specify a fine grid of $\delta^{*}$ between 0 and 1, multiply the off-diagonal elements of $\boldsymbol C_{0}$ by $\delta^{*}$, and compute the determinant of the resulting matrix.
\clearpage
\bibliographystyle{apalike}
\bibliography{Bib}